\documentclass[aps,prd,groupaddress,floatfix,tighten,nofootinbib,showpacs,amsfonts,superscriptaddress]{revtex4}
\usepackage{url}
\usepackage{xcolor}
\usepackage{hyperref}

\colorlet{shadecolor}{gray!15}
\definecolor{greenLinks}{rgb}{0, 0.6, 0}
\definecolor{blueLinks}{rgb}{0, 0, 0.6}
\definecolor{redLinks}{rgb}{0.6, 0, 0}
\definecolor{tempText}{rgb}{0.55, 0.10,0.67}
\definecolor{eprintLinks}{rgb}{0.4, 0.4, 0.4}
\definecolor{journalLinks}{rgb}{0.6, 0, 0}
\newcommand{\MYhref}[3][redLinks]{\href{#2}{\color{#1}{#3}}}%
\usepackage{amssymb}
\usepackage{amsmath,color}
\usepackage{epsfig}
\usepackage{graphicx,epsf,epsfig}
\usepackage{footnote}
\usepackage{multirow}
\usepackage{enumitem}
\usepackage{color}
\def\slc#1{\setbox0=\hbox{$#1$}                  
    \dimen0=\wd0                                 
    \setbox1=\hbox{/} \dimen1=\wd1               
    \ifdim\dimen0>\dimen1                        
       \rlap{\hbox to \dimen0{\hfil/\hfil}}      
       #1                                        
    \else                                        
       \rlap{\hbox to \dimen1{\hfil$#1$\hfil}}   
       /                                         
    \fi}

\def\be{\begin{equation}}
\def\ee{\end{equation}}
\def\gs{\mathrel{
   \rlap{\raise 0.511ex \hbox{$>$}}{\lower 0.511ex \hbox{$\sim$}}}}
\def\ls{\mathrel{
   \rlap{\raise 0.511ex \hbox{$<$}}{\lower 0.511ex \hbox{$\sim$}}}}

\newcommand{\onbb}{neutrinoless double beta decay }
\newcommand{\ba}{\begin{array}{c}}
\newcommand{\baz}{\begin{array}{cc}}
\newcommand{\barrr}{\begin{array}{rrr}}
\newcommand{\bad}{\begin{array}{ccc}}
\newcommand{\bav}{\begin{array}{cccc}}
\newcommand{\baf}{\begin{array}{ccccc}}
\newcommand{\bea}{\begin{equation} \begin{array}{c}}
\newcommand{\eea}{ \end{array} \end{equation}}
\newcommand{\ea}{\end{array}}

\usepackage[normalem]{ulem}
\def\lnv{lepton number violating }

\def \znbb {$\rm 0\nu\beta\beta$ }

\def\21{$\mathrm{SU(2)_L \otimes U(1)_Y}$ }

\newcommand {\ignore}[1]{}
\allowdisplaybreaks \allowdisplaybreaks[2]
\newcommand{\AddrDmp}{Department of Modern Physics, University of Science and
 Technology of China\\ Hefei, Anhui 230026, CHINA}
\newcommand{\AddrAHEP}{AHEP Group, Institut de F\'{i}sica Corpuscular --
  C.S.I.C./Universitat de Val\`{e}ncia, Parc Cientific de Paterna.\\
  C/Catedratico Jos\'e Beltr\'an, 2 E-46980 Paterna (Val\`{e}ncia) - SPAIN}

\newcommand{\AddrCINVESTAV}{Departamento de F\'{\i}sica, Centro de
 Investigaci\'on y de Estudios Avanzados del Instituto Polit\'ecnico
 Nacional, Apartado Postal 14-740, 07000 M\'exico D.F., M\'exico}

\begin{document}
\title{Classifying CP transformations according to their texture  zeros:
 \\theory and implications}
\author{Peng Chen}
 \email{pche@mail.ustc.edu.cn}
 \affiliation{\AddrDmp}
\author{Gui-Jun Ding}
 \email{dinggj@ustc.edu.cn}
 \affiliation{\AddrDmp}
\author{Felix Gonzalez-Canales}
 \email{felix.gonzalez@ific.uv.es}
 \affiliation{\AddrAHEP}\affiliation{\AddrCINVESTAV}
\author{J. W. F. Valle}
 \email{valle@ific.uv.es}
 \homepage[URL:]{http://astroparticles.es/}
 \affiliation{\AddrAHEP}

\begin{abstract}
  We provide a classification of generalized CP symmetries preserved by the
  neutrino mass matrix according to the number of zero entries in the
  associated transformation matrix. We determine the corresponding constrained
  form of the lepton mixing matrix, characterized by correlations between the
  mixing angles and the CP violating phases. We compare with the corresponding
  restrictions from current neutrino oscillation global fits and show that, in
  some cases, the Dirac CP phase characterizing oscillations is highly
  constrained. Implications for current and upcoming long baseline neutrino
  oscillation experiments T2K, NO$\nu$A and DUNE, as well as neutrinoless
  double beta decay experiments are discussed. We also study the cosmological
  implications of such schemes for leptogenesis.
\end{abstract}
\maketitle
\section{Introduction}\label{sec:introduction}
%
Non-Abelian symmetries provide an attractive framework in terms of
which to tackle the long-standing flavor problem in particle
physics~\cite{Altarelli:2010gt,ishimori2012introduction,King:2013eh,King:2014nza,King:2015aea}.
Assuming light neutrinos to be Majorana particles, as suggested on
general grounds~\cite{Schechter:1980gr}, we consider the case where
the neutrino and charged lepton mass matrices have remnant symmetries,
both of flavour and CP types. These would ultimately reflect some
unspecified flavour symmetry of the underlying gauge theory, providing
also a more general framework to describe mu-tau flavor symmetry in
neutrino physics~\cite{Chen:2015siy,Xing:2015fdg}. One can show that
flavor symmetries can be generated by performing two successive CP
transformations~\cite{Chen:2014wxa,Chen:2015nha}. Conversely the
remnant CP transformation of the lepton mass matrices can constrain
the lepton flavor mixing in a quite efficient way. In particular, the
Majorana and Dirac leptonic CP violating phases can be
predicted~\cite{Chen:2015siy,Chen:2014wxa,Chen:2015nha}. If one
remnant CP transformation ${\bf X}$ is preserved by the neutrino mass
matrix the ${\bf X}$ matrix should be symmetric and unitary. By
performing Takagi factorization we have
${\bf X} = {\bf \Sigma}{\bf \Sigma}^{T}$, where ${\bf \Sigma}$ is a unitary
matrix. Without loss of generality, we shall work in the charged lepton
diagonal basis so that the lepton flavor mixing completely arises from the
neutrino sector. The invariance of the neutrino mass matrix under the action
of $\mathbf{X}$ implies that the lepton mixing matrix $\mathbf{U}$ should
fulfill~\cite{Chen:2014wxa,Chen:2015nha,Chen:2015siy}
\begin{equation}\label{eq:cp_pmns}
 \mathbf{U}^{-1} \mathbf{X} \mathbf{U}^{\ast} =
 \text{diag}\left(\pm1, \pm1, \pm1\right)\,.
\end{equation}
Then the lepton mixing matrix is determined to be of the form:
\begin{equation}\label{eq:PMNS_master}
 \mathbf{U} = \mathbf{\Sigma} \mathbf{O}_{3\times3} \hat{ X}^{-1/2}\,,
\end{equation}
where $\hat{ X}^{-1/2}$ is a diagonal matrix with non-vanishing entries
equal to $\pm 1$ or $\pm i$, which is necessary for making neutrino masses
positive~\cite{Schechter:1980gr}. Without loss of generality, this matrix can
be parametrized as
\begin{equation}\label{eq:X_nu}
 \hat{ X}^{-1/2} = \textrm{diag} \left(1 , i^{k_1},  i^{k_2} \right) \, ,
\end{equation}
with $k_{1,2}=0, 1, 2, 3$. The Majorana phases may be changed by $\pi$
due to $\hat{\bf X}^{-1/2}$. On the other hand, the ${\bf O}_{3\times3}$ is a
generic three dimensional real orthogonal matrix, and it can be
parametrized as
\begin{equation}\label{eq:Orthogonal_matrix}
 {\bf O}_{3\times3} =
 \left( \begin{array}{ccc}
  1 &  0 & 0 \\
  0 &  \cos\theta_1  & \sin\theta_1 \\
  0 & -\sin\theta_1  & \cos\theta_1
 \end{array} \right)
 \left( \begin{array}{ccc}
  \cos\theta_2 &  0  &  \sin\theta_2 \\
  0  &  1 &  0 \\
  - \sin\theta_2  &  0  &  \cos\theta_2
 \end{array}\right)
 \left(\begin{array}{ccc}
  \cos\theta_3  &  \sin\theta_3  &  0  \\
  - \sin\theta_3  &  \cos\theta_3  &  0  \\
  0  &  0  &  1
 \end{array} \right)\,,
\end{equation}
where $\theta_{1,2,3}$ are real parameters, and a possible overall
minus sign of ${\bf O}_{3\times3}$ is omitted since it is irrelevant
to flavor mixing parameters. Therefore the lepton mixing matrix is
predicted to depend on three free parameters $\theta_{1,2,3}$ besides
the parameters characterizing the residual CP transformation
${\bf X}$. In the following, we shall classify all possible forms of the
remnant CP transformations according to the number of zero elements in
${\bf X}$ and investigate the corresponding predictions for the lepton
flavor mixing parameters and their implications for the lepton CP
violation in conventional neutrino oscillations, neutrinoless double
beta decay as well as leptogenesis.
%
\subsection{The angle-phase parametrization}
For three light Majorana neutrinos the leptonic mixing matrix can be
expressed in terms of three rotation angles and three CP violation
phases~\cite{Schechter:1980gr}. Here we will take these six
independent parameters expressed within the PDG prescription given
as~\cite{Agashe:2014kda}~\footnote{For a description in the original
  symmetric form of the lepton mixing matrix
  see~\cite{Chen:2015siy}. As discussed in~\cite{Rodejohann:2011vc}
  the symmetric presentation is more transparent in describing \znbb
  while the equivalent PDG prescription is more convenient of
  describing neutrino oscillations.}:
\begin{equation}\label{eq:symmetric_para}
{\bf U}_{\rm PDG}=\left( \begin{array}{ccc}
   c_{12} c_{13} &
   s_{12} c_{13} &
   s_{13} e^{ -i \delta_{\text{CP}} } \\
  -s_{12} c_{23} - c_{12} s_{13} s_{23} e^{i\delta_{\text{CP}}} &
   c_{12} c_{23} - s_{12} s_{13} s_{23} e^{i\delta_{\text{CP}}} &
   c_{13} s_{23}  \\
   s_{12} s_{23} - c_{12} s_{13} c_{23} e^{i\delta_{\text{CP}}} &
   -c_{12} s_{23}
   -s_{12}s_{13} c_{23} e^{i\delta_{\text{CP}}} &
   c_{13} c_{23}
 \end{array} \right)
 {\bf {\cal K}} \,,
\end{equation}
where $c_{ij}\equiv\cos\theta_{ij}$ and $s_{ij}\equiv\sin\theta_{ij}$
and ${\bf {\cal K}}$ is a diagonal matrix of phases chosen as
${\bf {\cal K} } = \text{diag} \left( 1, e^{i \alpha_{21}/2 },
e^{i \alpha_{31}/2 } \right)$. The $\delta_{\text{CP}}$ is the Dirac
CP violation phase and $\alpha_{21}$, $\alpha_{31}$ are two Majorana CP
violation phases. In this parametrization the mixing angles and magnitudes of
the entries of the ${\bf U}_{\rm PDG}$ matrix are related as:
\begin{equation}\label{eq:UU}
 \sin^{2} \theta_{13} =
  \left| \left( {\bf U}_{\rm PDG } \right)_{13} \right|^{2} \, , \quad
 \sin^{2} \theta_{12} =
  \frac{
   \left| \left( {\bf U}_{\rm PDG } \right)_{12} \right|^{2}
  }{
   1 - \left| \left( {\bf U}_{\rm PDG } \right)_{13} \right|^{2} }
 \quad \textrm{and} \quad
 \sin^{2} \theta_{23} =
  \frac{
   \left| \left( {\bf U}_{\rm PDG } \right)_{23} \right|^{2}
  }{
   1 - \left| \left( {\bf U}_{\rm PDG } \right)_{13} \right|^{2}
  } \,.
\end{equation}
The well known Jarlskog-like invariant is defined as
$J_{\rm CP}=\Im\left \{ \left( {\bf U}_{\rm PDG } \right)_{11}^{*}
\left( {\bf U}_{\rm PDG } \right)_{23}^{*} \left( {\bf U}_{\rm PDG }
\right)_{13} \left( {\bf U}_{\rm PDG } \right)_{21} \right \}$.
It describes CP violation in conventional neutrino oscillations, and takes the
form
\begin{equation}\label{INV:JCP}
 J_{\rm CP} = \frac{1}{8}
  \sin 2 \theta_{12} \, \sin 2 \theta_{23} \, \sin 2 \theta_{13}\,
  \cos\theta_{13} \,\sin \delta_{ \text{CP} } \,.
\end{equation}
This expression gives a very transparent interpretation of the Dirac
leptonic CP invariant. There are two additional rephase invariants
$I_{1} = \Im \left \{ \left( {\bf U}_{\rm PDG } \right)_{12}^{2}
\left( {\bf U}_{\rm PDG } \right)_{11}^{* 2} \right\}$ and
$I_{2} = \Im \left \{ \left( {\bf U}_{\rm PDG } \right)_{13}^{2}
\left( {\bf U}_{\rm PDG } \right)_{11}^{*2} \right\}$,
associated with the Majorana
phases~\cite{Branco:1986gr,Jenkins:2007ip,Branco:2011zb} they take
the following form
\begin{equation}\label{INV:I1-I2}
 \begin{array}{l}
  I_{1} = \frac{1}{4} \sin^{2} 2\theta_{12} \cos^{4} \theta_{13}
   \sin \alpha_{12}
  \quad \textrm{and} \quad
  I_{2} = \frac{1}{4} \sin^{2} 2 \theta_{13} \cos^{2} \theta_{12}
   \sin \alpha_{13}'\,,
 \end{array}
\end{equation}
where $\alpha_{13}'= \alpha_{13} - 2 \delta_{\text{CP}}$. These
invariants appear in \lnv processes such as \onbb which do not depend, as
expected, on the Dirac invariant $J_{\rm CP}$.
%
\section{Texture zeros of the remnant CP transformations}\label{sec:texture_zeros}
%
In order to perform the classification of the remnant CP transformations in
terms of texture zeros, it is necessary establish the way of counting these
zero entries in the ${\bf X}$ matrix: two zeros off-diagonal counts as one
texture zero, while one zero on the diagonal counts as one texture
zero~\cite{Fritzsch:1999ee}. The CP transformations are unitary-symmetric
matrices and consequently have non-zero determinant.

Hence, the ${\bf X}$ matrix cannot be represented by a matrix with six or five
zero elements. In other words, the maximum number of zero entries in the CP
transformation matrices must be four. The CP transformation matrices with four
texture zeros and the corresponding ${\bf \Sigma}$ matrices are given in the
Table~\ref{tab:four_zeros}. For this case the lepton mixing matrix is obtained
through Eq.~\eqref{eq:PMNS_master}, and the mixing parameters are given in the
fourth column of Table~\ref{tab:four_zeros}.

The type-I ${\bf X}$ matrix with four texture zeros corresponds to the
so-called $\mu-\tau$
reflection~\cite{Harrison:2002kp,Grimus:2003yn,Farzan:2006vj,Xing:2015fdg}.
It is remarkable that both the atmospheric mixing angle $\theta_{23}$
and the CP violation phase $\delta_{\text{CP}}$ are predicted to be
maximal for any values of the free parameters $\theta_{i}$ while
Majorana phases take on CP--conserving values.

For the type-II ${\bf X}$ matrix one sees that reactor and atmospheric angles
are strongly correlated each other,
\begin{equation}
 \sin^2 \theta_{23} = 1 - \tan^2 \theta_{13}, \qquad
 \sin^2 \theta_{12} = \frac{1}{2} \left( 1 + \tan^2 \theta_{13}
  \cos 2\theta_{3} \right) \,.
\end{equation}
Given the measured value of reactor angle $\theta_{13}$, we have
$\sin^2 \theta_{12} \simeq \frac{1}{2}$ and $\sin^2 \theta_{23} \simeq 1$. The
measured values of the solar and atmospheric mixing angles can
not be achieved in this case.

For the residual CP transformation with four texture zeros type-III,
the lepton mixing matrix is related to the type-II case through the
exchange of the second and third rows,
\begin{equation}
 {\bf \Sigma}_{III} =
 \begin{pmatrix}
  1 & 0 & 0 \\
  0 & 0 & 1 \\
  0 & 1 & 0
 \end{pmatrix}
 {\bf \Sigma}_{II}\,.
\end{equation}
As a consequence, except for the atmospheric angle $\theta_{23}$ and Jarlskog
$J_{\text CP}$, the expressions for the rephasing invariants and mixing angles
are the same as those obtained in the previous case. For the measured value of
$\theta_{13}$, the other two angles are
$\sin^2 \theta_{12} \simeq \frac{1}{2}$ and $\sin^2 \theta_{23} \simeq 0$. The
measured values of the three mixing angles $\theta_{12}$, $\theta_{13}$ and
$\theta_{23}$ can not be accommodated simultaneously.
\begin{table}[!htbp]\addtolength{\tabcolsep}{-2pt}
 \begin{center}
  {\footnotesize
  \begin{tabular}{|c|c|c|c| } \hline \hline
   \multicolumn{4}{|c|}{ Four Texture Zeros } \\ \hline \hline
    ~Type~ & ${\bf X}$ & ${\bf \Sigma}$ & mixing parameters \\ \hline
    \texttt{I} &
     $\left( \begin{array}{ccc}
      e^{i \alpha} & 0 & 0 \\
      0 & 0 & e^{i \beta} \\
      0 & e^{i \beta} & 0
     \end{array}\right)$ &
     $\begin{pmatrix}
      e^{i\alpha/2} & 0 & 0 \\
      0 & \frac{ e^{i \beta/2} }{ \sqrt{2} } &
       i \frac{ e^{i\beta/2}  }{ \sqrt{2} } \\
      0 & \frac{ e^{i \beta/2} }{ \sqrt{2} } &
      -i\frac{ e^{i\beta/2} }{ \sqrt{2} }
     \end{pmatrix}$  &
     $\begin{array}{c}
      \sin^2 \theta_{13} = \sin^2\theta_2,\;
      \sin^2 \theta_{23} = \frac{1}{2},\\
      \sin^2\theta_{12} = \sin^2\theta_3,\\
      \sin \delta_{CP} = \text{sign}(\sin\theta_2 \sin2\theta_3) ,\\
      \alpha_{21} = k_1\pi, \;
      \alpha^{\prime}_{31} = k_2\pi, \\
     \end{array}$  \\ \hline
    \texttt{II} &
     $\begin{pmatrix}
      0 & 0   &  e^{i\beta}\\
      0 & e^{i\alpha}    &  0  \\
      e^{i\beta}   &  0  &  0
     \end{pmatrix}$ &
     $\begin{pmatrix}
      0 & \frac{ e^{i\beta/2} }{ \sqrt{2} } &
       i \frac{ e^{i\beta/2} }{ \sqrt{2} } \\
      e^{i\alpha/2} & 0 & 0 \\
      0 & \frac{ e^{i\beta/2} }{ \sqrt{2} } &
       -i \frac{ e^{i\beta/2} }{ \sqrt{2} }
     \end{pmatrix}$ &
     $\begin{array}{c}
      \sin^2 \theta_{13} =
       \frac{1}{2} \cos^2 \theta_{2}, \;
      \sin^2 \theta_{23} =
       \frac{
        2 - 2 \cos 2\theta_{2}
       }{
        3 - \cos 2\theta_{2} }\,,\\
      \sin^2 \theta_{12} =
       \frac{1}{2} + \frac{ \cos^2 \theta_2
       \cos 2\theta_3 }{ 3 - \cos2 \theta_2 },\\
      J_{\text CP} = - \frac{1}{4} \sin \theta_2 \sin 2\theta _3
       \cos^2 \theta_2,\\
      I_1 = \frac{1}{4} (-1)^{k_1 + 1} \sin \theta_2 \sin 2\theta_3
       \cos^2 \theta_2, \\
      I_2 = \frac{1}{4} (-1)^{k_2} \sin \theta_2 \sin 2\theta_3
       \cos^2 \theta_2.
     \end{array}$  \\ \hline
     \texttt{III} &
     $\begin{pmatrix}
      0 & e^{i\beta} & 0 \\
      e^{i\beta} & 0 & 0 \\
      0 & 0 & e^{i\alpha}
     \end{pmatrix}$  &
     $\begin{pmatrix}
      0 & \frac{ e^{i\beta/2} }{\sqrt{2}} &
       i \frac{ e^{i\beta/2} }{\sqrt{2}} \\
      0 & \frac{ e^{i\beta/2} }{\sqrt{2}} &
       -i \frac{ e^{i\beta/2} }{\sqrt{2}} \\
      e^{i\alpha/2} & 0 & 0
     \end{pmatrix}$ &
     $\begin{array}{c}
      \sin^2 \theta_{13} =
       \frac{1}{2} \cos^2 \theta_{2}, \;
      \sin^2 \theta_{23} =
       \frac{ 1 + \cos2 \theta_{2} }{ 3 - \cos2 \theta_{2} },\\
      \sin^2 \theta_{12} =
       \frac{1}{2} + \frac{ \cos^2 \theta_2 \cos 2\theta_3 }{ 3
       - \cos 2\theta_2 },\\
      J_{\text CP} =
       \frac{1}{4} \sin \theta_2 \sin 2\theta_3 \cos^2 \theta_2,\\
      I_1 =
       \frac{1}{4} (-1)^{k_1 + 1} \sin \theta_2 \sin 2\theta_3
        \cos^2 \theta_2,\\
      I_2 =
      \frac{1}{4} (-1)^{k_2} \sin \theta_2 \sin 2\theta_3
       \cos^2 \theta_2.
     \end{array}$  \\ \hline \hline
  \end{tabular}
  }
 \end{center}
 \renewcommand{\arraystretch}{1.0}
 \caption{The CP transformation matrices with four texture zeros and the
  corresponding ${\bf \Sigma}$ matrices, where $\alpha$, $\beta$ and $\gamma$
  are real. The resulting lepton mixing matrix is obtained through the
  Eq.~\eqref{eq:PMNS_master}. Type-I is the conventional $\mu-\tau$ reflection
  while for type-II and -III ${\bf \Sigma}$ matrices are related by a
  permutation of the second and third row. The experimentally measured values
  of the three mixing angles $\theta_{12}$, $\theta_{23}$ and $\theta_{13}$
  can not be reproduced.}~\label{tab:four_zeros}
\end{table}
There is only one CP transformation matrix with three texture zeros that we
will denote as type-IV CP transformation matrix and whose explicit form is:
\begin{equation}
 \texttt{type-IV}:~~{\bf X} =
 \textrm{diag} \left( e^{i \alpha}, e^{i\beta}, e^{i \gamma} \right) \, ,
\end{equation}
with $\alpha, \beta, \gamma\in \mathbb{R}$. Its Takagi factorization reads as
${\bf \Sigma}
= \textrm{diag} \left( e^{i \alpha/2}, e^{i \beta/2}, e^{i \gamma/2} \right)$.
The lepton mixing parameters are given by
$\sin^2 \theta_{13} = \sin^2 \theta_{2}$, $\sin^2 \theta_{12} =
\sin^2 \theta_{3}$, $\sin^2 \theta_{23} = \sin^2 \theta_{1}$, and
$\sin \delta_{CP} = \sin\alpha_{21} = \sin \alpha_{31} = 0$. We see that all
the three CP phases are zero or $\pi$ in this case while the lepton mixing
angles are unconstrained.

The CP transformation matrices with two texture zeros and the corresponding
${\bf \Sigma}$ matrices are given in Table~\ref{tab:two_zeros}. The explicit
form of the lepton matrix can be obtained from Eq.~\eqref{eq:PMNS_master} and
the mixing parameters are given in the fourth column of
Table~\ref{tab:two_zeros}.
\begin{table}[!htbp] \footnotesize \addtolength{\tabcolsep}{-2pt}
 \begin{center}
  \begin{tabular}{|c|c|c|c| } \hline \hline
   \multicolumn{4}{|c|}{Two Texture Zeros} \\ \hline \hline
   ~Type~ & ${\bf X}$ & ${\bf \Sigma}$ & mixing parameters \\ \hline
   \texttt{ \uppercase \expandafter{\romannumeral5} } &
   $\begin{pmatrix}
    e^{i\alpha} & 0 & 0 \\
    0 &  e^{i\beta} \cos \Theta &
    i e^{ i ( \beta + \gamma )/2 } \sin \Theta \\
    0 & ie^{i ( \beta + \gamma )/2 } \sin \Theta &
    e^{i\gamma} \cos \Theta
   \end{pmatrix}$ &
   $\begin{pmatrix}
    e^{\frac{i \alpha}{2}} & 0 & 0 \\
    0 & e^{\frac{i\beta}{2}} \cos \frac{\Theta}{2} &
     i e^{\frac{i\beta}{2}} \sin \frac{\Theta}{2} \\
    0 & i e^{\frac{i\gamma}{2}} \sin \frac{\Theta}{2} &
     e^{\frac{i\gamma}{2}} \cos \frac{\Theta}{2}
   \end{pmatrix}$ &
   $\begin{array}{c}
    \sin^2 \theta_{13} = \sin^2 \theta_{2},\;
    \sin^2 \theta_{12} = \sin^2 \theta_{3},\\
    \sin^2 \theta_{23} = \frac{1}{2} \left( 1 - \cos \Theta
     \cos 2\theta_1 \right), \\
    \sin\delta_{\text CP} = \frac{ \text{sign} \left( \sin \theta_{2}
     \sin 2\theta_{3} \right) \sin \Theta }{
     \sqrt{ 1 - \cos^2 \Theta \cos^2 2\theta_{1} } },\\
    \alpha_{21} = k_1 \pi, \; \alpha^{\prime}_{31} = k_2\pi,\\
   \end{array}$ \\ \hline
   \texttt{\uppercase\expandafter{\romannumeral6}} &
   $\begin{pmatrix}
    e^{i\alpha} \cos \Theta & 0 &
     i e^{i ( \alpha + \gamma )/2 } \sin \Theta \\
    0 & e^{i\beta} & 0 \\
    i e^{ ( \alpha + \gamma )/2 } \sin \Theta & 0 &
     e^{i\gamma} \cos \Theta
   \end{pmatrix}$ &
   $\begin{pmatrix}
    0 & e^{ \frac{i\alpha}{2} } \cos \frac{\Theta}{2} &
     i e^{ \frac{i\alpha}{2} } \sin \frac{\Theta}{2} \\
    e^{ \frac{i\beta}{2} } & 0 & 0 \\
    0 & i e^{ \frac{i\gamma}{2} } \sin \frac{\Theta}{2} &
     e^{\frac{i\gamma}{2}} \cos \frac{\Theta}{2}
   \end{pmatrix}$ & See
   Eqs.~(\ref{eq:mixing_angles_two_zeroBII},~\ref{eq:invariants_two_zeroBII})
    \\ \hline
   \texttt{\uppercase\expandafter{\romannumeral7}} &
   $\begin{pmatrix}
    e^{i\alpha} \cos \Theta & i e^{i( \alpha + \beta )/2} \sin \Theta & 0 \\
    ie^{i( \alpha + \beta )/2} \sin \Theta & e^{i\beta} \cos \Theta & 0 \\
    0 & 0 & e^{i\gamma}
   \end{pmatrix}$ &
   $\begin{pmatrix}
    0 & e^{\frac{i\alpha}{2}} \cos \frac{\Theta}{2} &
     i e^{\frac{i\alpha}{2}} \sin \frac{\Theta}{2} \\
    0 & ie^{\frac{i\beta}{2}} \sin \frac{\Theta}{2} &
    e^{\frac{i\beta}{2}} \cos\frac{\Theta}{2} \\
    e^{\frac{i\gamma}{2}} & 0 & 0
   \end{pmatrix}$ &
   --- \\ \hline \hline
  \end{tabular}
 \end{center} \renewcommand{\arraystretch}{1.0}
 \caption{The CP transformation matrices with two texture zeros and
   the corresponding ${\bf \Sigma}$ matrices. The predictions for the
   lepton mixing parameters are displayed in the last column. The lepton
   mixing matrices for type-VI and -VII are related through the exchange
   of the second and third rows.\label{tab:two_zeros}}
\end{table}
The type-V CP transformation with with two texture zeros is exactly the
generalized $\mu-\tau$ reflection symmetry~\cite{Chen:2015siy}. The
atmospheric angle and the Dirac CP phase $\delta_{CP}$ only depend on two
parameters $\theta_1$ and $\Theta$, as they are related by
\begin{equation}
 \sin^2 \delta_{\text CP} \sin^2 2\theta_{23} = \sin^2 \Theta\,.
\end{equation}
Moreover the Majorana phases $\alpha_{21}$ and $\alpha_{31}$ take on
CP--conserving values. The phenomenological implications of this
interesting mixing pattern for neutrinoless double beta decay as well as
conventional neutrino oscillations have been discussed in detail in
Ref.~\cite{Chen:2015siy}.

\begin{figure}[!htbp]
 \begin{center}
  \begin{tabular}{cc} \hskip-0.6in
   \includegraphics[width=0.40\linewidth]{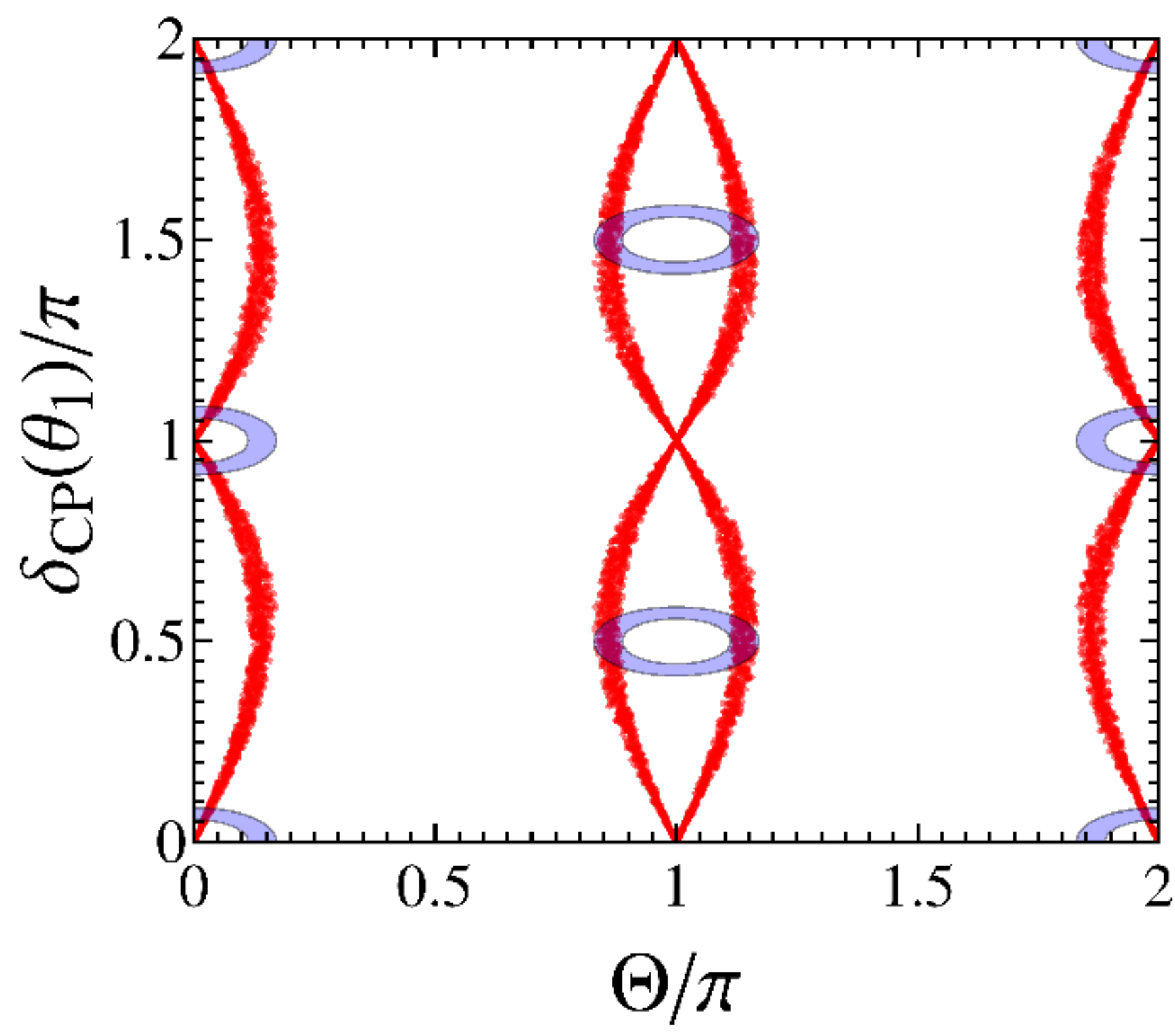}  
  \end{tabular}
  \caption{\label{fig:correlation_par} The correlations between the
   parameters $\Theta$, $\delta_{\text{CP}}$ (red points) and $\theta_1$
   (blue rings) for the case of type-VI CP transformation, where the three
   lepton mixing angles are required to lie in the experimentally preferred
   $3\sigma$ regions.}
 \end{center}
\end{figure}

For the ${\bf X}$ matrix with two texture zeros of type-VI, the mixing
angles are given by
\begin{eqnarray}
 \nonumber &&
 \sin^2 \theta_{13} =
  \frac{1}{2} \left( 1 - \cos \Theta \cos 2\theta_1 \right)
  \cos^2 \theta_2,\quad
 \sin^2 \theta_{23} =
  \frac{ 2 \sin^2 \theta_2 }{ 2 - \left( 1 - \cos \Theta \cos 2\theta_1
  \right) \cos^2 \theta_2 }\,,\\ \label{eq:mixing_angles_two_zeroBII}
 && \sin^2 \theta_{12} = \frac{ ( 1 + \cos \Theta \cos 2\theta_1 )
  \cos^2 \theta_3 + \sin \theta_2 \left[ ( 1 - \cos \Theta \cos 2\theta_1)
  \sin \theta_2 \sin^2 \theta_3 - \cos \Theta \sin 2\theta_1 \sin 2\theta_3
  \right] }{ 2 - \left( 1 - \cos \Theta \cos 2\theta_1 \right) \cos^2 \theta_2
  } \,.
\end{eqnarray}
We easily see that the following sum rules are fulfilled,
\begin{eqnarray}
 \nonumber &&
 \sin^2 \theta_{23} \cos^2 \theta_{13} = \sin^2 \theta_{2} \,, \qquad
 \frac{ \cos^2 \theta_{13} \cos^2 \theta_{23} - \sin^2 \theta_{13} }{ 1 -
  \sin^2 \theta_{23} \cos^2\theta_{13} } = \cos \Theta \cos 2\theta_1 \,, \\
 \nonumber &&
 \sin^2 \theta_{12} \left( 1 - \sin^2 \theta_{23} \cos^2 \theta_{13} \right)
  - \cos^2 \theta_{23} \cos^2 \theta_3 - \sin^2 \theta_{13}
  \sin^2 \theta_{23} \sin^2 \theta_3 \\ \label{eq:correlation_two_zeroBII}
 && = \pm \frac{1}{2} \sqrt{ 1 - \sin^2 \theta_{23} \cos^2 \theta_{13} }
  \left( \frac{ \cos 2\theta_{13} }{ \cos^2 \theta_{13} } - \sin^2 \theta_{23}
  \right) \tan 2 \theta_1 \sin 2 \theta_3 \,.
\end{eqnarray}
The Jarlskog-like invariant and the invariants associated with the Majorana
phases take the following form
\begin{eqnarray}
 \nonumber &&
 J_{\text{CP}} = - \frac{1}{4} \sin \Theta \sin \theta_2 \cos^2 \theta_2
  \sin 2\theta_3 \, , \\
 \nonumber &&
 I_1 = \frac{1}{4} (-1)^{ k_1 + 1 } \sin \Theta \sin \theta_2 \left[
  2 \cos \Theta \sin 2\theta_1 \sin \theta_2 \cos 2\theta_3 + \left(
  \cos^2 \theta_2 + \cos \Theta \cos2 \theta_1 ( 1 + \sin^2 \theta_2 )
  \right) \sin 2\theta_3 \right] \,,\\ \label{eq:invariants_two_zeroBII} &&
 I_2 = \frac{1}{2} (-1)^{k_2} \sin \Theta \cos^2 \theta_2 \sin \theta_3
  \left[ ( 1 - \cos \Theta \cos 2\theta_1 ) \sin \theta_2 \cos \theta_3
  + \cos \Theta \sin 2\theta_1 \sin \theta_3 \right]\,.
\end{eqnarray}
Notice that the three phases $\alpha$, $\beta$ and $\gamma$ can be absorbed
into the charged lepton fields, therefore they are unphysical and hence do not
appear in the above mixing parameters. The correlations between the mixing
parameters $\Theta$, $\delta_{\rm{CP}}$ and $\theta_1$ are displayed in
Fig.~\ref{fig:correlation_par}.
The blue rings are allowed parameter
regions of $\Theta$ and $\theta_1$ predicted by
Eq.~\eqref{eq:correlation_two_zeroBII},
where $\theta_{13}$ and $\theta_{23}$ are compatible with the preferred values
from the neutrino oscillations global fit in~\cite{Forero:2014bxa} at the
$3\sigma$ level. The red points are obtained from a random numerical scan over
the parameters $\Theta$ and $\theta_{1,2,3}$. We can see the allowed regions
of $\Theta$ from the two different approaches are compatible with each other.

The lepton mixing matrix corresponding to the type-VII ${\bf X}$
matrix with two texture zeros, is related to the one of type-VI by the
exchange of the second and the third rows. As a result, the lepton mixing
parameters for the type-VII case are the same as those of type-VI except that
$\theta_{23}$ and $\delta_{\text CP}$ becomes $\pi/2 - \theta_{23}$ and
$\pi + \delta_{\text CP}$ respectively.  A detailed analysis of theoretical
predictions of the CP transformation matrices with two texture zeros will be
given in the next section.

Finally, the CP transformation matrices with one texture zero and the
corresponding ${\bf \Sigma}$ matrices are given in Table~\ref{tab:one_zero}.
The corresponding expressions for the mixing matrix are obtained by using
Eq.~\eqref{eq:PMNS_master} and the mixing parameters can be extracted
straightforwardly.
\begin{table}[hptb] \addtolength{\tabcolsep}{-2pt}
 {\footnotesize
 \begin{center}
  \begin{tabular}{|c|c|c| } \hline \hline
   \multicolumn{3}{|c|}{One Texture Zero} \\ \hline \hline
   ~Type~  & ${\bf X}$ & ${\bf \Sigma}$ \\ \hline
   \texttt{\uppercase\expandafter{\romannumeral8}} &
   $\begin{pmatrix}
    0 & e^{i\alpha} c_\Theta & e^{i\beta} s_\Theta \\ \vspace{2mm}
    e^{i\alpha} c_\Theta & e^{i\gamma} s^2_\Theta &
     -e^{i( -\alpha + \beta + \gamma) } c_\Theta s_\Theta  \\
    e^{i\beta} s_\Theta &
     -e^{i( -\alpha + \beta + \gamma) } c_\Theta s_\Theta    &
     e^{i(-2 \alpha + 2\beta + \gamma) } c^2_\Theta
   \end{pmatrix}$ &
   $\frac{1}{\sqrt{2}}
    \begin{pmatrix}\vspace{2mm}
     - i e^{ i\left(\alpha - \gamma/2 \right) } &
      e^{ i\left(\alpha - \gamma/2 \right) } & 0 \\ \vspace{2mm}
     i e^{ i \gamma/2  } c_\Theta &
      e^{i\gamma/2}c_\Theta &
      \sqrt{2} e^{ i\gamma/2 } s_\Theta \\
     i e^{ i \left( - \alpha + \beta + \gamma/2 \right)} s_\Theta &
      e^{ i \left( -\alpha +\beta + \gamma/2 \right)} s_\Theta &
      - \sqrt{2} e^{i \left( -\alpha +\beta + \gamma/2 \right) }
      c_\Theta
   \end{pmatrix}$ \\ \hline
   \texttt{\uppercase\expandafter{\romannumeral9}} &
   $\begin{pmatrix}
    e^{i\alpha}c^2_\Theta & e^{i\beta} s_\Theta &
     e^{i\gamma} c_\Theta s_\Theta \\
    e^{i\beta} s_\Theta & 0 &
     - e^{ i( - \alpha + \beta + \gamma ) } c_\Theta \\
    e^{i\gamma} c_\Theta s_\Theta &
     - e^{i( -\alpha + \beta + \gamma) } c_\Theta &
     e^{i( -\alpha + 2\gamma ) } s^2_\Theta
   \end{pmatrix}$ &
   $\frac{1}{\sqrt{2}}
   \begin{pmatrix}
    -i e^{i\alpha/2}s_\Theta & - e^{i\alpha/2} s_\Theta &
     \sqrt{2} e^{ i\alpha/2 } c_\Theta \\
    i e^{ i \left( \beta - \alpha/2 \right)} &
     - e^{i \left( \beta - \alpha/2 \right)} &  0 \\
    i e^{ i \left( \gamma - \alpha/2 \right)} c_\Theta &
     e^{ i \left( \gamma - \alpha/2 \right) } c_\Theta &
    \sqrt{2}e^{i\left( \gamma - \alpha/2 \right)} s_\Theta
   \end{pmatrix}$ \\ \hline
   \texttt{\uppercase\expandafter{\romannumeral10}} &
   $\begin{pmatrix}
    e^{i\alpha} c^2_\Theta & e^{i\gamma} c_\Theta s_\Theta &
     e^{i\beta} s_\Theta \\
    e^{i\gamma} c_\Theta s_\Theta &
     e^{i( -\alpha + 2\gamma) } s^2_\Theta &
     - e^{ i( - \alpha + \beta + \gamma ) } c_\Theta  \\
    e^{i\beta} s_\Theta &
     - e^{ i( - \alpha + \beta + \gamma ) } c_\Theta & 0
   \end{pmatrix}$ &
   $\frac{1}{\sqrt{2}}
   \begin{pmatrix}
    -ie^{i\alpha/2} s_\Theta &
     -e^{i\alpha/2}s_\Theta &
     \sqrt{2} e^{ i\alpha/2 } c_\Theta \\
    ie^{i \left( \gamma - \alpha/2 \right) } c_\Theta &
     e^{i \left( \gamma - \alpha/2 \right)} c_\Theta &
     \sqrt{2} e^{i \left( \gamma - \alpha/2 \right)} s_\Theta \\
    ie^{i \left( \beta - \alpha/2 \right) } &
    - e^{ i \left( \beta - \alpha/2 \right) } & 0
   \end{pmatrix}$ \\ \hline\hline
  \end{tabular}
 \end{center} } \renewcommand{\arraystretch}{1.0}
 \caption{The CP transformation matrices with one texture zeros and the
  corresponding ${\bf \Sigma}$ matrices with $c_{\Theta} \equiv \cos \Theta$
  and $s_{\Theta} \equiv \sin \Theta$. The lepton mixing matrix is obtained
  through Eq.~\eqref{eq:PMNS_master}. The lepton mixing matrix for type-IX and
  type-X are related by the exchange of the second and third rows.
  \label{tab:one_zero}}
\end{table}
For the type-VIII CP transformation with one zero element, the mixing
angles are given by:
\begin{eqnarray} \nonumber
 \sin^2 \theta_{13} & = & \frac{1}{8} \left( 3 - \cos 2\theta_1 - 2
  \cos^2 \theta_1 \cos 2\theta_2 \right),\\
 \nonumber
 \sin^2 \theta_{12} & = &\frac{ 4 \cos^2 \theta_1 \cos^2 \theta_3 - 2
  \sin 2\theta_1 \sin \theta_2 \sin 2 \theta_3 + 4 \left(
  \sin^2 \theta_1 \sin^2 \theta_2 + \cos^2 \theta_2 \right) \sin^2 \theta_3
 }{
  5 + \cos 2\theta_1 + 2 \cos^2 \theta_1 \cos 2\theta_2 },   \\
 \label{eq:mix_par_10I}
 \sin^2 \theta_{23} & = & \frac{ 4 \cos^2 \Theta \left( \sin^2 \theta_2 +
  \sin^2 \theta_1 \cos^2 \theta_2 \right) + 8 \cos^2 \theta_1 \cos^2 \theta_2
  \sin^2 \Theta + 2 \sqrt{2} \sin 2\theta_1 \cos^2 \theta_2 \sin 2\Theta
 }{
  5 + \cos 2\theta_1 + 2 \cos^2 \theta_1 \cos 2\theta_2 }\,.
\end{eqnarray}
The smallness of $\theta_{13}$ requires $\theta_{1} \simeq 0, \pi$,
and $\theta_{2} \simeq 0, \pi$. Consequently, the solar mixing angles
would be $\sin^2\theta_{12}\simeq\frac{1}{2}$ which is outside the
experimentally allowed ranges~\cite{Forero:2014bxa}. As a result, the
measured values of $\theta_{12}$ and $\theta_{13}$ can not be
accommodated simultaneously in this case, and this mixing pattern is
not viable. This observation is indeed confirmed in our numerical analysis.
\begin{figure}[hptb]
 \begin{center}
 \begin{tabular}{cc} \hskip-0.6in
  \includegraphics[width=0.36\linewidth]{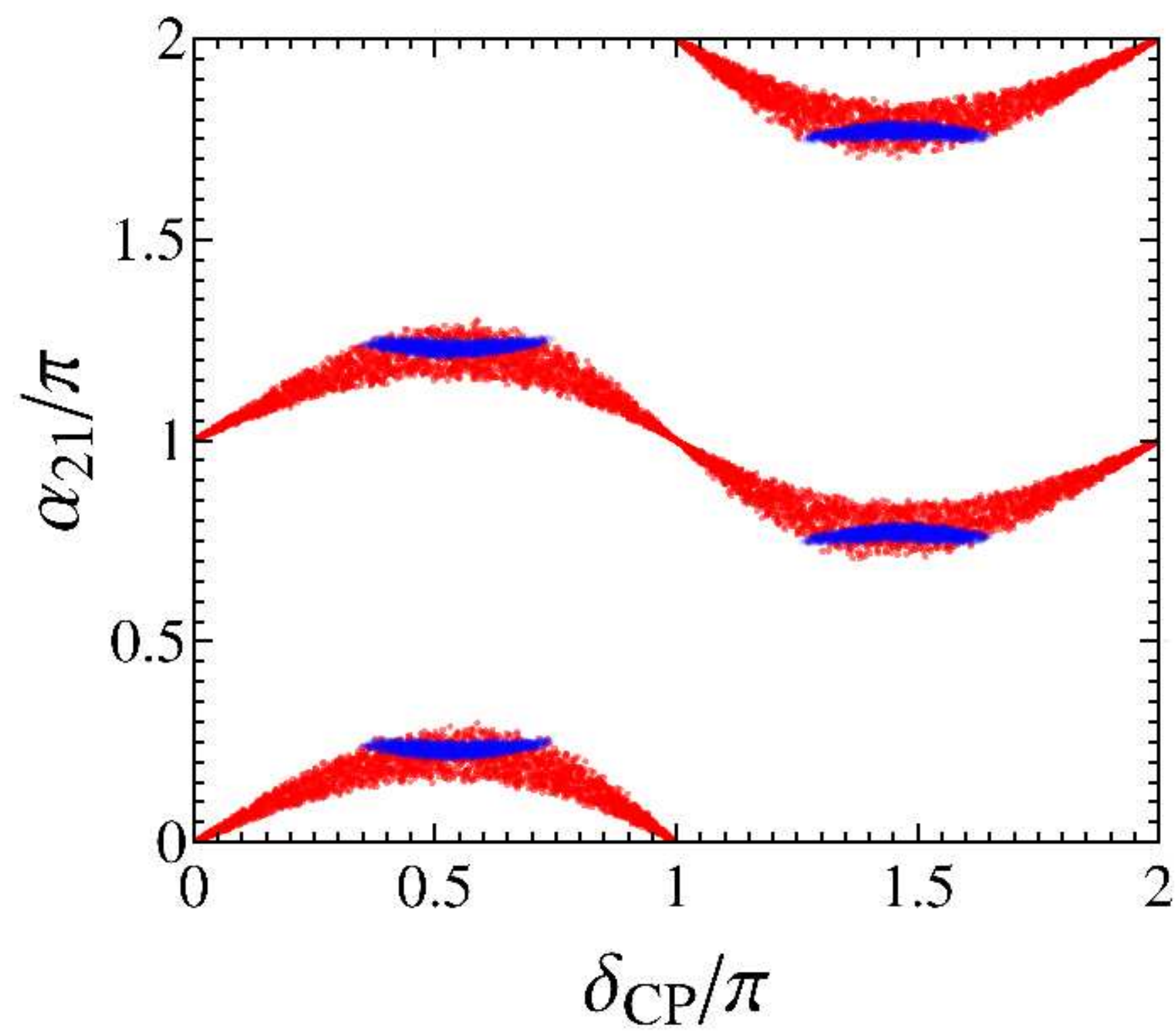} &
  \includegraphics[width=0.36\linewidth]{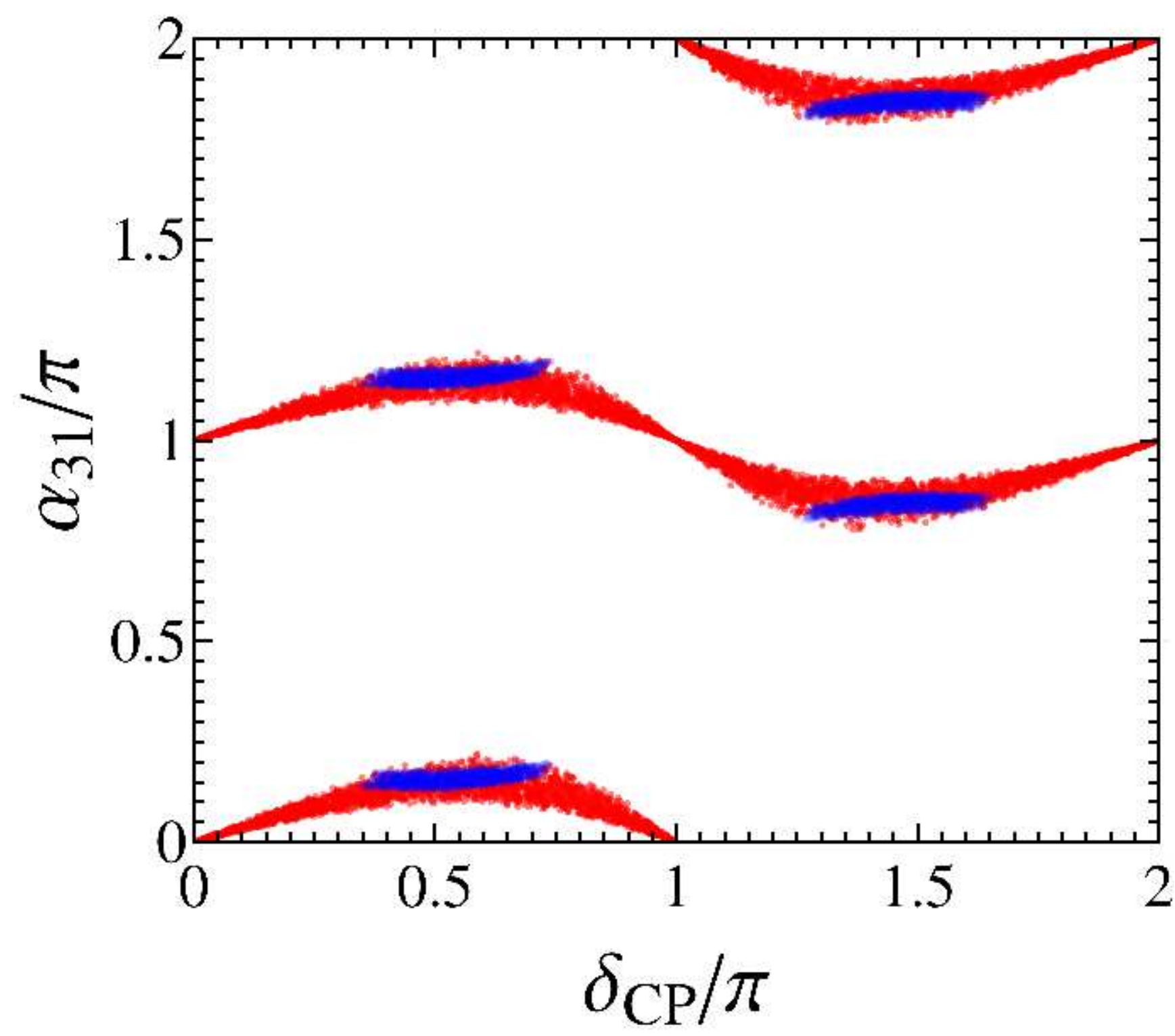} \\  \hskip-0.6in
  \includegraphics[width=0.36\linewidth]{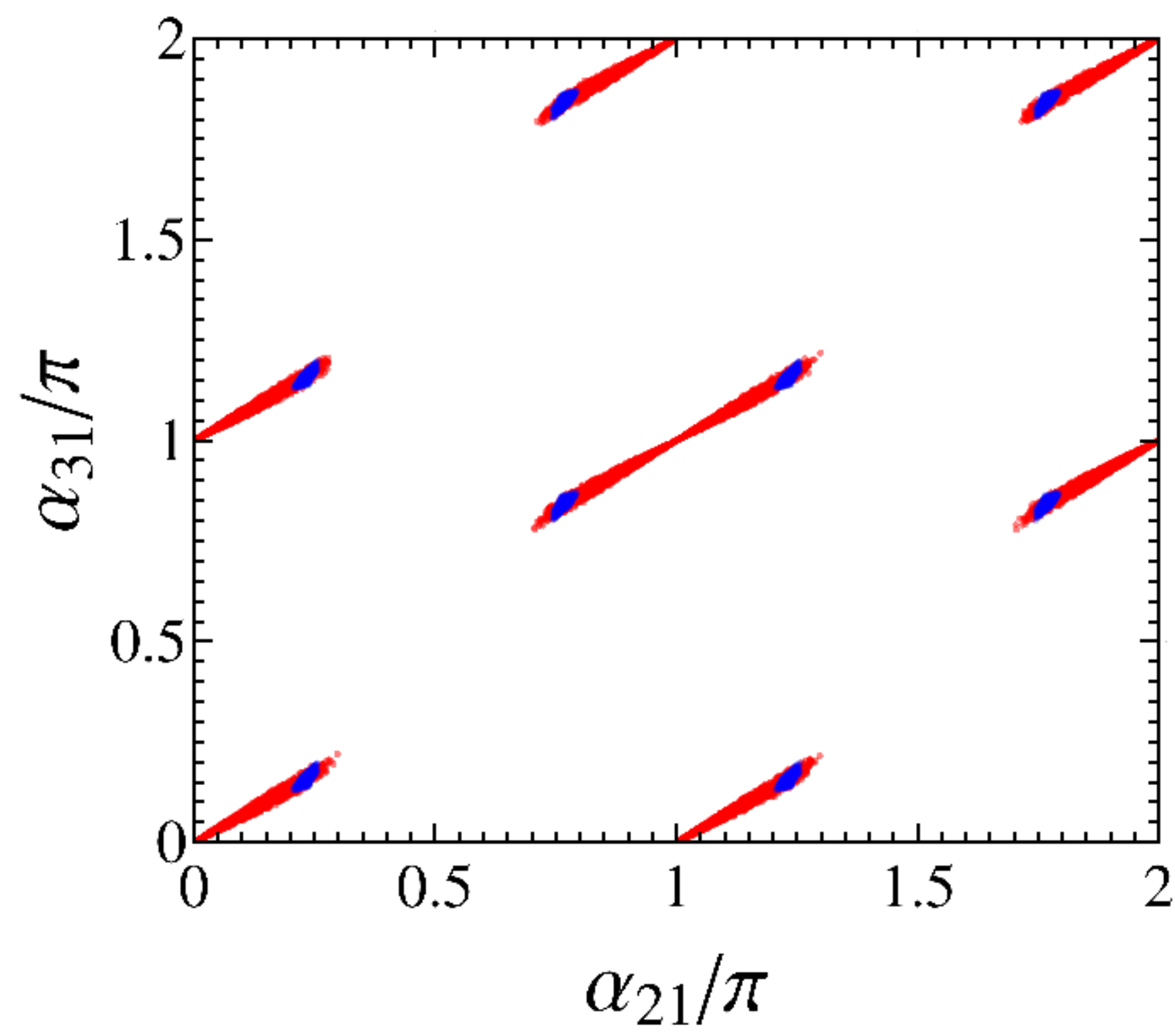} &
  \includegraphics[width=0.36\linewidth]{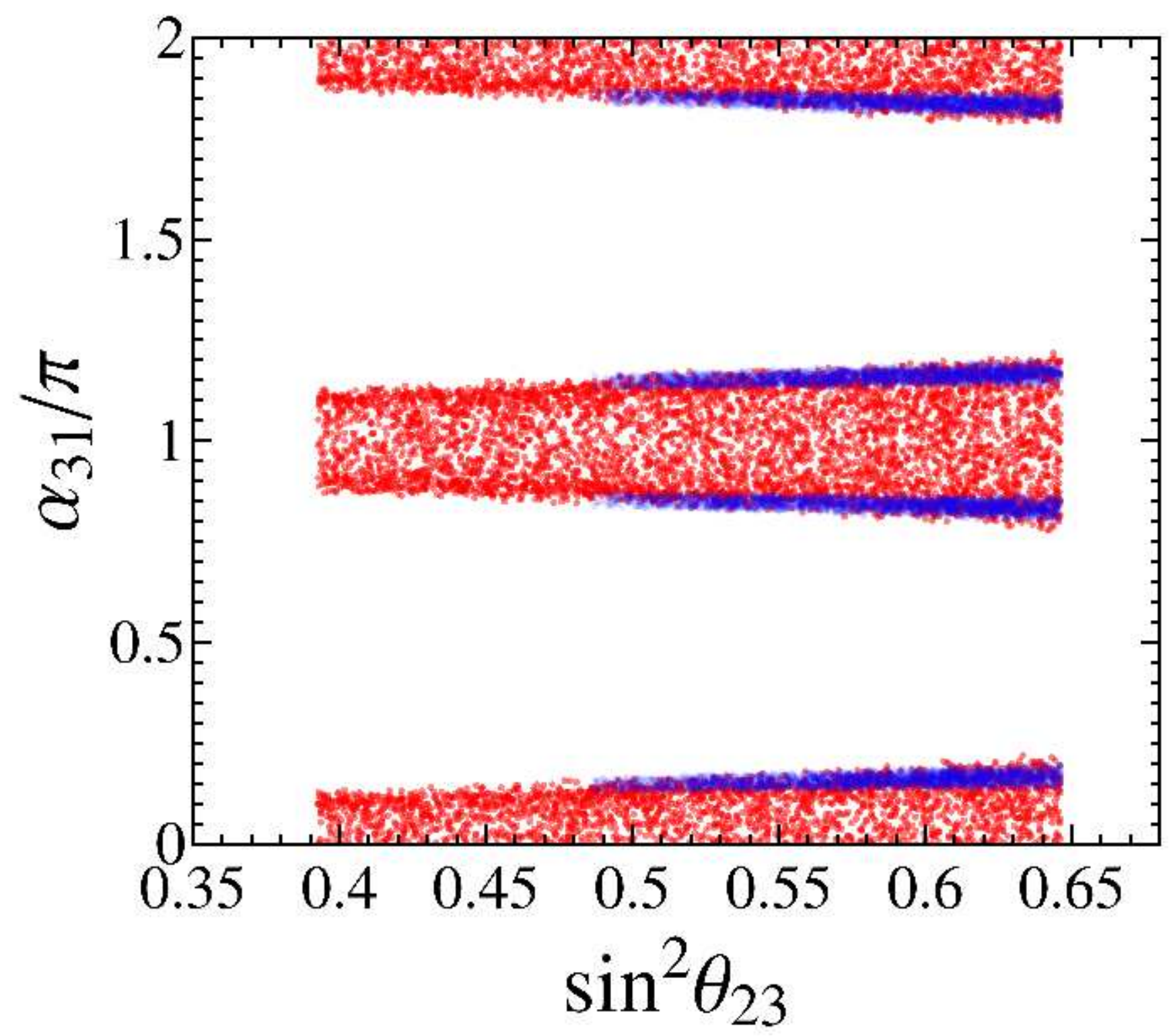}
 \end{tabular}
 \caption{\label{fig:correlation_two_zero2}The correlation between the
  mixing parameters predicted in the case of type-VI residual CP
  transformation. All the parameters $\theta_{1,2,3}$ and $\Theta$ are
  taken to be random numbers in the interval of $[0, 2\pi]$. The three
  mixing angles $\theta_{12}$, $\theta_{13}$ and $\theta_{23}$ are
  required to be within the experimentally preferred $3\sigma$
  intervals~\cite{Forero:2014bxa}. The blue points indicate the
  numerical results obtained by fixing $\Theta=\frac{\pi}{7}$ as a
  benchmark example}.
 \end{center}
\end{figure}

For the type-IX remnant CP transformation with one zero, we find the
lepton mixing angles are
\begin{eqnarray} \nonumber
 \sin^2 \theta_{13} & = & \frac{1}{4} \left[ 1 + \cos^2 \theta_1
  \cos^2 \theta_2 + \left( 3 \cos^2 \theta_1 \cos^2 \theta_2 - 1 \right)
  \cos 2\Theta - \sqrt{2} \sin 2\theta_1 \cos^2 \theta_2 \sin 2\Theta
  \right], \\ \nonumber
 \sin^2 \theta_{12} & = & 2 \Big[ \left( \sqrt{2} \cos \Theta \left(
  \sin \theta_1 \cos \theta_3 + \cos \theta_1 \sin \theta_2 \sin \theta_3
  \right) + \sin \Theta \left( \cos \theta_1 \cos \theta_3 - \sin \theta_1
  \sin \theta_2 \sin \theta_3 \right) \right)^2 \\ \nonumber
 & & \hskip-1.5cm
  + \cos^2 \theta_2 \sin^2 \theta_3 \sin^2 \Theta \Big] \Big/
  \left[ 3 - \cos^2 \theta_1 \cos^2 \theta_2 + \left( 1 - 3
  \cos^2 \theta_1 \cos^2 \theta_2 \right) \cos 2\Theta + \sqrt{2}
  \sin 2\theta_1 \cos^2 \theta_2 \sin 2\Theta \right],\\
  \label{eq:mix_par_10II}
 \sin^2\theta_{23} & = &
  \frac{ 2 \left( \sin^2 \theta_2 + \sin^2 \theta_1 \cos^2 \theta_2 \right)
  }{
   3 - \cos^2 \theta_1 \cos^2 \theta_2 + \left( 1 - 3 \cos^2 \theta_1
   \cos^2 \theta_2 \right) \cos 2\Theta + \sqrt{2} \sin 2\theta_1
   \cos^2 \theta_2 \sin 2\Theta },
\end{eqnarray}
Also, we see that the mixing angles are correlated as follows,
\begin{eqnarray} \nonumber
 & & 2 \cos^2 \theta_{13} \cos^2 \theta_{23} = 1 - \cos^2 \theta_1
  \cos^2 \theta_2 \,, \\
 & & \cos 2\theta_{13} = 2 \cos^2 \theta_{13} \cos^2 \theta_{23}
  \cos^2 \Theta + ( 1 - 2 \cos^2 \theta_{13} \cos^2 \theta_{23} )
  ( \sqrt{2} \tan \theta_1 \sin 2\Theta - \cos 2\Theta )\,.
\end{eqnarray}
The Jarlskog-like invariant associated with the Dirac CP phase has the
form
\begin{eqnarray} \nonumber
 & & J_{\text CP} = \frac{1}{16} \cos \theta_1 \cos \theta_2 \left[ 4
  \sin 2\theta_1 \sin \theta_2 \cos 2\theta_3 - \sin 2\theta_3 \left( 1 -
  3 \cos 2\theta_1 + 2 \cos^2 \theta_1 \cos 2\theta_2 \right) \right]
  \cos 2\Theta \\ \nonumber
 & & \quad + \frac{1}{ 128 \sqrt{2} } \big[ \left( 12 \cos^2 \theta_1
  \sin \theta_1 \cos 3\theta_2 + \left( \sin \theta_1 - 15 \sin 3\theta_1
  \right) \cos \theta_2 \right) \sin 2\theta_3 \\
  \label{eq:mix_par_10II-1}
 & & \quad + 4 \left( \cos\theta_1 + 3\cos 3\theta_1 \right) \sin 2\theta_2
  \cos 2\theta_3 \big] \sin 2\Theta\,,
\end{eqnarray}
\begin{figure}[!hptb]
 \begin{center}
  \begin{tabular}{cc} \hskip-0.5in
   \includegraphics[width=0.36\linewidth]{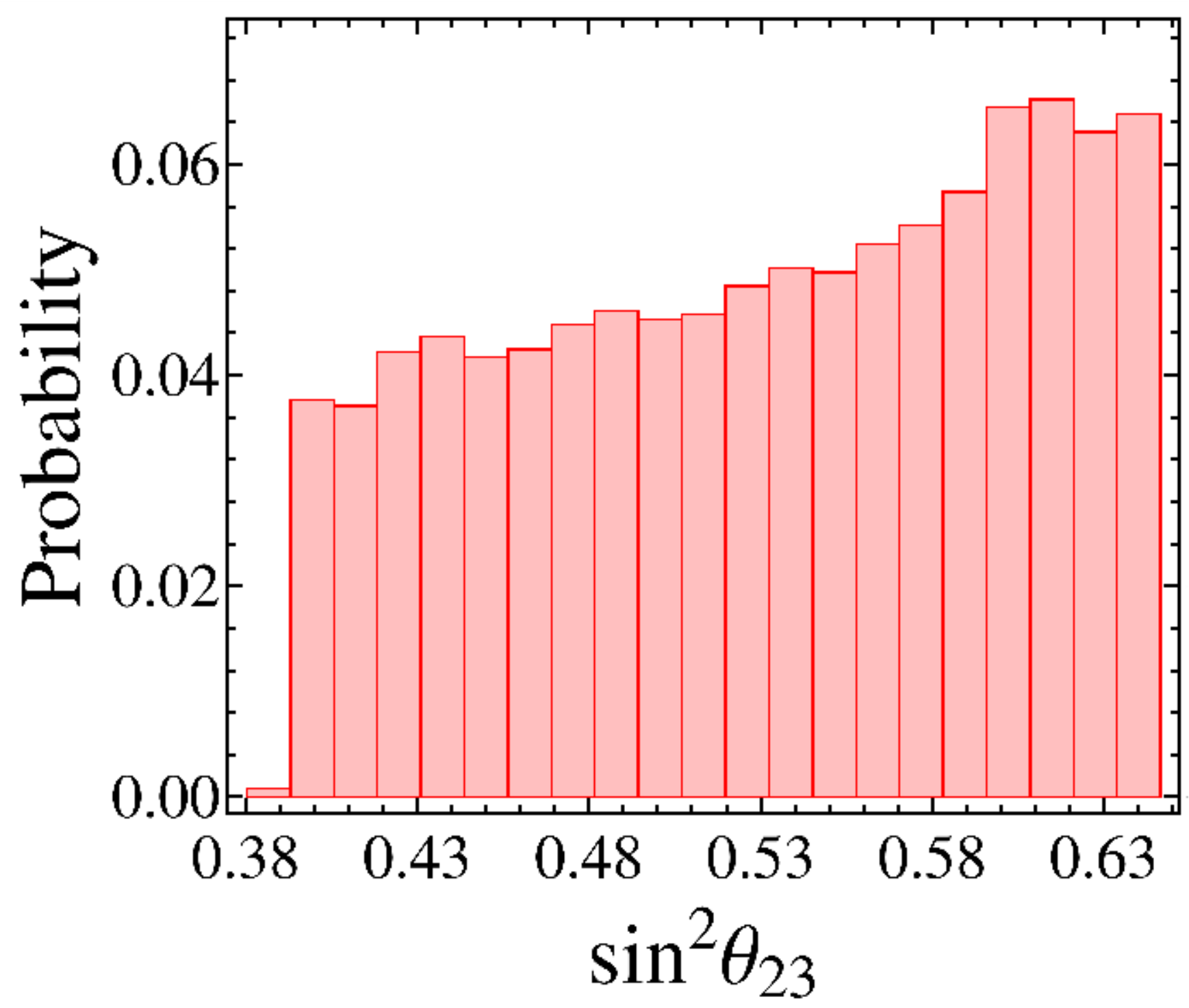} &~
   \includegraphics[width=0.36\linewidth]{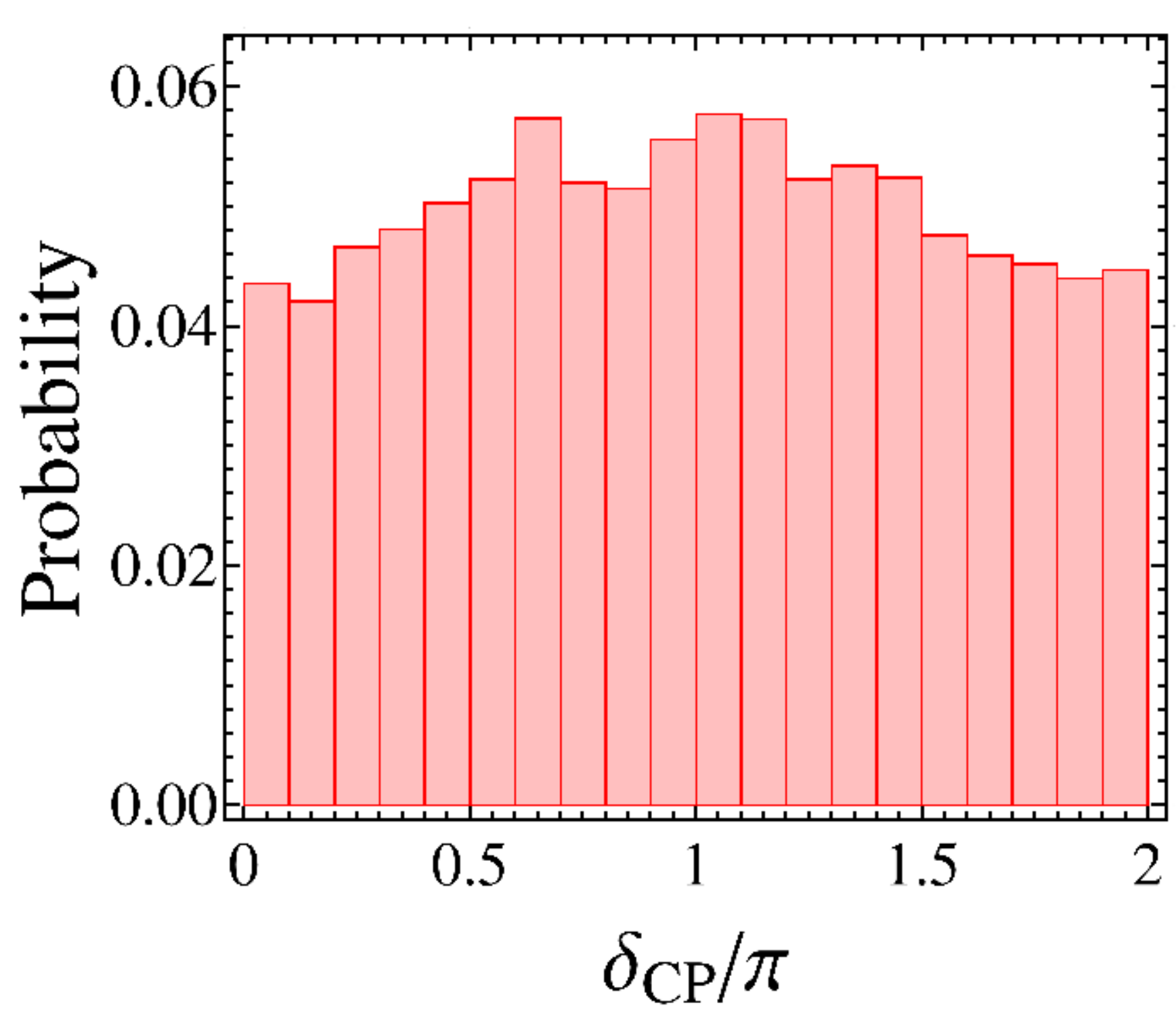} \\ \hskip-0.5in
   \includegraphics[width=0.36\linewidth]{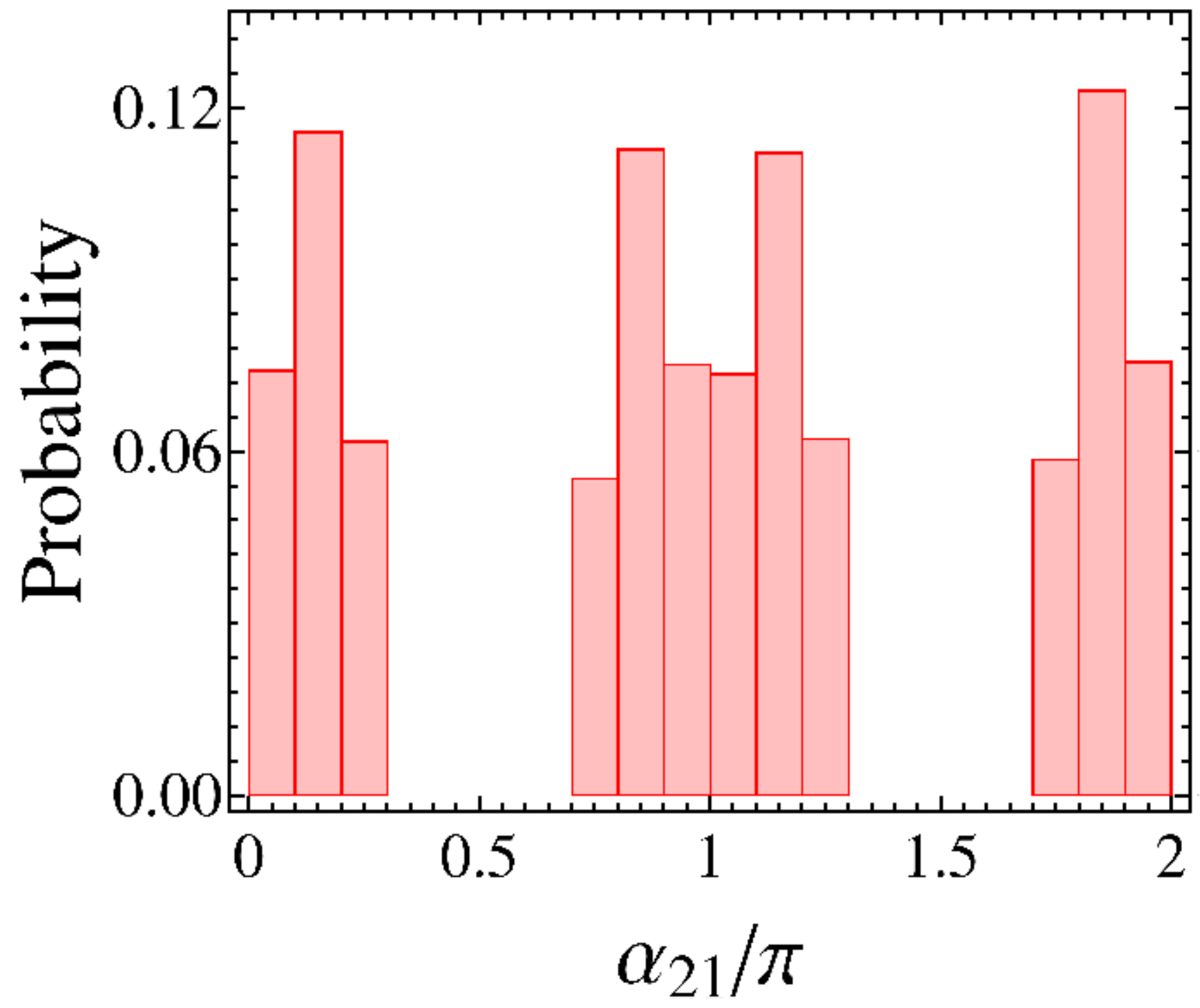} &~
   \includegraphics[width=0.36\linewidth]{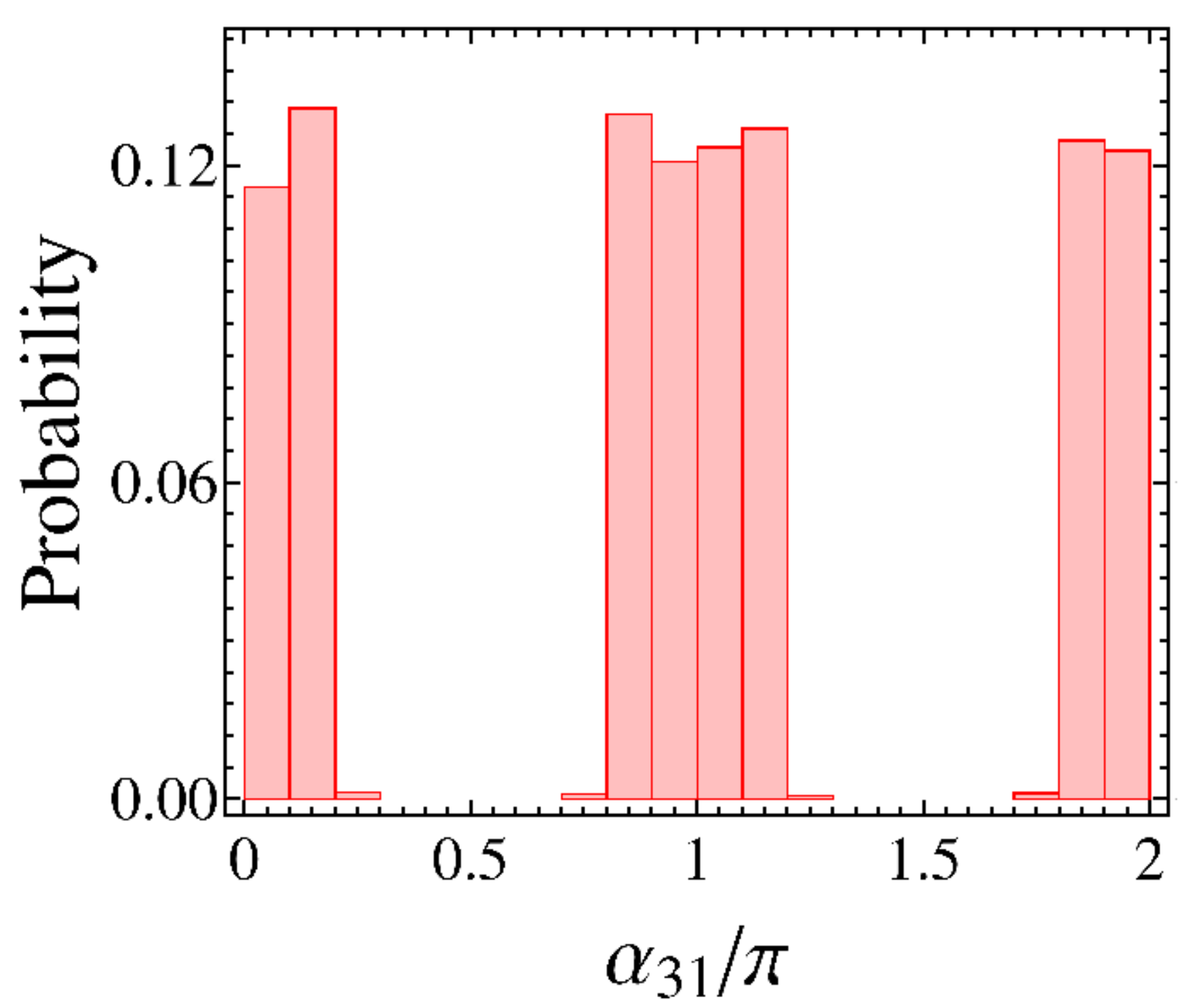}
  \end{tabular}
  \caption{\label{fig:distribution_two_zero2} The probability
   distribution of the atmospheric mixing angle $\sin^2\theta_{23}$,
   and the CP violating phases $\delta_{CP}$, $\alpha_{21}$ and
   $\alpha_{31}$ predicted for the case of type-VI residual CP
   transformation. All the parameters $\theta_{1,2,3}$ and $\Theta$ are
   taken to be random numbers in the range $[0, 2\pi]$, and the three
   lepton mixing angles are required to be compatible with experimental
   data at $3\sigma$ level~\cite{Forero:2014bxa}}.
 \end{center}
\end{figure}
while the invariants associated with the Majorana phases are
\begin{eqnarray} \nonumber
 & &
 I_1 = \bigg\{ \frac{1}{ 64 \sqrt{2} } \Big[ 4 \left( 3 \cos 3\theta_1 - 7
  \cos \theta_1 \right) \sin \theta_2  \cos 2\theta_3 + \left( 31
  \sin \theta_1 - 9 \sin 3\theta_1 + 6 \cos \theta_1 \sin 2\theta_1
  \cos 2\theta_2  \right) \sin 2\theta_3 \Big] \\ \nonumber
 & &
 \quad \times \cos \theta_2 \cos \Theta \sin \Theta + \frac{1}{ 64 \sqrt{2} }
  \Big[ \left( 9 \sin \theta_1 - 15 \sin 3\theta_1 + 10 \cos \theta_{1}
  \sin 2\theta_{1} \cos 2\theta_{2} \right) \sin 2\theta_3 \\ \nonumber
 & &
 \quad - 4  \left( \cos \theta_1 - 5 \cos 3\theta_1 \right) \sin \theta_2
  \cos 2\theta_3 \Big] \cos \theta_2 \cos 3 \Theta \sin \Theta + \frac{1}{128}
  \Big[ 4 \left( 5 \sin \theta_1 - 3 \sin 3\theta_1 \right) \sin \theta_2
  \cos 2\theta_3 \\ \nonumber
 & & \quad
 + \left( 13 \cos \theta_1 - 9 \cos 3\theta_1 - 2 \left( 5 - 3 \cos 2\theta_1
  \right) \cos \theta_1 \cos 2\theta_2 \right) \sin 2\theta_3 \Big]
  \cos \theta_2 \sin^2 \Theta \\ \nonumber
 & & \quad
 + \frac{1}{128} \Big[ \left( 9 \cos \theta_1-21\cos 3\theta_1 - 2 \left( 1 -
  7 \cos 2\theta_1 \right) \cos \theta_1 \cos 2\theta_2 \right) \sin 2\theta_3
  \\ \label{eq:mix_par_IX_I1}
 & & \quad
 + 4 \left( \sin \theta_1 - 7 \sin 3\theta_1 \right) \sin \theta_2
  \cos 2\theta_3 \Big] \cos \theta_2 \sin \Theta \sin 3\Theta \bigg\}
  \cos(k_1\pi),\\ \nonumber
 & &
 I_2 = \bigg\{ \frac{1}{128 \sqrt{2} } \Big[ \left( \left( 31 \sin \theta_1
  + 15 \sin 3\theta_1 \right) \cos \theta_2 - 12 \cos^2 \theta_1
  \sin \theta_1 \cos 3\theta_2 \right) \sin 2\theta_3 \\ \nonumber
 & & \quad
 - 8 \left( 6 + \left( 1 + 3 \cos 2\theta_1 \right) \cos 2\theta_3 \right)
  \cos \theta_1 \sin 2\theta_2 \Big] \cos \Theta \sin \Theta \\ \nonumber
 & & \quad
 + \frac{1}{128 \sqrt{2} } \Big[ \left( \left( 9 \sin \theta_1 + 25
  \sin 3\theta_1 \right) \cos \theta_2 - 20 \cos^2 \theta_1 \sin \theta_1
  \cos 3\theta_2 \right) \sin 2\theta_3 \\ \nonumber
 & & \quad
  - 8 \left( 2 - \left( 1 - 5 \cos 2\theta_1 \right) \cos 2\theta_3 \right)
  \cos \theta_1 \sin 2\theta_2 \Big] \cos 3\Theta \sin \Theta \\ \nonumber
 & & \quad
 + \frac{1}{256} \Big[ \left[ 2 \left( -1 + 15 \cos 2\theta_1\right)
  \cos \theta_2 + \left( 10 - 6 \cos 2\theta_1 \right) \cos 3\theta_2 \right]
  \cos \theta_1 \sin 2\theta_3 \\ \nonumber
 & & \quad
  + 8 \left( 2 + \left( 1 + 3 \cos 2\theta_1 \right) \cos 2\theta_3 \right)
  \sin \theta_1 \sin 2\theta_2 \Big] \sin^2 \Theta \\ \nonumber
 & & \quad
 + \frac{1}{128} \Big[ \left( \left( -13 + 35 \cos 2\theta_1 \right)
  \cos \theta_2 + \left( 1 - 7 \cos 2\theta_1 \right) \cos 3\theta_2 \right)
  \cos \theta_1 \sin 2\theta_3 \\ \label{eq:mix_par_IX_I2}
 & & \quad
  + 4 \left( 2 + \left( 5 + 7 \cos 2\theta_1 \right) \cos 2\theta_3 \right)
  \sin \theta_1 \sin 2\theta_2 \Big] \sin \Theta \sin 3\Theta \bigg\}
  \cos(k_2\pi)\,.
\end{eqnarray}
\begin{figure}[hptb]
 \begin{center}
 \begin{tabular}{cc}\hskip-0.6in
  \includegraphics[width=0.36\linewidth]{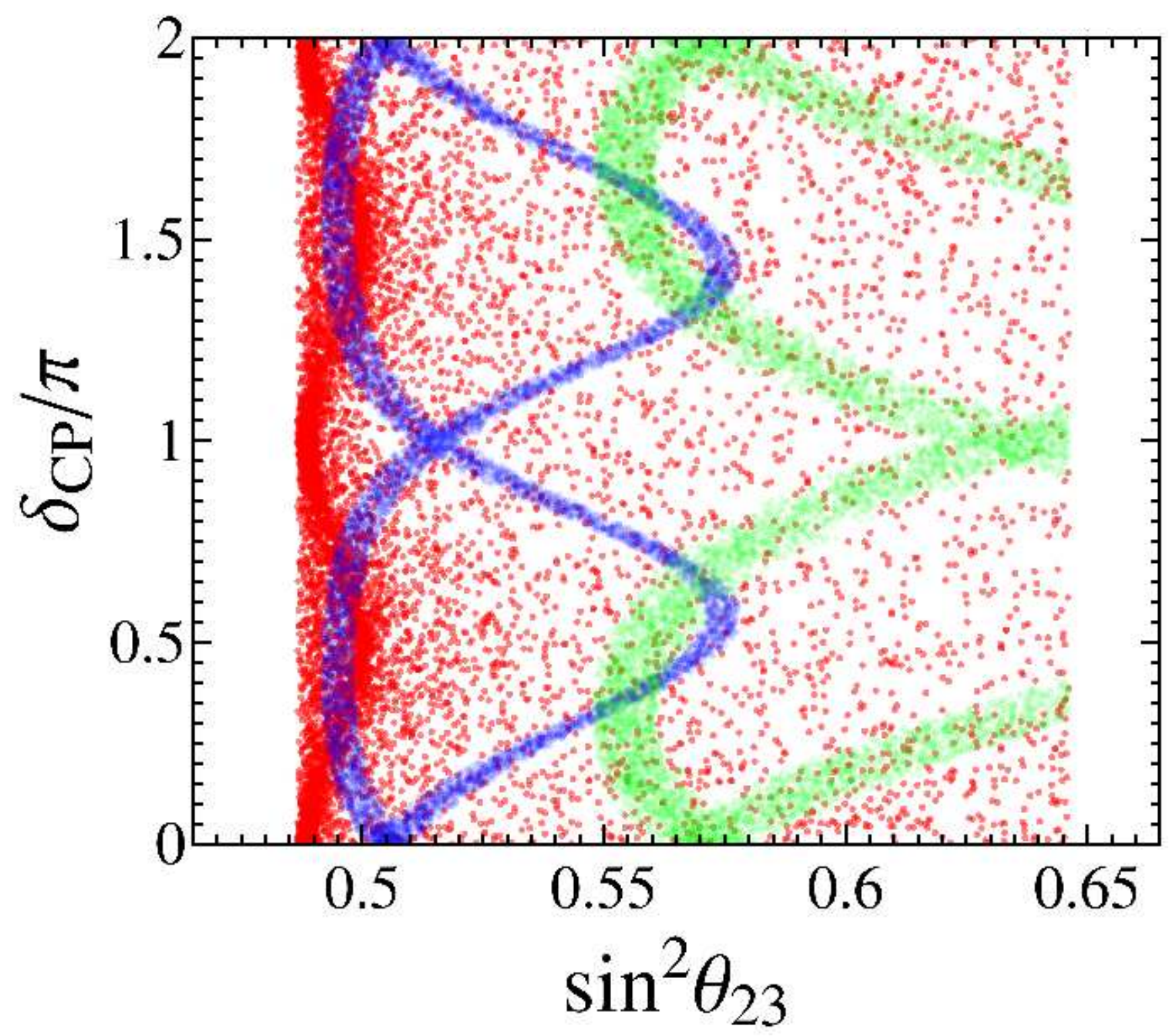} &
  \includegraphics[width=0.36\linewidth]{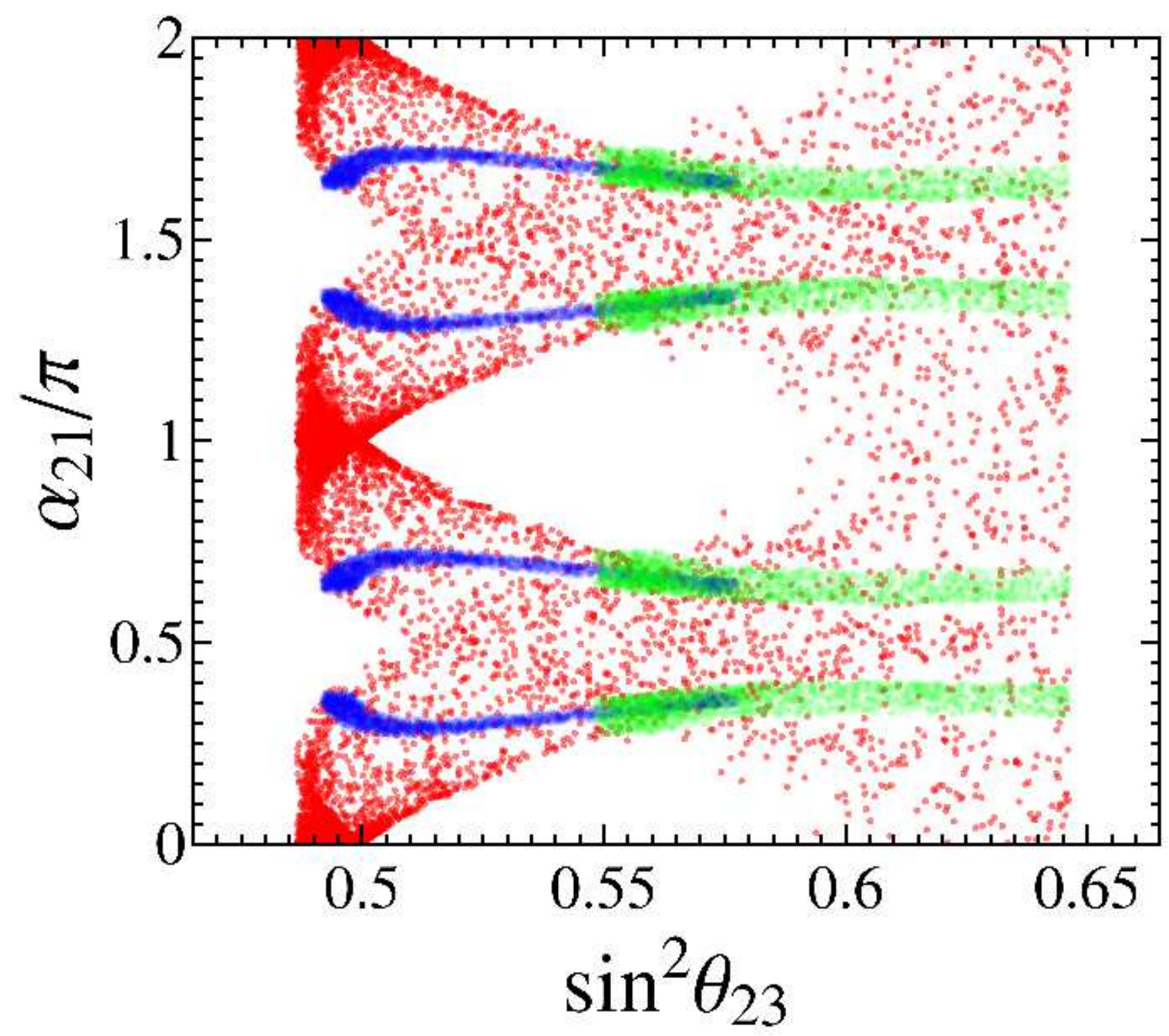} \\ \hskip-0.6in
  \includegraphics[width=0.36\linewidth]{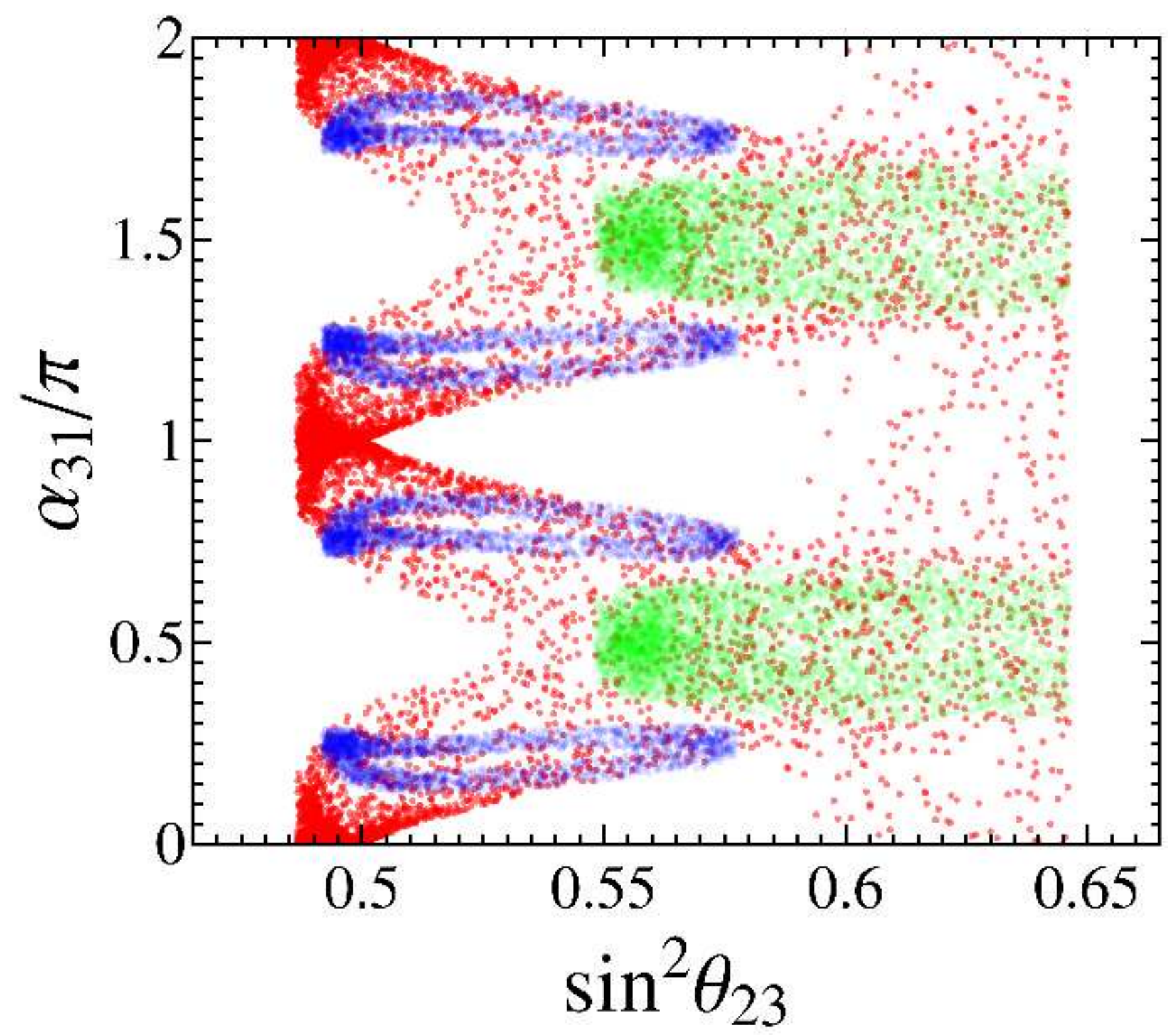} &
  \includegraphics[width=0.36\linewidth]{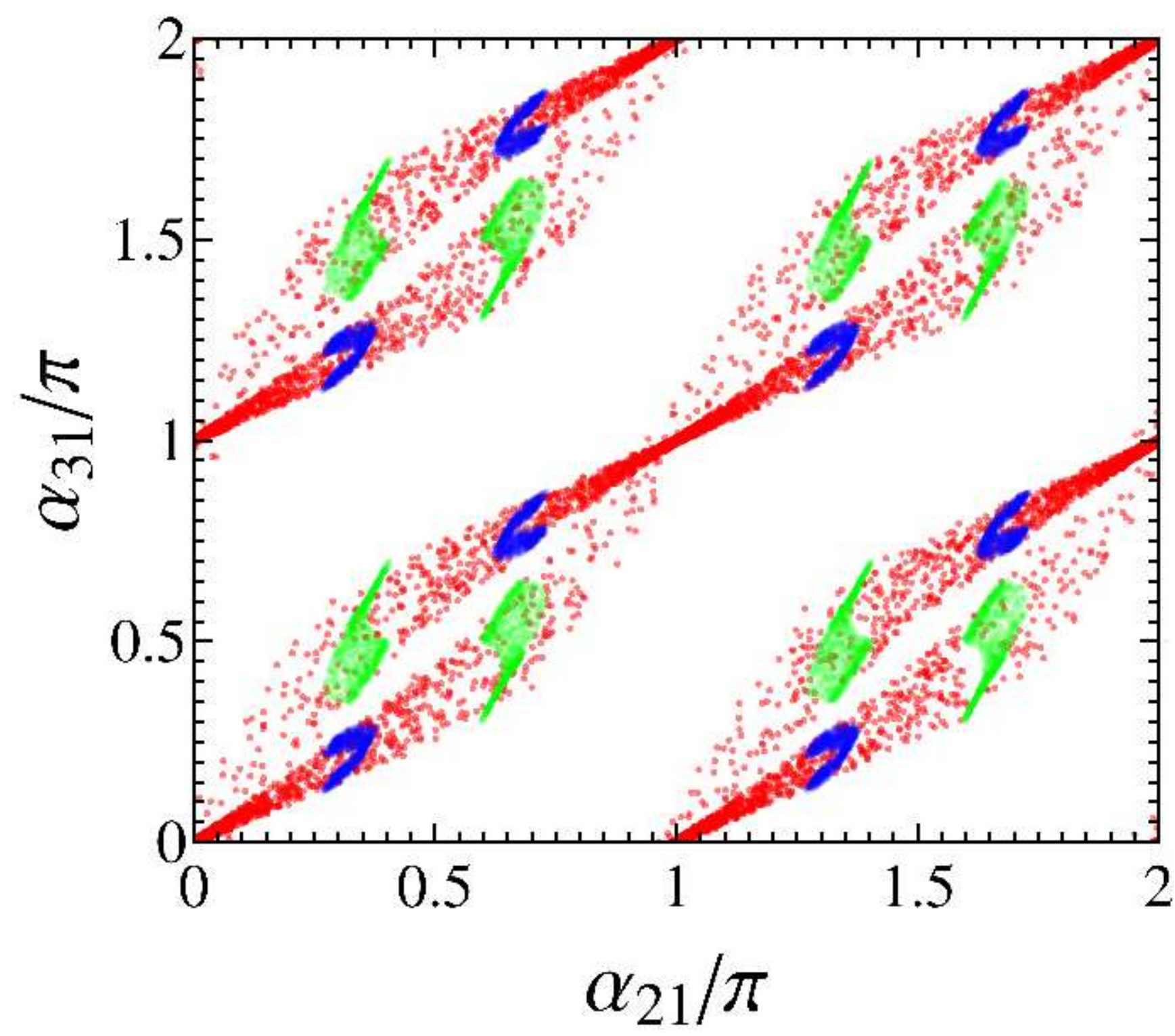}
 \end{tabular}
 \caption{\label{fig:correlation_one_zero2} The correlation between the
  mixing parameters predicted in the case of type-X residual CP
  transformation. All the parameters $\theta_{1,2,3}$ and $\Theta$ are
  vatied randomly in the range $[0, 2\pi]$. The three mixing angles
  $\theta_{12}$, $\theta_{13}$ and $\theta_{23}$ are required to be
  within the $3\sigma$ allowed ranges~\cite{Forero:2014bxa}. The blue
  (green) points are obtained by fixing $\Theta=\frac{2\pi}{17}
  (\frac{2\pi}{9})$ as a benchmark example.}
 \end{center}
\end{figure}
\begin{figure}[hptb]
 \begin{center}
 \begin{tabular}{cc}\hskip-0.5in
  \includegraphics[width=0.36\linewidth]{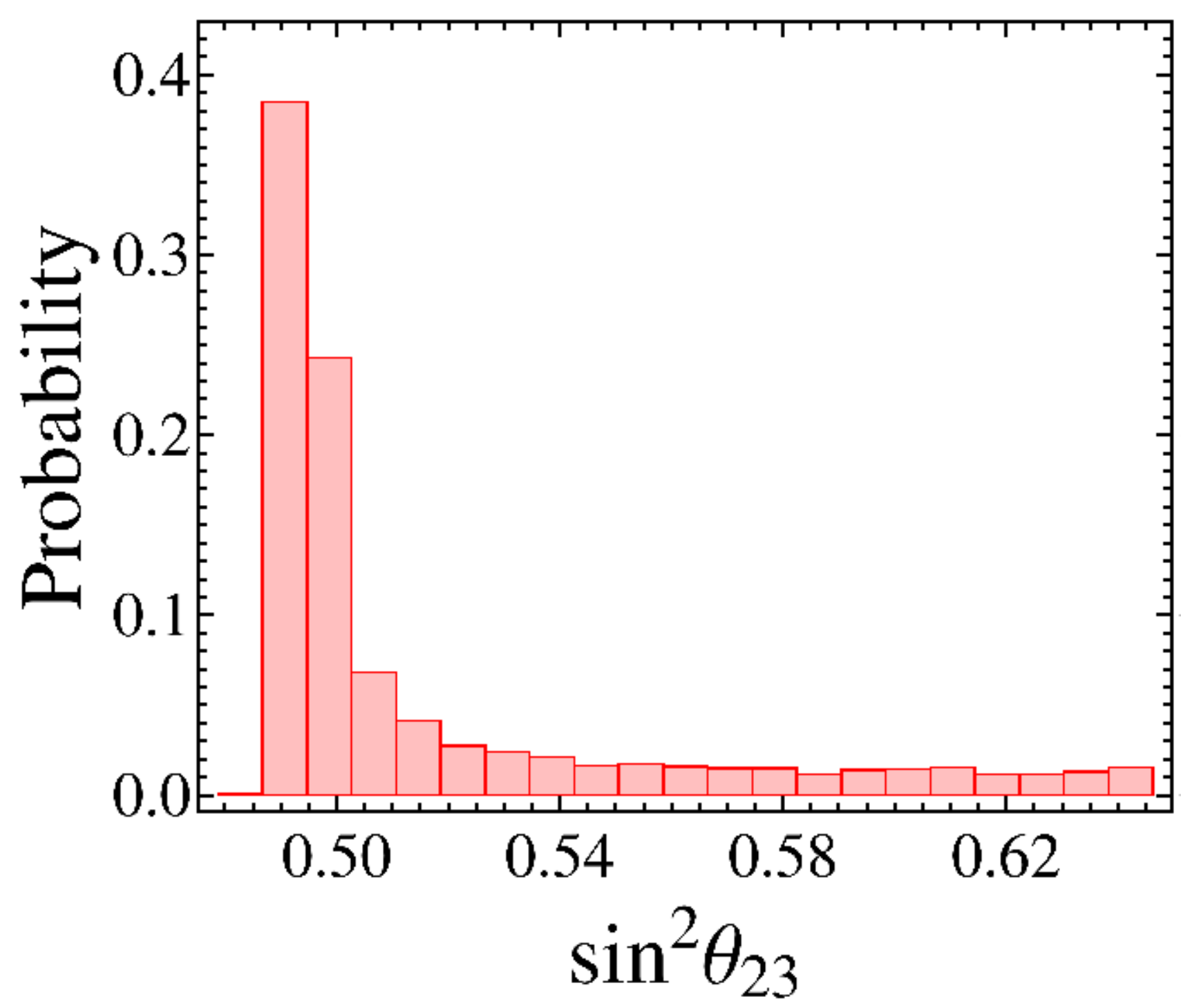} &~
  \includegraphics[width=0.36\linewidth]{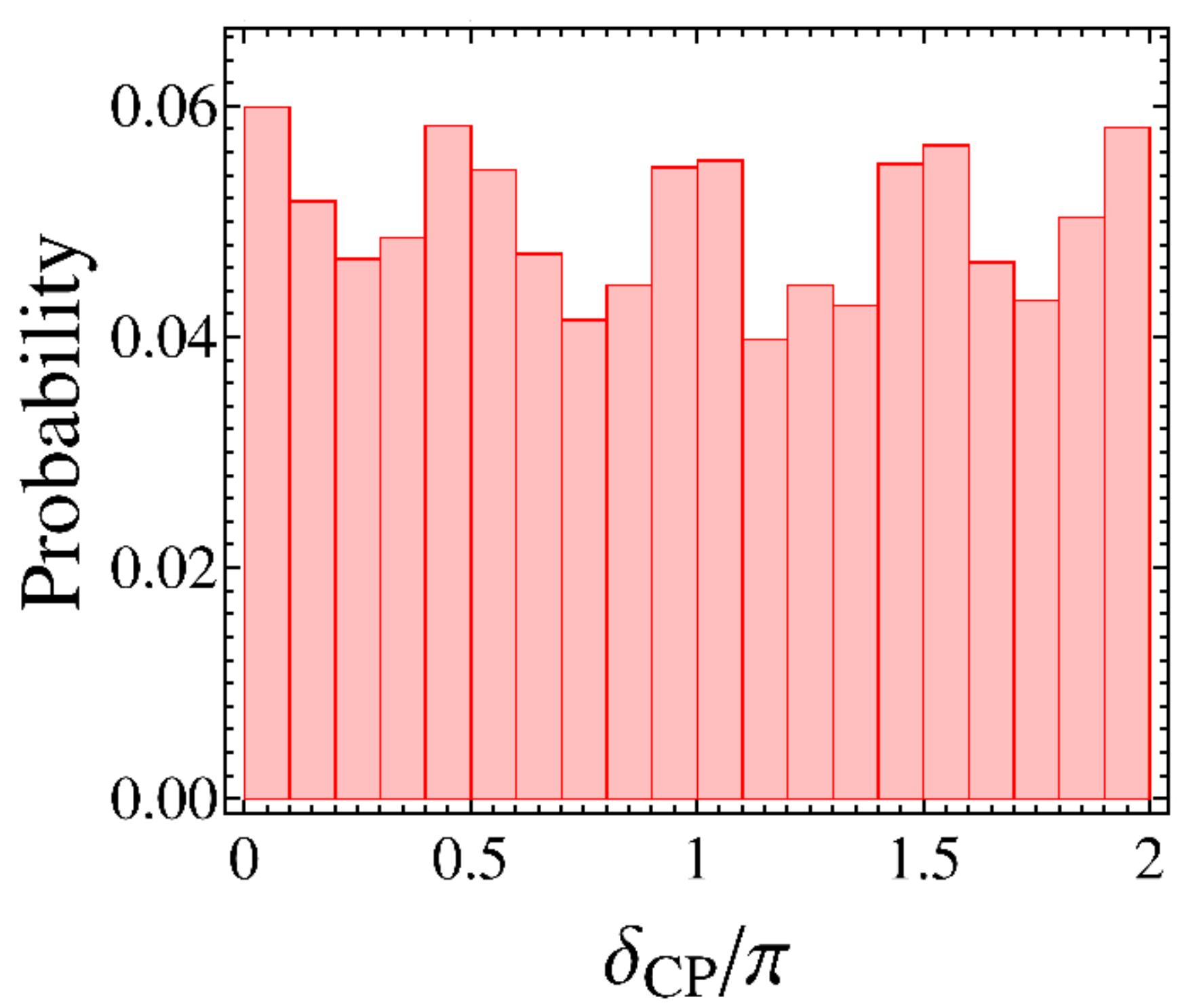} \\ \hskip-0.5in
  \includegraphics[width=0.36\linewidth]{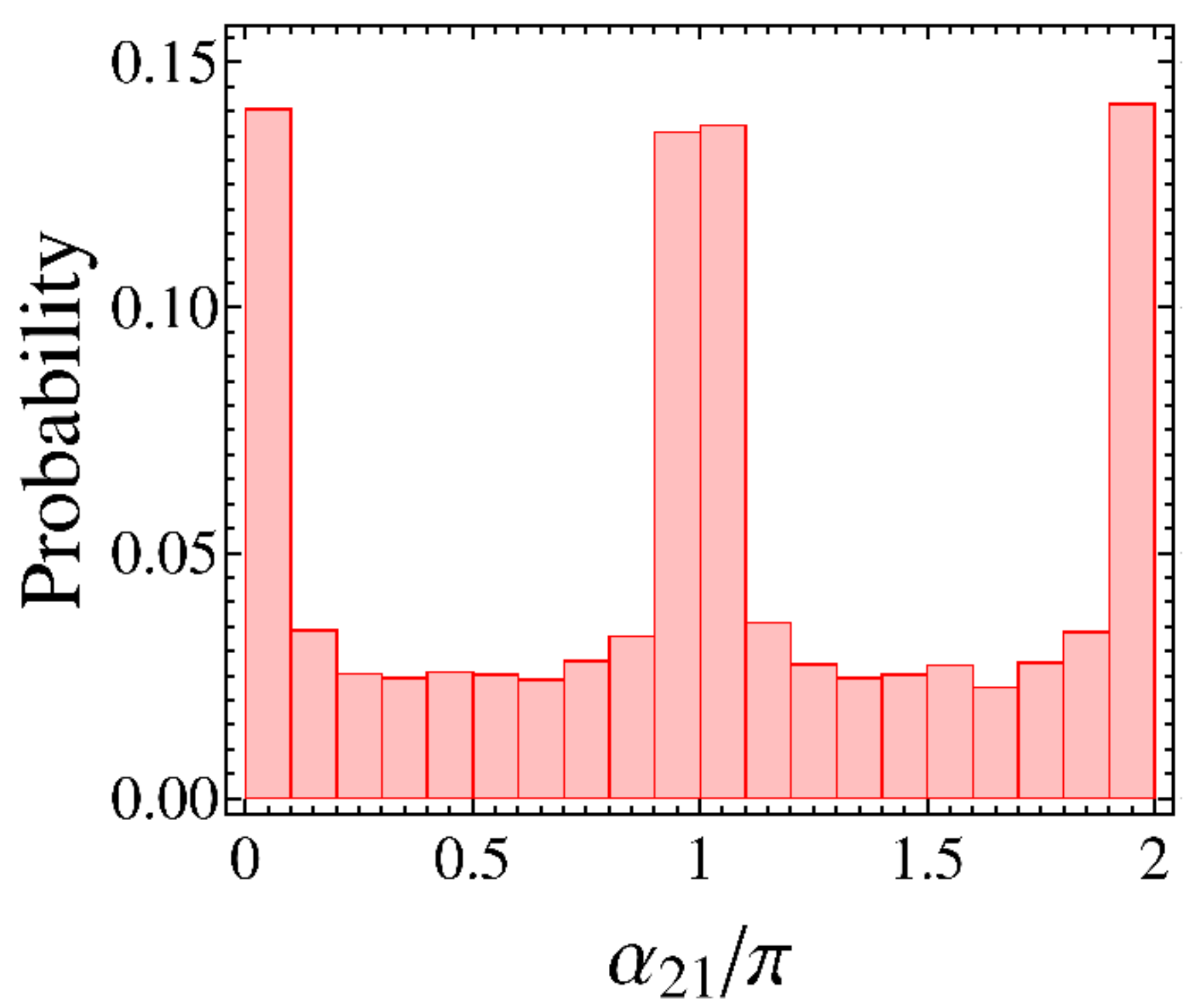} &~
  \includegraphics[width=0.36\linewidth]{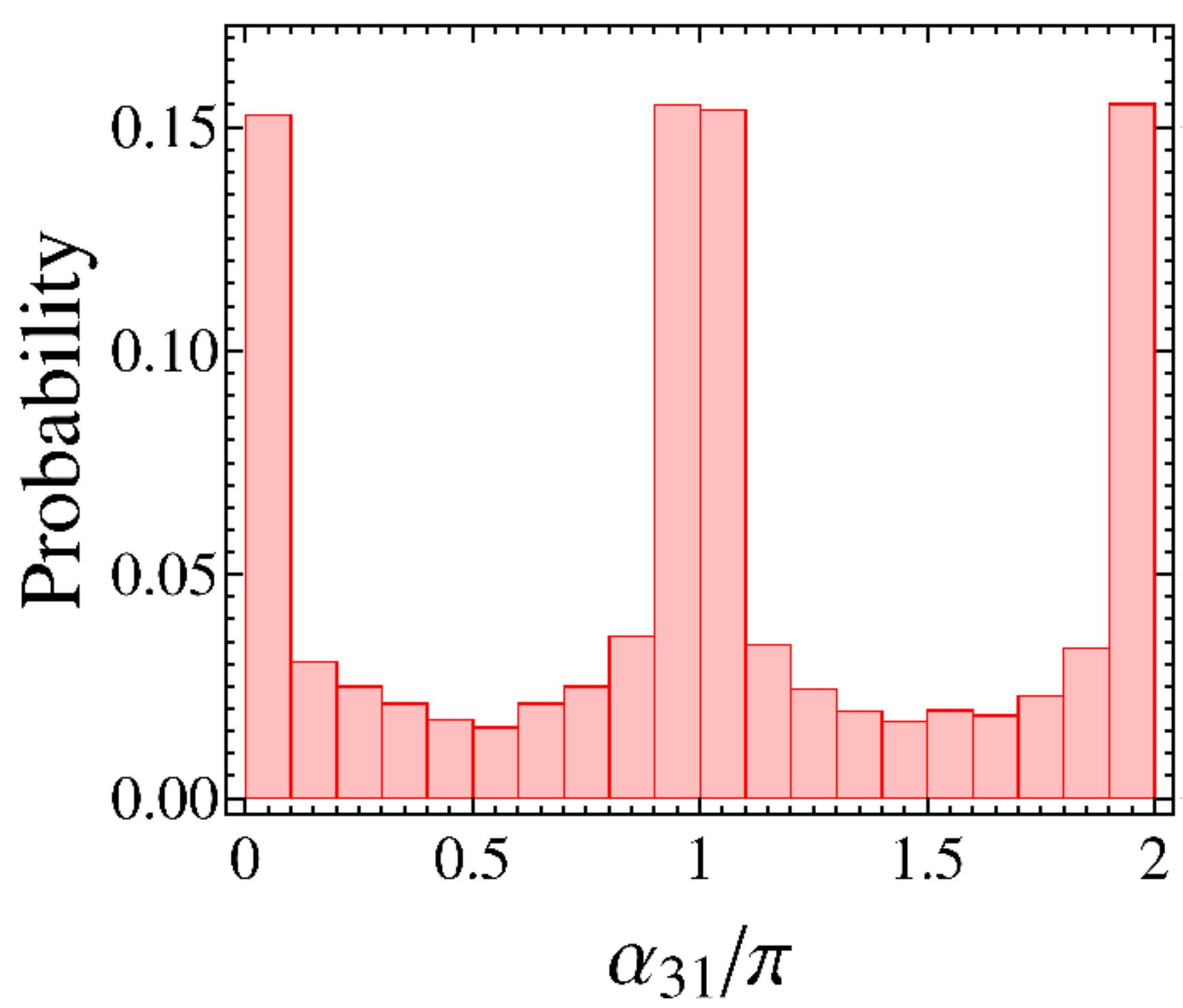}
 \end{tabular}
 \caption{\label{fig:distribution_one_zero2}The probability
  distribution of the lepton mixing parameters $\sin^2\theta_{23}$,
  $\delta_{CP}$, $\alpha_{21}$ and $\alpha_{31}$ predicted for the
  case of type-X residual CP transformation. All the parameters
  $\theta_{1,2,3}$ and $\Theta$ are taken to be random numbers in the
  range $[0, 2\pi]$, and the three lepton mixing angles are required
  to be compatible with experimental data at $3\sigma$
  level~\cite{Forero:2014bxa}.}
 \end{center}
\end{figure}
For the type-X remnant CP transformation with one zero, we find that
the resulting lepton mixing matrix is related with the previous case
through the exchange of the second and third rows. The expressions for
the reactor and solar mixing angles, as well as for the $I_1$ and
$I_2$ invariants, are the same as those of
Eqs.~(\ref{eq:mix_par_10II},~\ref{eq:mix_par_IX_I1},
\ref{eq:mix_par_IX_I2}), while the atmospheric mixing angle
$\sin^2 \theta_{23}$ becomes $1-\sin^2\theta_{23}$. The Dirac CP phase
$\delta_{\text CP}$ becomes $\pi + \delta_{\text CP}$ so that the
overall sign of the Jarlskog invariant is reversed.
%
\section{Democratic CP symmetry as an example without zero elements}
%
If no entry of the residual CP transformation $\mathbf{X}$ vanishes,
the explicit form of $\mathbf{X}$ can not be fixed uniquely. For illustration,
in this section we shall study a particular CP symmetry whose elements have
the same absolute value. That is to say, the absolute value of each element of
$\mathbf{X}$ is equal to $1/\sqrt{3}$.
In what follows it will be dubbed as democratic CP symmetry. In this case, the
most general form of $\mathbf{X}$ can be written as
\begin{equation}
 \mathbf{X} = \frac{1}{\sqrt{3}}
 \left( \begin{array}{ccc}
  e^{ i \alpha } & e^{ i ( \frac{ \alpha + \beta }{2} + \beta_3 ) } &
   e^{ i ( \frac{ \alpha + \gamma }{2} + \beta_2 ) } \\
  e^{ i ( \frac{ \alpha + \beta }{2} + \beta_3 ) } & e^{i\beta}  &
   e^{ i ( \frac{ \beta + \gamma }{2} + \beta_1 ) } \\
  e^{ i ( \frac{ \alpha + \gamma }{2} + \beta_2 ) } &
   e^{ i ( \frac{ \beta + \gamma }{2} + \beta_1 ) } & e^{i\gamma}
 \end{array}\right)\,.
\end{equation}
The unitary condition of $\mathbf{X}$ implies that $\beta_1$, $\beta_2$ and
$\beta_3$ should satisfy the following equalities
$e^{ i \beta_1 } + e^{ -i \beta_1 } + e^{ i ( \beta_2 - \beta_3 ) } =0$,
$e^{ i \beta_2 } + e^{ -i \beta_2 } + e^{ i ( \beta_3 - \beta_1 ) } =0$, and
$e^{ i \beta_3 } + e^{ -i \beta_3 } + e^{ i ( \beta_1 - \beta_2 ) } = 0$.
It can be easily checked that these equations have four pairs of solutions,
\begin{eqnarray}
 \nonumber && \beta_1 = \beta_2 = \beta_3 = \pm\frac{2\pi}{3} \,, \qquad\qquad
 \mathbf{X}^\pm_1 = \frac{1}{\sqrt{3}}
 \left( \begin{array}{ccc}
  e^{i \alpha } & e^{ i ( \frac{ \alpha + \beta }{2} \pm \frac{2\pi}{3} ) } &
   e^{i ( \frac{ \alpha + \gamma }{2} \pm \frac{2\pi}{3} ) } \\
  e^{i ( \frac{ \alpha + \beta }{2} \pm \frac{2\pi}{3} ) } & e^{i\beta} &
   e^{ i ( \frac{ \beta + \gamma }{2} \pm \frac{2\pi}{3} ) } \\
  e^{i ( \frac{ \alpha + \gamma }{2} \pm \frac{2\pi}{3} ) } &
   e^{i ( \frac{ \beta + \gamma }{2} \pm \frac{2\pi}{3} ) } & e^{i \gamma }
 \end{array} \right)\,,\\
 \nonumber && \beta_1 = \pm \frac{2\pi}{3},
  \beta_2 = \beta_3 = \mp \frac{\pi}{3},\qquad
 \mathbf{X}^\pm_2 = \frac{1}{\sqrt{3}}
 \left(\begin{array}{ccc}
  e^{i \alpha } & e^{i ( \frac{ \alpha + \beta }{2} \mp \frac{\pi}{3} ) } &
   e^{i ( \frac{ \alpha + \gamma }{2} \mp \frac{\pi}{3} ) } \\
  e^{i ( \frac{ \alpha + \beta }{2} \mp \frac{\pi}{3} ) } & e^{i\beta}  &
   e^{i ( \frac{ \beta + \gamma }{2} \pm \frac{2\pi}{3} ) } \\
  e^{i ( \frac{ \alpha + \gamma }{2} \mp \frac{\pi}{3} ) } &
   e^{i ( \frac{ \beta + \gamma }{2} \pm \frac{2\pi}{3} ) } & e^{i\gamma}
 \end{array}\right)\,,\\
 \nonumber && \beta_2 = \pm \frac{2\pi}{3},
  \beta_1 = \beta_3 = \mp\frac{\pi}{3}\,,\qquad
 \mathbf{X}^\pm_3 = \frac{1}{\sqrt{3}}
 \left(\begin{array}{ccc}
  e^{i \alpha } & e^{i ( \frac{ \alpha + \beta }{2} \mp \frac{\pi}{3} ) } &
   e^{i ( \frac{ \alpha + \gamma }{2} \pm \frac{2\pi}{3} ) } \\
  e^{i ( \frac{ \alpha + \beta}{2} \mp \frac{\pi}{3} ) } & e^{i \beta } &
   e^{i ( \frac{ \beta + \gamma }{2} \mp \frac{\pi}{3} ) } \\
  e^{i ( \frac{ \alpha + \gamma }{2} \pm \frac{2\pi}{3} ) } &
   e^{i ( \frac{ \beta + \gamma }{2} \mp \frac{\pi}{3} ) } & e^{i\gamma}
 \end{array} \right)\,,\\
 && \beta_3 = \pm \frac{2\pi}{3},
 \beta_1 = \beta_2 = \mp \frac{\pi}{3}\,,\qquad
 \mathbf{X}^\pm_4 = \frac{1}{\sqrt{3}}
 \left(\begin{array}{ccc}
  e^{i \alpha } & e^{i ( \frac{ \alpha + \beta }{2} \pm \frac{2\pi}{3} ) } &
   e^{i ( \frac{ \alpha + \gamma }{2} \mp \frac{\pi}{3} ) } \\
  e^{i ( \frac{ \alpha + \beta }{2} \pm \frac{2\pi}{3} ) } & e^{i\beta} &
   e^{ i ( \frac{ \beta + \gamma }{2} \mp \frac{\pi}{3} ) } \\
  e^{i ( \frac{ \alpha + \gamma }{2} \mp \frac{\pi}{3} ) } &
   e^{i ( \frac{ \beta + \gamma }{2} \mp \frac{\pi}{3} ) } & e^{i\gamma}
 \end{array}\right)\,.
\end{eqnarray}
We can see that the four admissible CP transformations $\mathbf{X}^\pm_1$,
$\mathbf{X}^\pm_2$, $\mathbf{X}^\pm_3$ and $\mathbf{X}^\pm_4$ are related to
each other as follows
$\mathbf{X}^\pm_1 = \text{diag}(1,-1,-1) \mathbf{X}^\pm_2 \text{diag}(1,-1,-1)
= \text{diag}(-1,1,-1) \mathbf{X}^\pm_3 \text{diag}(-1,1,-1)
= \text{diag}(-1,-1,1) \mathbf{X}^\pm_4 \text{diag}(-1,-1,1)$.
Therefore the Takagi factorization matrix $\mathbf{\Sigma}^\pm_i$ for
$\mathbf{X}^{\pm}_i$ ($i=1, 2, 3, 4$) are related with each other as well,
$\mathbf{\Sigma}^\pm_1
 = \text{diag}(1,-1,-1) \mathbf{\Sigma}^\pm_2
 = \text{diag}(-1,1,-1) \mathbf{\Sigma}^\pm_3
 = \text{diag}(-1,-1,1)\mathbf{\Sigma}^\pm_4$.
As a result, we conclude that the four CP transformations $\mathbf{X}^\pm_1$,
$\mathbf{X}^\pm_2$, $\mathbf{X}^\pm_3$ and $\mathbf{X}^\pm_4$ give rise to the
same lepton mixing matrix up to a phase factor which can be absorbed by
redefining the charged lepton fields. Furthermore, it can be easily checked
that $\mathbf{\Sigma}^+_i$ and $\mathbf{\Sigma}^-_i$ can be related by
$\mathbf{\Sigma}^-_i = \text{diag}( e^{i\alpha}, e^{i\beta}, e^{i\gamma} )
\mathbf{\Sigma}^{+\ast}_i$. Therefore the predicted PMNS matrix by
$\mathbf{X}^+_i$ and $\mathbf{X}^-_i$ are complex conjugate of the each other
up to the phase factor $\text{diag}(e^{i\alpha}, e^{i\beta}, e^{i\gamma})$
which can also be absorbed by the charged leptons. Hence it is sufficient to
only discuss the CP transformation $\mathbf{X}_{XI}\equiv\mathbf{X}^+_1$ which
corresponds to $\beta_1=\beta_2=\beta_3=2\pi/3$ with
\begin{equation}\label{eq:PMNS_XI}
 \mathbf{X}_{XI} = \frac{1}{\sqrt{3}}
 \left( \begin{array}{ccc}
  e^{i \alpha } & e^{i ( \frac{ \alpha + \beta }{2} + \frac{2\pi}{3} ) } &
   e^{i ( \frac{ \alpha + \gamma}{2} + \frac{2\pi}{3} ) } \\
  e^{i ( \frac{ \alpha + \beta }{2} + \frac{2\pi}{3} ) } & e^{i\beta} &
   e^{i ( \frac{ \beta + \gamma }{2} + \frac{2\pi}{3} ) } \\
  e^{i ( \frac{ \alpha + \gamma }{2} + \frac{2\pi}{3} ) } &
   e^{i ( \frac{ \beta + \gamma }{2} + \frac{2\pi}{3} ) } & e^{i\gamma}
 \end{array}\right)\,,
\end{equation}
The corresponding Takagi factorization and the prediction for the PMNS
matrix can be straightforwardly obtained
\begin{equation}
 \mathbf{\Sigma}_{XI} =
 \text{diag}( e^{i\alpha}, e^{i\beta}, e^{i\gamma} ) \, e^{-\frac{i\pi}{12}}
 \left( \begin{array}{ccc}
  \sqrt{\frac{2}{3}}  & \frac{1}{\sqrt{3}} & 0 \\
  \frac{-1}{\sqrt{6}} & \frac{1}{\sqrt{3}} & \frac{1}{\sqrt{2}}  \\
  \frac{-1}{\sqrt{6}} & \frac{1}{\sqrt{3}} & \frac{-1}{\sqrt{2}} \\
 \end{array} \right) \text{diag}(1, e^{i\pi/3}, 1) \,,\qquad
 {\bf U} = {\bf \Sigma}_{XI} {\bf O}_{3 \times 3} \hat{\bf X}_{\nu}^{-1/2} .
\end{equation}
We can read out the lepton mixing angles as
\begin{eqnarray} \nonumber
 \sin^2 \theta_{13} & = &
  \frac{1}{6} \left[ 4 \sin^2 \theta_2 + \sqrt{2} \sin 2\theta_2 \sin \theta_1
  + 2 \sin^2 \theta_1 \cos^2 \theta_2 \right]\,,\\ \nonumber
 \sin^2 \theta_{12} & = &
  \Big\{ \big[ \sin \theta_1 \big( \sqrt{2} \sin 2\theta_2 - 2 \sin \theta_1
  \sin^2 \theta_2 \big) - 4 \cos^2 \theta_2 \big] \sin^2 \theta_3
  - 2 \cos^2 \theta_1 \cos^2 \theta_3 + \big[ \sin 2\theta_1 \sin \theta_2 \\
 \nonumber \qquad &&
  - \sqrt{2} \cos \theta_1 \cos \theta_2 \big] \sin 2\theta_3 \Big\} /
  \big( 4 \sin^2 \theta_2 + \sqrt{2} \sin 2\theta_2 \sin \theta_1
  + 2 \sin^2 \theta_1 \cos^2 \theta_2 - 6 \big)\,,\\
 \sin^2 \theta_{23} & = &
  \frac{ \sin 2\theta_2 \left( \sqrt{2} \sin \theta_1 + 2 \sqrt{3}
  \cos \theta_1 \right) - 2 \sin^2 \theta_2 - \cos^2 \theta_2 \left( \sqrt{6}
  \sin 2\theta_1 + \cos 2\theta_1 + 5 \right) }{ 2 \left( 4 \sin^2 \theta_2
  + \sqrt{2} \sin 2\theta_2 \sin \theta_1 + 2 \sin^2 \theta_1 \cos^2 \theta_2
  - 6 \right)}\,.
\end{eqnarray}
For the CP invariants we get
\begin{eqnarray} \nonumber
 J_{CP} & = &
  \frac{-1}{ 48 \sqrt{2} } \Big\{ \left[ 4 \sqrt{2} \sin 2\theta_2
  \sin^2 \theta_1 \cos \theta_1 + 4 \sin 2\theta_1 \cos 2\theta_2 \right]
  \cos 2\theta_3 + \Big[ 5 \sin \theta_2 \sin^2 \theta_1 \\ \nonumber
 & & \qquad
  + \sqrt{2} \sin \theta_1 \left( 5 \cos^2 \theta_1 - 1 \right) \cos \theta_2
  + \left( 3 \cos^2 \theta_1 + 1 \right) \sin 3 \theta_2 + \sqrt{2}
  \sin^3 \theta_1 \cos 3\theta_2 \Big] \sin 2\theta_3 \Big\}\,,\\ \nonumber
 I_1 & = &
  \frac{ (-1)^{k_1} }{ 12\sqrt{3} }
  \cos \theta_1 \cos \theta_2 \Big\{ \left[ 3 \sqrt{2} \sin^2 \theta_1
  - 2 \sin 2\theta_2 \sin \theta_1 + \sqrt{2} \left( \cos^2 \theta_1
  + 1 \right) \cos 2\theta_2 \right] \sin 2\theta_3 \\ \nonumber
 & & \qquad
 + 2 \left[ 2 \cos \theta_1 \cos \theta_2 - \sqrt{2} \sin 2\theta_1
  \sin \theta_2 \right] \cos 2\theta_3 \Big\}\,,\\ \nonumber
 I_2 & = &
 \frac{ (-1)^{k_2} }{ 6 \sqrt{3} } \left( \sin \theta_1 \cos \theta_3 +
  \sin \theta_2 \sin \theta_3
  \cos \theta_1 \right) \Big\{ \left[ 2 \sin \theta_1 \cos 2\theta_2 +
  \sqrt{2} \sin 2\theta_2 \left( \cos^2 \theta_1 + 1 \right) \right]
  \cos \theta_3 \\
 & & \qquad
 - \left[ \sqrt{2} \sin 2\theta_1 \cos \theta_2 + 2 \sin \theta_2
  \cos \theta_1 \right] \sin \theta_3 \Big \}\,.
\end{eqnarray}
\begin{figure}[!h]
 \begin{center}
 \begin{tabular}{cc} \hskip-0.6in
  \includegraphics[width=0.36\linewidth]{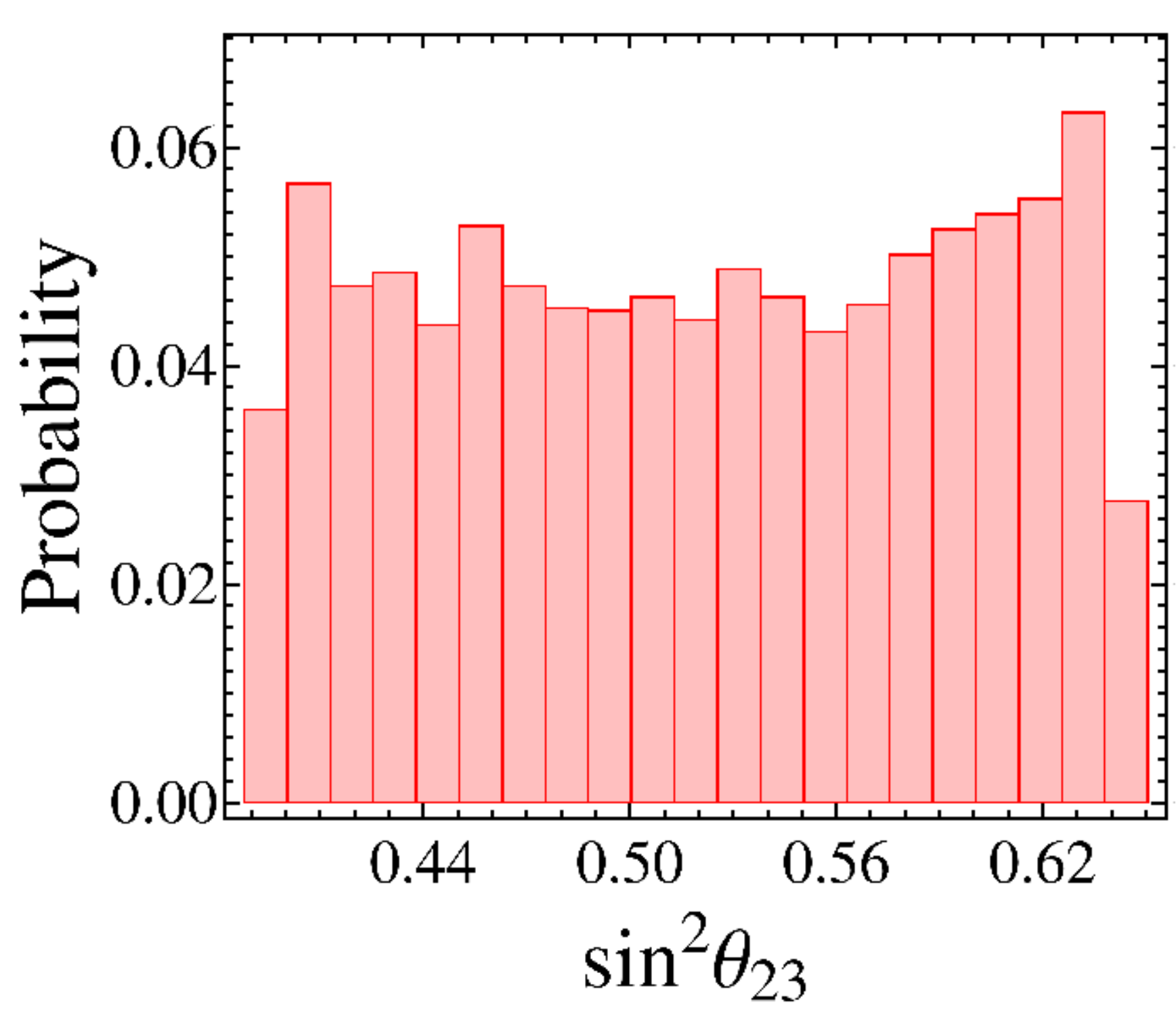} &
  \includegraphics[width=0.36\linewidth]{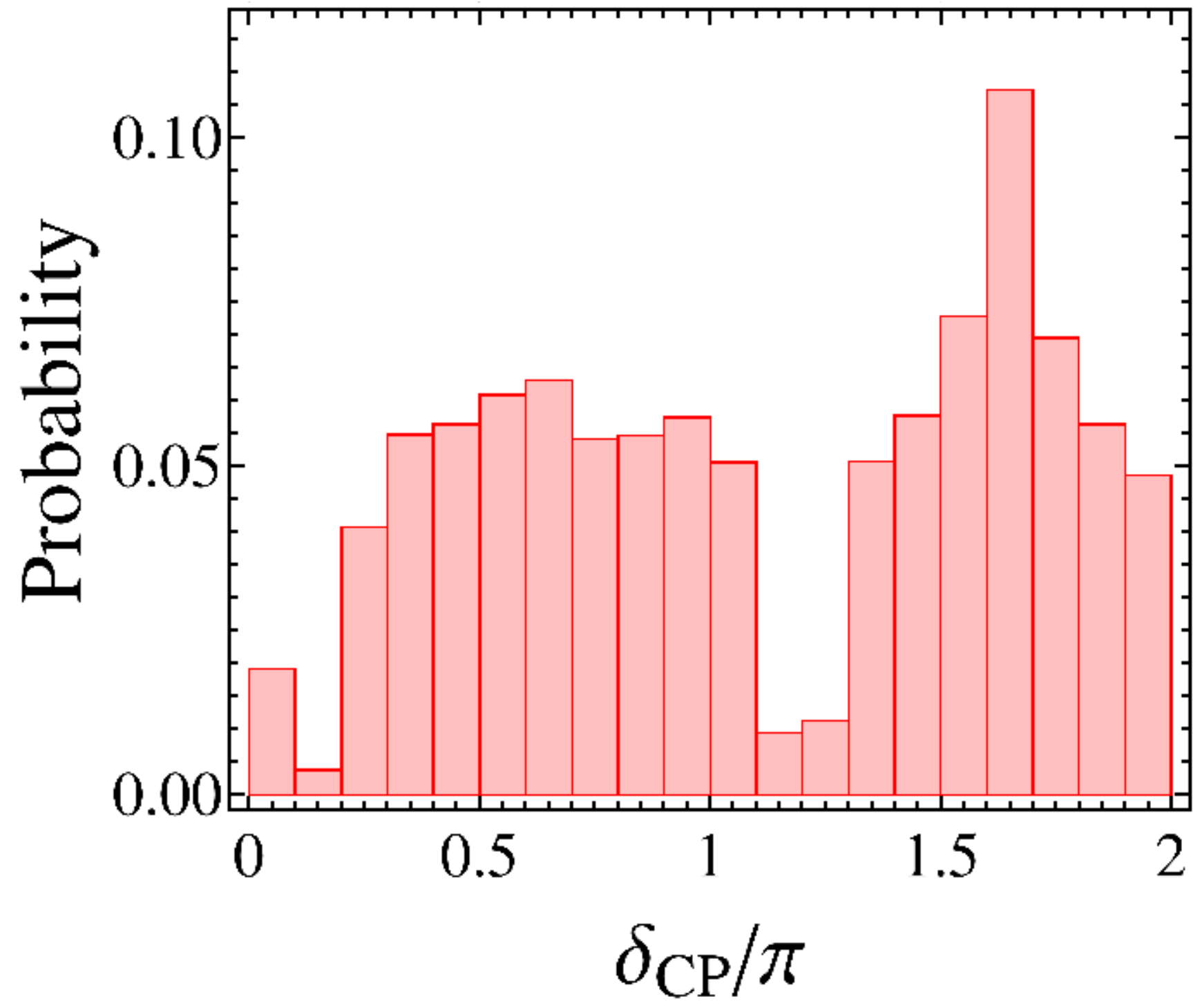} \\ \hskip-0.6in
  \includegraphics[width=0.36\linewidth]{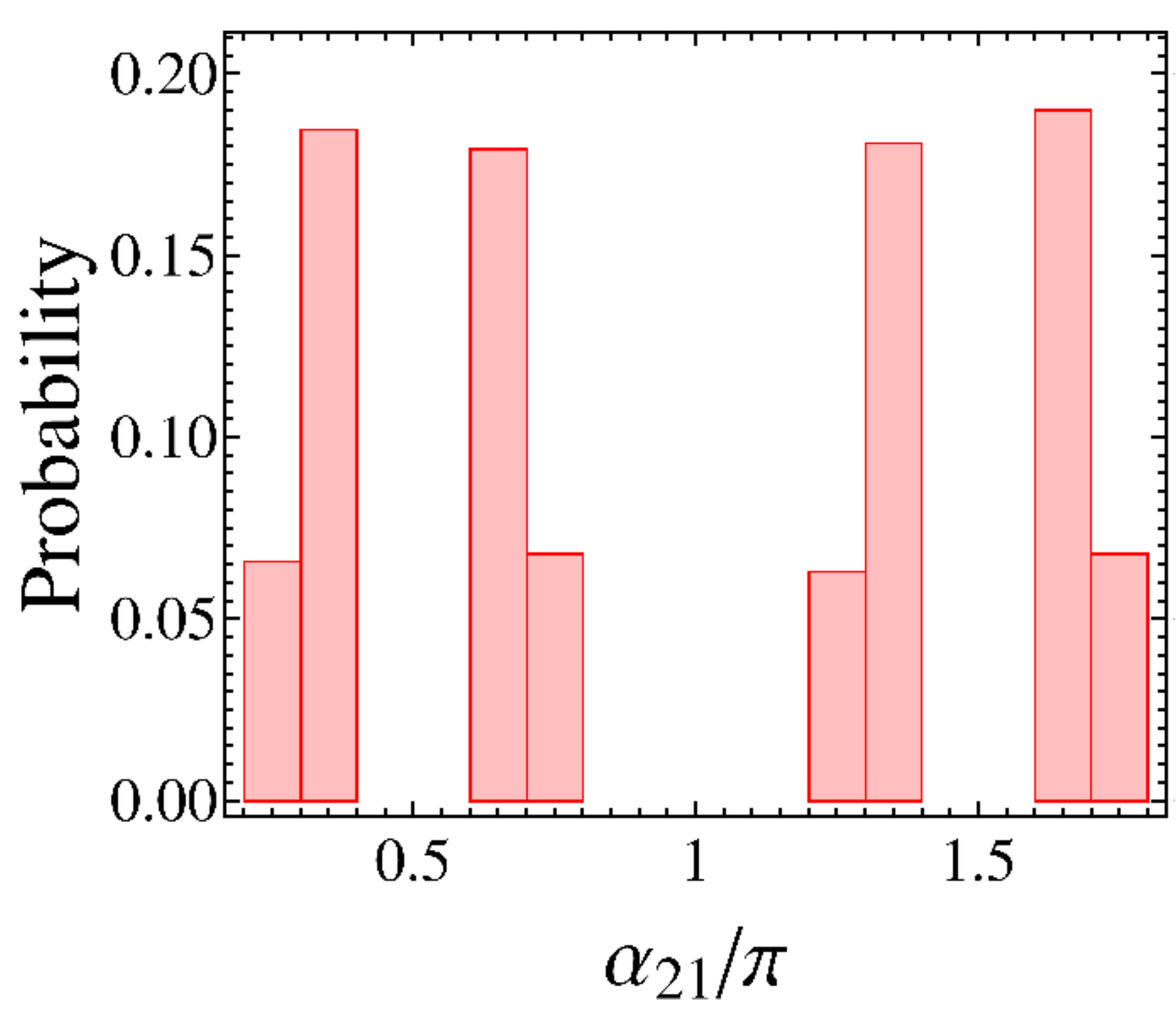} &
  \includegraphics[width=0.36\linewidth]{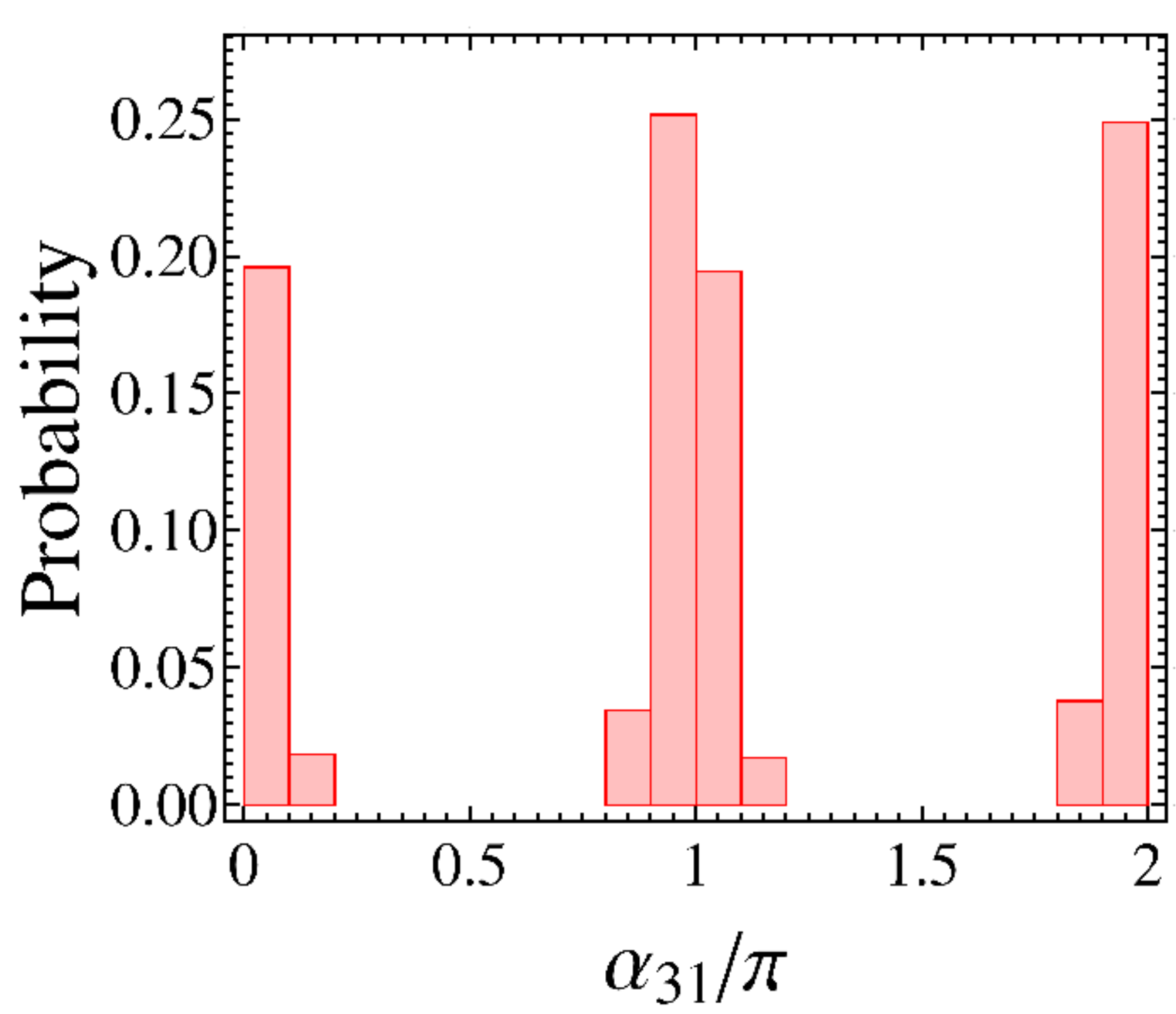}
 \end{tabular}
 \caption{\label{fig:distribution_XI} The probability distribution of
  the lepton mixing parameters $\sin^2\theta_{23}$, $\delta_{CP}$,
  $\alpha_{21}$ and $\alpha_{31}$ predicted for the case of democratic
  CP symmetry. The parameters $\theta_{1,2,3}$ are taken to be random
  numbers in the range $[0, 2\pi]$, and the three lepton mixing angles
  are required to be compatible with experimental data at $3\sigma$
  level~\cite{Forero:2014bxa}.}
 \end{center}
\end{figure}

\section{Numerical analysis}
%
Summarizing the above discussion, we see that type-I, -IV, -V, -VI,
-VII, -IX, -X and -XI residual CP transformations with zero elements
can accommodate the current experimental neutrino oscillation
data~\cite{Forero:2014bxa}. In all these cases, the three mixing
angles $\theta_{12}$, $\theta_{13}$ and $\theta_{23}$ as well as three
CP phases $\delta_{CP}$, $\alpha_{21}$ and $\alpha_{31}$ are found to
depend on just four free independent parameters $\Theta$,
$\theta_{1}$, $\theta_{2}$ and $\theta_{3}$, where $\Theta$
characterizes the shape of the residual CP transformations. This
characterizes the degree of predictivity of our present framework.
\begin{table}[!hptb] \addtolength{\tabcolsep}{-2pt}
 \begin{center}
 \footnotesize{
 \begin{tabular}{|c|c|c|c|c|c|c| } \hline \hline
  \multicolumn{7}{|c|}{Type-VI,~~NH} \\ \hline \hline
   ~~$\Theta$~~ & ~~~~~$\theta_1$~~~~~ &
   ~~~~~~~~~~~~~$\theta_2$~~~~~~~~~~~~~ & ~~~~~$\theta_3$~~~~~ &
   ~~~~$\delta_{CP}$~~~~ & ~$\alpha_{21}$ (mod $\pi$)~ & ~$\alpha_{31}$
   (mod $\pi$)~ \\ \hline
  \multirow{4}{*}{$\frac{\pi }{9}$ }
   & $8.413^{\circ}$ & $48.167^{\circ}$ or $131.833^{\circ}$ &
    $49.591^{\circ}$ &$304.258^{\circ}$ &$147.618^{\circ}$ &
    $159.573^{\circ}$\\\cline{2-7}
   & $8.413^{\circ}$ & $48.167^{\circ}$ or $131.833^{\circ}$ &
    $118.010^{\circ}$ & $136.032^{\circ}$ & $32.382^{\circ}$ &
    $23.728^{\circ}$\\\cline{2-7}
   & $171.587^{\circ}$ & $48.167^{\circ}$ or $131.833^{\circ}$ &
    $61.990^{\circ}$ &$223.968^{\circ}$ &$147.618^{\circ}$ &
    $156.272^{\circ}$\\\cline{2-7}
   & $171.587^{\circ}$ & $48.167^{\circ}$ or $131.833^{\circ}$ &
    $130.409^{\circ}$ &$55.742^{\circ}$ &$32.382^{\circ}$ &
    $20.427^{\circ}$\\\hline
   \multirow{4}{*}{$\frac{2 \pi }{17}$ }
   & $7.645^{\circ}$ & $48.167^{\circ}$ or $131.833^{\circ}$ &
    $50.218^{\circ}$ &$299.581^{\circ}$ &$145.744^{\circ}$ &
    $158.210^{\circ}$\\\cline{2-7}
   & $7.645^{\circ}$ & $48.167^{\circ}$ or $131.833^{\circ}$ &
    $118.541^{\circ}$ &$132.071^{\circ}$ &$34.256^{\circ}$ &
    $24.956^{\circ}$\\\cline{2-7}
   & $172.355^{\circ}$ & $48.167^{\circ}$ or $131.833^{\circ}$ &
    $61.459^{\circ}$ &$227.929^{\circ}$ &$145.744^{\circ}$ &
    $155.044^{\circ}$\\\cline{2-7}
   & $172.355^{\circ}$ & $48.167^{\circ}$ or $131.833^{\circ}$ &
    $129.782^{\circ}$ &$60.419^{\circ}$ &$34.256^{\circ}$ &
    $21.790^{\circ}$\\\hline
   \multirow{4}{*}{$\frac{\pi }{8}$ }
   & $6.601^{\circ}$ & $48.167^{\circ}$ or $131.833^{\circ}$ &
    $51.057^{\circ}$ &$293.664^{\circ}$ &$143.642^{\circ}$ &
    $156.616^{\circ}$\\\cline{2-7}
   & $6.601^{\circ}$ & $48.167^{\circ}$ or $131.833^{\circ}$ &
    $119.265^{\circ}$ &$126.967^{\circ}$ &$36.358^{\circ}$ &
    $26.276^{\circ}$\\\cline{2-7}
   & $173.399^{\circ}$ & $48.167^{\circ}$ or $131.833^{\circ}$ &
    $60.735^{\circ}$ &$233.033^{\circ}$ &$143.642^{\circ}$ &
    $153.724^{\circ}$\\\cline{2-7}
   & $173.399^{\circ}$ & $48.167^{\circ}$ or $131.833^{\circ}$ &
    $128.943^{\circ}$ &$66.336^{\circ}$ &$36.358^{\circ}$ &
    $23.384^{\circ}$\\\hline
   \multirow{4}{*}{$\frac{2 \pi }{15}$ }
   & $5.036^{\circ}$ & $48.167^{\circ}$ or $131.833^{\circ}$ &
    $52.287^{\circ}$ &$285.502^{\circ}$ &$141.268^{\circ}$ &
    $154.690^{\circ}$\\\cline{2-7}
   & $5.036^{\circ}$ & $48.167^{\circ}$ or $131.833^{\circ}$ &
    $120.355^{\circ}$ &$119.731^{\circ}$ &$38.732^{\circ}$ &
    $27.651^{\circ}$\\\cline{2-7}
   & $174.964^{\circ}$ & $48.167^{\circ}$ or $131.833^{\circ}$ &
    $59.645^{\circ}$ &$240.269^{\circ}$ &$141.268^{\circ}$ &
    $152.349^{\circ}$\\\cline{2-7}
   & $174.964^{\circ}$ & $48.167^{\circ}$ or $131.833^{\circ}$ &
    $127.713^{\circ}$ &$74.498^{\circ}$ &$38.732^{\circ}$ &
    $25.310^{\circ}$\\\hline
   \multirow{4}{*}{$\frac{\pi }{7}$ }
   & $1.655^{\circ}$ & $48.167^{\circ}$ or $131.833^{\circ}$ &
    $54.848^{\circ}$ &$269.598^{\circ}$ &$138.569^{\circ}$ &
    $152.045^{\circ}$\\\cline{2-7}
   & $1.655^{\circ}$ & $48.167^{\circ}$ or $131.833^{\circ}$ &
    $122.745^{\circ}$ &$104.895^{\circ}$ &$41.431^{\circ}$ &
    $28.774^{\circ}$\\\cline{2-7}
   & $178.345^{\circ}$ & $48.167^{\circ}$ or $131.833^{\circ}$ &
    $57.255^{\circ}$ &$255.105^{\circ}$ &$138.569^{\circ}$ &
    $151.226^{\circ}$\\\cline{2-7}
   & $178.345^{\circ}$ & $48.167^{\circ}$ or $131.833^{\circ}$ &
    $125.152^{\circ}$ &$90.402^{\circ}$ &$41.431^{\circ}$ &
    $27.955^{\circ}$ \\ \hline \hline
  \end{tabular} }
\end{center} \renewcommand{\arraystretch}{1.0}
\caption{\label{tab:Theta_spe_VI_NH}The predictions for the Dirac and
  Majorana CP phases in the case of type-VI residual CP transformation.
  The parameter $\Theta$ is set to the representative values of $\pi/9$,
  $2\pi/17$, $\pi/8$, $2\pi/15$ and $\pi/7$. The parameters $\theta_1$,
  $\theta_2$ and $\theta_3$ are fixed by the requirement of reproducing
  the best fit values of the three lepton mixing angles for NH neutrino
mass spectrum~\cite{Forero:2014bxa}.}
\end{table}
For example, the type-I CP transformation corresponds to the widely
studied $\mu-\tau$ reflection, and leads to $\theta_{23}=45^{\circ}$,
$\delta_{CP} = \pm 90^{\circ}$ and $\alpha_{21}, \alpha_{31}= 0$ or
$\pi$ while the solar and reactor mixing angles are not
constrained. On the other hand, the type-IV CP transformation with
three texture zeros is diagonal, and corresponds to the conventional
CP transformation. As expected, in this case all three CP phases are
predicted to vanish.
%
The CP transformation of type-V is the same as the
generalized $\mu-\tau$ reflection which has been discussed by us in
Ref.~\cite{Chen:2015siy}. For the case of $\Theta = \pi/2$, our
generalized $\mu-\tau$ reflection reduces to the standard $\mu-\tau$
reflection symmetry.  This would provide an interesting new starting
point for model building if either $\theta_{23}$ or $\delta_{CP}$ were
established to be non-maximal by future neutrino oscillation
experiments.
\begin{table}[hptb] \addtolength{\tabcolsep}{-2pt}
 \begin{center}
 \footnotesize{
  \begin{tabular}{|c|c|c|c|c|c|c| } \hline \hline
   \multicolumn{7}{|c|}{Type-VI,~~IH} \\ \hline \hline
    ~~$\Theta$~~ & ~~~~~$\theta_1$~~~~~ &
    ~~~~~~~~~~~~~$\theta_2$~~~~~~~~~~~~~ & ~~~~~$\theta_3$~~~~~ &
    ~~~~$\delta_{CP}$~~~~ &~$\alpha_{21}$ (mod $\pi$)~ & ~$\alpha_{31}$
    (mod $\pi$)~ \\ \hline
   \multirow{4}{*}{$\frac{\pi }{9}$ }
   & $8.769^{\circ}$ & $48.442^{\circ}$ or $131.558^{\circ}$ &
    $49.303^{\circ}$ &$305.441^{\circ}$ &$147.458^{\circ}$ &
    $159.550^{\circ}$\\\cline{2-7}
   & $8.769^{\circ}$ & $48.442^{\circ}$ or $131.558^{\circ}$ &
    $117.713^{\circ}$ &$137.276^{\circ}$ &$32.542^{\circ}$ &
    $23.926^{\circ}$\\\cline{2-7}
   & $171.231^{\circ}$ & $48.442^{\circ}$ or $131.558^{\circ}$ &
    $62.287^{\circ}$ &$222.724^{\circ}$ &$147.458^{\circ}$ &
    $156.074^{\circ}$\\\cline{2-7}
   & $171.231^{\circ}$ & $48.442^{\circ}$ or $131.558^{\circ}$ &
    $130.697^{\circ}$ &$54.559^{\circ}$ &$32.542^{\circ}$ &
    $20.450^{\circ}$\\\hline
   \multirow{4}{*}{$\frac{2 \pi }{17}$ }
   & $8.037^{\circ}$ & $48.442^{\circ}$ or $131.558^{\circ}$ &
    $49.907^{\circ}$ &$300.956^{\circ}$ &$145.574^{\circ}$ &
    $158.190^{\circ}$\\\cline{2-7}
   & $8.037^{\circ}$ & $48.442^{\circ}$ or $131.558^{\circ}$ &
    $118.220^{\circ}$ &$133.513^{\circ}$ &$34.426^{\circ}$ &
    $25.172^{\circ}$\\\cline{2-7}
   & $171.963^{\circ}$ & $48.442^{\circ}$ or $131.558^{\circ}$ &
    $61.780^{\circ}$ &$226.487^{\circ}$ &$145.574^{\circ}$ &
    $154.828^{\circ}$\\\cline{2-7}
   & $171.963^{\circ}$ & $48.442^{\circ}$ or $131.558^{\circ}$ &
    $130.093^{\circ}$ &$59.044^{\circ}$ &$34.426^{\circ}$ &
    $21.810^{\circ}$\\\hline
   \multirow{4}{*}{$\frac{\pi }{8}$ }
   & $7.054^{\circ}$ & $48.442^{\circ}$ or $131.558^{\circ}$ &
    $50.706^{\circ}$ &$295.348^{\circ}$ &$143.460^{\circ}$ &
    $156.605^{\circ}$\\\cline{2-7}
   & $7.054^{\circ}$ & $48.442^{\circ}$ or $131.558^{\circ}$ &
    $118.903^{\circ}$ &$128.721^{\circ}$ &$36.540^{\circ}$ &
    $26.517^{\circ}$\\\cline{2-7}
   & $172.946^{\circ}$ & $48.442^{\circ}$ or $131.558^{\circ}$ &
    $61.097^{\circ}$ &$231.279^{\circ}$ &$143.460^{\circ}$ &
    $153.483^{\circ}$\\\cline{2-7}
   & $172.946^{\circ}$ & $48.442^{\circ}$ or $131.558^{\circ}$ &
    $129.294^{\circ}$ &$64.652^{\circ}$ &$36.540^{\circ}$ &
    $23.395^{\circ}$\\\hline
   \multirow{4}{*}{$\frac{2 \pi }{15}$ }
   & $5.621^{\circ}$ & $48.442^{\circ}$ or $131.558^{\circ}$ &
    $51.846^{\circ}$ &$287.814^{\circ}$ &$141.074^{\circ}$ &
    $154.704^{\circ}$\\\cline{2-7}
   & $5.621^{\circ}$ & $48.442^{\circ}$ or $131.558^{\circ}$ &
    $119.902^{\circ}$ &$122.120^{\circ}$ &$38.926^{\circ}$ &
    $27.936^{\circ}$\\\cline{2-7}
   & $174.379^{\circ}$ & $48.442^{\circ}$ or $131.558^{\circ}$ &
    $60.098^{\circ}$ &$237.880^{\circ}$ &$141.074^{\circ}$ &
    $152.064^{\circ}$\\\cline{2-7}
   & $174.379^{\circ}$ & $48.442^{\circ}$ or $131.558^{\circ}$ &
    $128.154^{\circ}$ &$72.186^{\circ}$ &$38.926^{\circ}$ &
    $25.296^{\circ}$\\\hline
   \multirow{4}{*}{$\frac{\pi }{7}$ }
   & $3.005^{\circ}$ & $48.442^{\circ}$ or $131.558^{\circ}$ &
    $53.862^{\circ}$ &$275.327^{\circ}$ &$138.360^{\circ}$ &
    $152.241^{\circ}$\\\cline{2-7}
   & $3.005^{\circ}$ & $48.442^{\circ}$ or $131.558^{\circ}$ &
    $121.745^{\circ}$ &$110.708^{\circ}$ &$41.640^{\circ}$ &
    $29.262^{\circ}$\\\cline{2-7}
   & $176.995^{\circ}$ & $48.442^{\circ}$ or $131.558^{\circ}$ &
    $58.255^{\circ}$ &$249.292^{\circ}$ &$138.360^{\circ}$ &
    $150.738^{\circ}$\\\cline{2-7}
   & $176.995^{\circ}$ & $48.442^{\circ}$ or $131.558^{\circ}$ &
    $126.138^{\circ}$ &$84.673^{\circ}$ &$41.640^{\circ}$ &
    $27.759^{\circ}$\\\hline \hline
 \end{tabular}}
 \end{center} \renewcommand{\arraystretch}{1.0}
 \caption{\label{tab:Theta_spe_VI_IH}The predictions for the Dirac and
  Majorana CP phases in the case of type-VI residual CP transformation.
  The parameter $\Theta$ is set to the representative values of $\pi/9$,
  $2\pi/17$, $\pi/8$, $2\pi/15$ and $\pi/7$. The parameters $\theta_1$,
  $\theta_2$ and $\theta_3$ are fixed by the requirement of reproducing
  the best fit values of the three lepton mixing angles for IH neutrino
  mass spectrum~\cite{Forero:2014bxa}.}
\end{table}

As we already mentioned, the lepton mixing matrices for the case of
two--zero texture type-VI and type-VII are related by the exchange of
the second and the third rows. The mixing angles and CP invariants are
given in Eq.~\eqref{eq:mixing_angles_two_zeroBII} and
Eq.~\eqref{eq:invariants_two_zeroBII} respectively. In order to
visualize the theoretical predictions in a more clear way, we perform
a numerical analysis where the free parameters $\theta_{1,2,3}$ and
the CP parameter $\Theta$ are scanned over the range of $[0, 2\pi]$,
while the mixing parameters are calculated for each point, retaining
only points that agree at $3\sigma$ level with experimentally
determined mixing angles~\cite{Forero:2014bxa}. The correlations
between the mixing parameters and distributions of the mixing
parameters are plotted in Fig.~\ref{fig:correlation_two_zero2} and
Fig.~\ref{fig:distribution_two_zero2}.

One sees that the three CP phases are strongly correlated with each
other, the Majorana phase $\alpha_{21}$ around $\pi/5$, $4\pi/5$,
$6\pi/5$ and $9\pi/5$ is preferred, and the Majorana phase
$\alpha_{31}$ around $3\pi/20$, $17\pi/20$, $23\pi/23$ and $37\pi/20$
is favored. If we set a value to the CP parameter $\Theta$ then the
explicit form of the CP transformation ${\bf X}$ is fixed, so that
definite predictions for the CP phases are obtained.
As examples, we consider the case that the parameter $\Theta$ takes
some specific values $\frac{\pi}{9}$, $\frac{2\pi}{17}$,
$\frac{\pi}{8}$, $\frac{2\pi}{15}$ and $\frac{\pi}{7}$, the values of
the parameters $\theta_1$, $\theta_2$ and $\theta_3$ are determined by
the experimental best fit values of the lepton mixing angles
from~\cite{Forero:2014bxa}. As a consequence, the lepton mixing matrix
is fully fixed up to the factor $\hat{X}^{-1/2}$, and the values of
the CP violating phases can be predicted, as are shown in
Table~\ref{tab:Theta_spe_VI_NH} for normal hierarchy (NH) and
Table~\ref{tab:Theta_spe_VI_IH} for inverted hierarchy (IH). We can
see that different values of the Dirac CP phase $\delta_{CP}$ can be
achieved. Note in particular that, for certain values of $\Theta$,
$\theta_1$, $\theta_2$ and $\theta_3$, the magnitude of $\delta_{CP}$
can be quite close to $270^{\circ}$ which is weakly favored by present
data~\cite{Abe:2015awa}.

The lepton mixing matrices for the one zero textures type-IX and -X
differ by a permutation of the second and the third rows. The
expressions for the mixing angles and CP invariants are given in
Eqs.~(\ref{eq:mix_par_10II}, \ref{eq:mix_par_10II-1},
\ref{eq:mix_par_IX_I1}, \ref{eq:mix_par_IX_I2}). The numerical results
for the correlation among the mixing parameters and probability
distributions of the mixing parameters are displayed in
Fig.~\ref{fig:correlation_one_zero2} and
Fig.~\ref{fig:distribution_one_zero2}. The strong correlations between
different CP phases emerge once the value of the parameter $\Theta$ is
fixed.

We see that $\theta_{23}$ close to maximal mixing is favored for the
type-X texture, and both Majorana phases $\alpha_{21}$ and
$\alpha_{31}$ tend to close to $0$, $\pi$ and $2\pi$. There appears to
be no preferred $\delta_{CP}$ phase within the viable parameter space.
Furthermore, we study some concrete benchmark cases in which the
parameters $\Theta$ take on certain representative values. The value
of the parameters $\theta_1$, $\theta_2$ and $\theta_3$ are fixed by
the best fit value of the lepton mixing angles. In this way the CP
violating phases can be predicted as listed in
Table~\ref{tab:Theta_spe_X_NH} and Table~\ref{tab:Theta_spe_X_IH} for
NH and IH mass spectrums respectively.  Future long baseline
facilities DUNE~\cite{Adams:2013qkq}, LBNO~\cite{::2013kaa},
T2HK~\cite{Kearns:2013lea} can bring us increased precision on the
Dirac phase $\delta_{CP}$.
If $\delta_{CP}$ was measured to be far from any of the values in
Table~\ref{tab:Theta_spe_X_NH} and Table~\ref{tab:Theta_spe_X_IH}, the
present proposal would be disfavored.

\begin{table}[!htbp] \addtolength{\tabcolsep}{-2pt}
 \begin{center}
 {\footnotesize
 \begin{tabular}{|c|c|c|c|c|c|c| } \hline \hline
  \multicolumn{7}{|c|}{Type-X,~~NH} \\ \hline\hline
   ~~$\Theta$~~ & ~~~~~$\theta_1$~~~~~ & ~~~~~$\theta_2$~~~~~ &
   ~~~~~$\theta_3$~~~~~ & ~~~~$\delta_{CP}$~~~~ &  ~$\alpha_{21}$
   (mod $\pi$)~ & ~$\alpha_{31}$ (mod $\pi$)~ \\ \hline
   \multirow{4}{*}{$\frac{\pi }{9}$ }
   & $66.743^{\circ}$ & $4.290^{\circ}$ & $57.595^{\circ}$ &
    $108.232^{\circ}$ &$60.800^{\circ}$ &$40.840^{\circ}$\\\cline{2-7}
   & $66.743^{\circ}$ & $4.290^{\circ}$ & $123.826^{\circ}$ &
    $264.695^{\circ}$ &$119.200^{\circ}$ &
    $136.503^{\circ}$\\\cline{2-7}
   & $66.743^{\circ}$ & $175.710^{\circ}$ & $57.595^{\circ}$ &
    $251.768^{\circ}$ &$119.200^{\circ}$ &
    $139.160^{\circ}$\\\cline{2-7}
   & $66.743^{\circ}$ & $175.710^{\circ}$ & $123.826^{\circ}$ &
    $95.305^{\circ}$ &$60.800^{\circ}$ &$43.497^{\circ}$\\\hline
   \multirow{4}{*}{$\frac{2 \pi }{17}$ }
   & $66.215^{\circ}$ & $12.491^{\circ}$ & $58.980^{\circ}$ &
    $122.397^{\circ}$ &$62.920^{\circ}$ &
    $39.600^{\circ}$\\\cline{2-7}
   & $66.215^{\circ}$ & $12.491^{\circ}$ & $124.976^{\circ}$ &
    $277.881^{\circ}$ &$117.080^{\circ}$ &
    $132.164^{\circ}$\\\cline{2-7}
   & $66.215^{\circ}$ & $167.509^{\circ}$ & $58.980^{\circ}$ &
    $237.603^{\circ}$ &$117.080^{\circ}$ &
    $140.400^{\circ}$\\\cline{2-7}
   & $66.215^{\circ}$ & $167.509^{\circ}$ & $124.976^{\circ}$ &
    $82.119^{\circ}$ &$62.920^{\circ}$ &$47.836^{\circ}$\\\hline
   \multirow{4}{*}{$\frac{2 \pi }{9}$ }
   & $66.060^{\circ}$ & $13.984^{\circ}$ & $60.881^{\circ}$ &
    $245.654^{\circ}$ &$117.249^{\circ}$ &$76.422^{\circ}$\\\cline{2-7}
   & $66.060^{\circ}$ & $13.984^{\circ}$ & $114.766^{\circ}$ &
    $5.228^{\circ}$ &$62.751^{\circ}$ &$78.747^{\circ}$\\\cline{2-7}
   & $66.060^{\circ}$ & $166.016^{\circ}$ & $60.881^{\circ}$ &
    $114.346^{\circ}$ &$62.751^{\circ}$ &$103.578^{\circ}$\\\cline{2-7}
   & $66.060^{\circ}$ & $166.016^{\circ}$ & $114.766^{\circ}$ &
    $354.772^{\circ}$ &$117.249^{\circ}$ &$101.253^{\circ}$\\\hline
   \multirow{4}{*}{$\frac{3 \pi }{13}$ }
   & $66.499^{\circ}$ & $9.096^{\circ}$ & $62.706^{\circ}$ &
    $268.830^{\circ}$ &$124.491^{\circ}$ &$87.338^{\circ}$\\\cline{2-7}
   & $66.499^{\circ}$ & $9.096^{\circ}$ & $113.502^{\circ}$ &
    $20.806^{\circ}$ &$55.509^{\circ}$ &$74.823^{\circ}$\\\cline{2-7}
   & $66.499^{\circ}$ & $170.904^{\circ}$ & $62.706^{\circ}$ &
    $91.170^{\circ}$ &$55.509^{\circ}$ &$92.662^{\circ}$\\\cline{2-7}
   & $66.499^{\circ}$ & $170.904^{\circ}$ & $113.502^{\circ}$ &
    $339.194^{\circ}$ &$124.491^{\circ}$ &$105.177^{\circ}$
    \\\hline \hline
  \end{tabular}
 }
 \end{center} \renewcommand{\arraystretch}{1.0}
 \caption{\label{tab:Theta_spe_X_NH}The predictions for the Dirac and Majorana
  CP phases in the case of type-X residual CP transformation. The parameter
  $\Theta$ is set to the representative values of $\pi/9$. $2\pi/17$, $2\pi/9$
  and $3\pi/13$. The parameters $\theta_1$, $\theta_2$ and $\theta_3$ are
  fixed by the requirement of reproducing the best fit values of the three
  lepton mixing angles for NH neutrino mass spectrum~\cite{Forero:2014bxa}.}
\end{table}
\begin{table}[h!]\addtolength{\tabcolsep}{-2pt}
 \begin{center}
 {\footnotesize
  \begin{tabular}{|c|c|c|c|c|c|c| } \hline\hline
   \multicolumn{7}{|c|}{Type-X,~~IH} \\\hline\hline
   ~~$\Theta$~~ & ~~~~~$\theta_1$~~~~~ & ~~~~~$\theta_2$~~~~~ &
   ~~~~~$\theta_3$~~~~~ & ~~~~$\delta_{CP}$~~~~ &
   ~$\alpha_{21}$ (mod $\pi$)~ & ~$\alpha_{31}$ (mod $\pi$)~ \\\hline
   \multirow{4}{*}{$\frac{2\pi}{17}$}
   & $65.802^{\circ}$ & $6.854^{\circ}$ & $58.218^{\circ}$ &
    $113.294^{\circ}$ &$64.070^{\circ}$ &$42.230^{\circ}$\\\cline{2-7}
   & $65.802^{\circ}$ & $6.854^{\circ}$ & $124.081^{\circ}$ &
    $268.239^{\circ}$ &$115.930^{\circ}$ &
    $133.106^{\circ}$\\\cline{2-7}
   & $65.802^{\circ}$ & $173.146^{\circ}$ & $58.218^{\circ}$ &
    $246.706^{\circ}$ &$115.930^{\circ}$ &
    $137.770^{\circ}$\\\cline{2-7}
   & $65.802^{\circ}$ & $173.146^{\circ}$ & $124.081^{\circ}$ &
    $91.761^{\circ}$ &$64.070^{\circ}$ &$46.894^{\circ}$\\\hline
   \multirow{4}{*}{$\frac{\pi }{8}$ }
   & $65.244^{\circ}$ & $13.626^{\circ}$ & $59.379^{\circ}$ &
    $126.150^{\circ}$ &$66.506^{\circ}$ &$41.401^{\circ}$\\\cline{2-7}
   & $65.244^{\circ}$ & $13.626^{\circ}$ & $124.950^{\circ}$ &
    $279.935^{\circ}$ &$113.494^{\circ}$ &
    $128.680^{\circ}$\\\cline{2-7}
   & $65.244^{\circ}$ & $166.374^{\circ}$ & $59.379^{\circ}$ &
    $233.850^{\circ}$ &$113.494^{\circ}$ &
    $138.599^{\circ}$\\\cline{2-7}
   & $65.244^{\circ}$ & $166.374^{\circ}$ & $124.950^{\circ}$ &
    $80.065^{\circ}$ &$66.506^{\circ}$ &$51.320^{\circ}$\\\hline
   \multirow{4}{*}{$\frac{2 \pi }{15}$ }
   & $64.729^{\circ}$ & $17.577^{\circ}$ & $59.985^{\circ}$ &
    $136.166^{\circ}$ &$69.451^{\circ}$ &$41.655^{\circ}$\\\cline{2-7}
   & $64.729^{\circ}$ & $17.577^{\circ}$ & $125.180^{\circ}$ &
    $288.512^{\circ}$ &$110.549^{\circ}$ &
    $124.550^{\circ}$\\\cline{2-7}
   & $64.729^{\circ}$ & $162.423^{\circ}$ & $59.985^{\circ}$ &
    $223.834^{\circ}$ &$110.549^{\circ}$ &
    $138.345^{\circ}$\\\cline{2-7}
   & $64.729^{\circ}$ & $162.423^{\circ}$ & $125.180^{\circ}$ &
    $71.488^{\circ}$ &$69.451^{\circ}$ &$55.450^{\circ}$\\\hline
   \multirow{4}{*}{$\frac{3 \pi }{13}$ }
   & $65.352^{\circ}$ & $12.624^{\circ}$ & $61.950^{\circ}$ &
    $253.073^{\circ}$ &$122.680^{\circ}$ &$82.112^{\circ}$\\\cline{2-7}
   & $65.352^{\circ}$ & $12.624^{\circ}$ & $113.570^{\circ}$ &
    $7.050^{\circ}$ &$57.320^{\circ}$ &$73.409^{\circ}$\\\cline{2-7}
   & $65.352^{\circ}$ & $167.376^{\circ}$ & $61.950^{\circ}$ &
    $106.927^{\circ}$ &$57.320^{\circ}$ &$97.888^{\circ}$\\\cline{2-7}
   & $65.352^{\circ}$ & $167.376^{\circ}$ & $113.570^{\circ}$ &
    $352.950^{\circ}$ &$122.680^{\circ}$ &$106.591^{\circ}$\\\hline
   \multirow{4}{*}{$\frac{4 \pi }{17}$ }
   & $65.587^{\circ}$ & $10.048^{\circ}$ & $63.026^{\circ}$ &
    $265.635^{\circ}$ &$126.593^{\circ}$ &$88.250^{\circ}$\\\cline{2-7}
   & $65.587^{\circ}$ & $10.048^{\circ}$ & $112.818^{\circ}$ &
    $15.195^{\circ}$ &$53.407^{\circ}$ &$71.217^{\circ}$\\\cline{2-7}
   & $65.587^{\circ}$ & $169.952^{\circ}$ & $63.026^{\circ}$ &
    $94.365^{\circ}$ &$53.407^{\circ}$ &$91.750^{\circ}$\\\cline{2-7}
   & $65.587^{\circ}$ & $169.952^{\circ}$ & $112.818^{\circ}$ &
    $344.805^{\circ}$ &$126.593^{\circ}$ &$108.783^{\circ}$
   \\\hline \hline
 \end{tabular}
 }
 \end{center} \renewcommand{\arraystretch}{1.0}
 \caption{\label{tab:Theta_spe_X_IH}The predictions for the Dirac and Majorana
  CP phases in the case of type-X residual CP transformation. The parameter
  $\Theta$ is set to the representative values of $2\pi/17$, $\pi/8$,
  $2\pi/15$, $3\pi/13$ and $4\pi/17$ . The parameters $\theta_1$, $\theta_2$
  and $\theta_3$ are fixed by the requirement of reproducing the best fit
  values of the three lepton mixing angles for IH neutrino mass
  spectrum~\cite{Forero:2014bxa}.}
\end{table}
We perform a numerical analysis by treating the parameters $\theta_{1,2,3}$ as
random real numbers scanned over the range $[0, 2\pi]$, with the three mixing
angles calculated for each point in the parameter space. Subsequently only
points which simultaneously are be compatible with experimental
data~\cite{Forero:2014bxa} are retained and from these points the CP violating
phases are calculated. The predicted distributions of the lepton mixing
parameters are shown in Fig.~\ref{fig:distribution_XI}.

Since no specific values of $\theta_{12}$ and $\theta_{13}$ are favored within
$3\sigma$, and hence they are not shown in the figure. We see that the
atmospheric mixing angle $\theta_{23}$ can be either the first octant or the
second octant. As regards the CP phases, there appears to be a slight
preference for $\delta_{CP}\sim\pi/2$ and $\delta_{CP}\sim3\pi/2$, and the
Majorana phase $\alpha_{21}$ around $\pi/4$, $3\pi/4$, $5\pi/4$ and $7\pi/4$
are favored while the values of $\alpha_{31}$ around $0$ and $\pi$ are
preferred. For certain values of $\theta_{1, 2, 3}$, the best fit values of
the mixing angles can be reproduced, and the corresponding predictions for CP
phases are listed in Table~\ref{tab:theta_spe_XI}.
\begin{table}[!htbp] \addtolength{\tabcolsep}{-2pt}
 \begin{center}
  \begin{tabular}{|c|c|c|c|c|c|c|} \hline\hline
  \multicolumn{7}{|c|}{Type-XI} \\\hline\hline
   & ~~~~~$\theta_1$~~~~~ & ~~~~$\theta_2$~~~~ & ~~~~~$\theta_3$~~~~~ &
   ~~~~$\delta_{CP}$~~~~ & ~$\alpha_{21}$ (mod $\pi$)~ &
   ~$\alpha_{31}$ (mod $\pi$)~  \\\hline
  \multirow{4}{*}{~NH~}
   & $13.015^{\circ}$ & $2.354^{\circ}$ & $127.433^{\circ}$ &$65.801^{\circ}$
   & $59.150^{\circ}$ & $172.892^{\circ}$ \\\cline{2-7}
  & $13.015^{\circ}$ & $2.354^{\circ}$ & $179.849^{\circ}$ &$309.873^{\circ}$
   & $120.850^{\circ}$ & $177.814^{\circ}$ \\\cline{2-7}
  & $176.556^{\circ}$ & $9.122^{\circ}$ & $2.170^{\circ}$ &$172.589^{\circ}$
   & $123.407^{\circ}$ & $6.067^{\circ}$ \\\cline{2-7}
  & $176.556^{\circ}$ & $9.122^{\circ}$ & $53.464^{\circ}$ &$285.771^{\circ}$
   & $56.593^{\circ}$ & $175.842^{\circ}$\\\hline
  \multirow{4}{*}{~IH~}
   & $13.424^{\circ}$ & $2.113^{\circ}$ & $127.462^{\circ}$ & $64.401^{\circ}$
    & $59.347^{\circ}$ & $172.721^{\circ}$ \\\cline{2-7}
  & $13.424^{\circ}$ & $2.113^{\circ}$ & $179.962^{\circ}$ & $308.266^{\circ}$
   & $120.653^{\circ}$ & $177.239^{\circ}$ \\\cline{2-7}
  & $176.886^{\circ}$ & $9.403^{\circ}$ & $2.220^{\circ}$ &$173.932^{\circ}$
   & $123.420^{\circ}$ & $6.171^{\circ}$ \\\cline{2-7}
  & $176.886^{\circ}$ & $9.403^{\circ}$ & $53.508^{\circ}$ & $287.100^{\circ}$
   & $56.580^{\circ}$ & $175.919^{\circ}$ \\ \hline \hline
  \end{tabular}
 \end{center} \renewcommand{\arraystretch}{1.0}
 \caption{\label{tab:theta_spe_XI}The predictions for the Dirac and Majorana
  CP phases for the democratic residual CP transformation. The values of
  $\theta_1$, $\theta_2$ and $\theta_3$ are fixed by the requirement of
  accommodating the best fit values of the three lepton mixing
  angles~\cite{Forero:2014bxa}. }
\end{table}
\section{Phenomenological implications}
Implications of the generalized $\mu-\tau$ reflection symmetry have
already been discussed in Ref.~\cite{Chen:2015siy}. In this section,
we shall consider the phenomenological implications of the residual CP
transformations as we have classified above in
Tables~\ref{tab:four_zeros},~\ref{tab:two_zeros} and
\ref{tab:one_zero}, focussing on the case of ``neutrino appearance''
oscillation experiments and neutrinoless double beta decay. The
cosmological implications for leptogenesis will be studied as well.
%
\subsection{CP violation in conventional neutrino oscillations}
%
\begin{figure}[!hptb]
 \begin{center}
 \begin{tabular}{ccc}
  \includegraphics[width=0.45\linewidth]{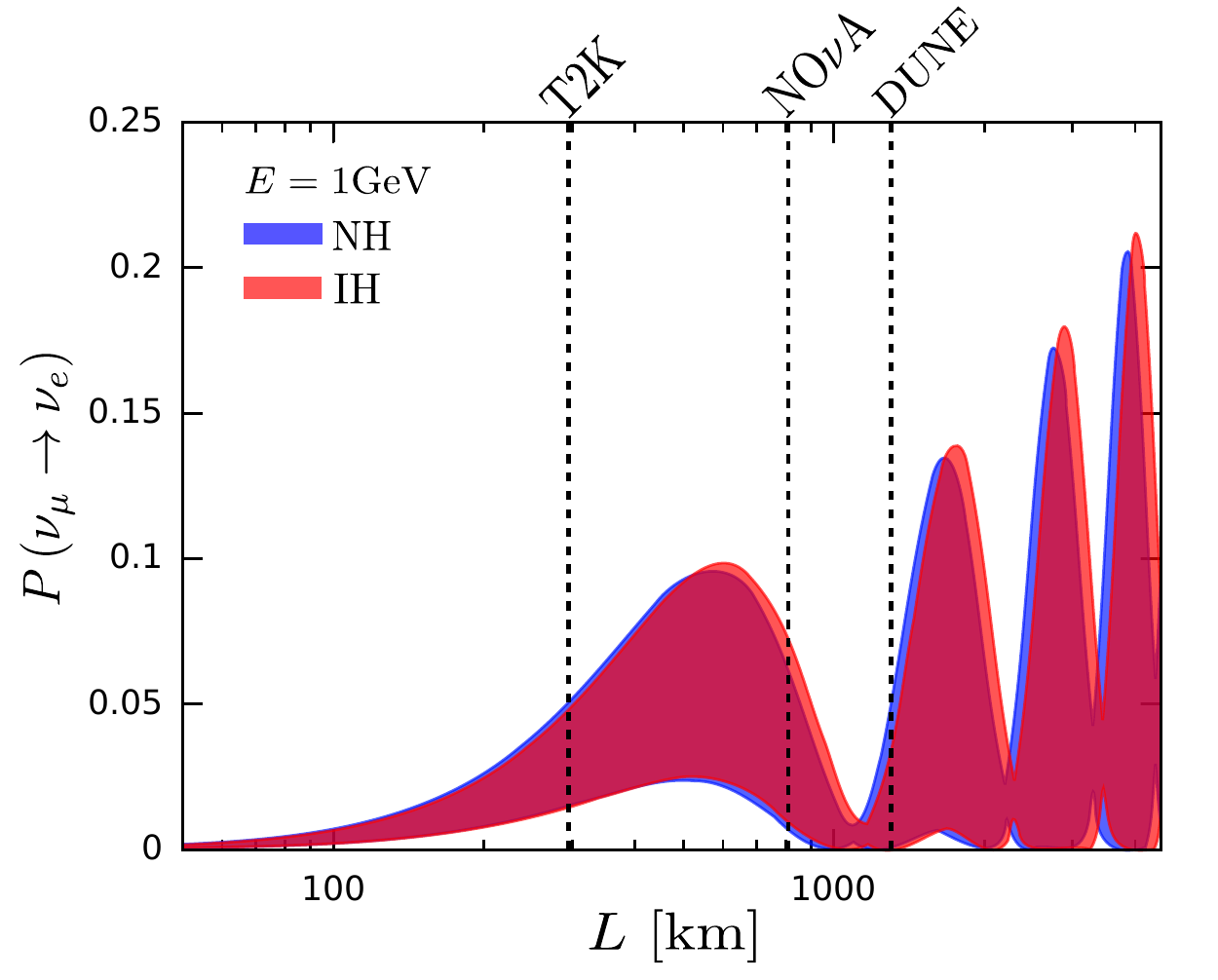} &
  \includegraphics[width=0.45\linewidth]{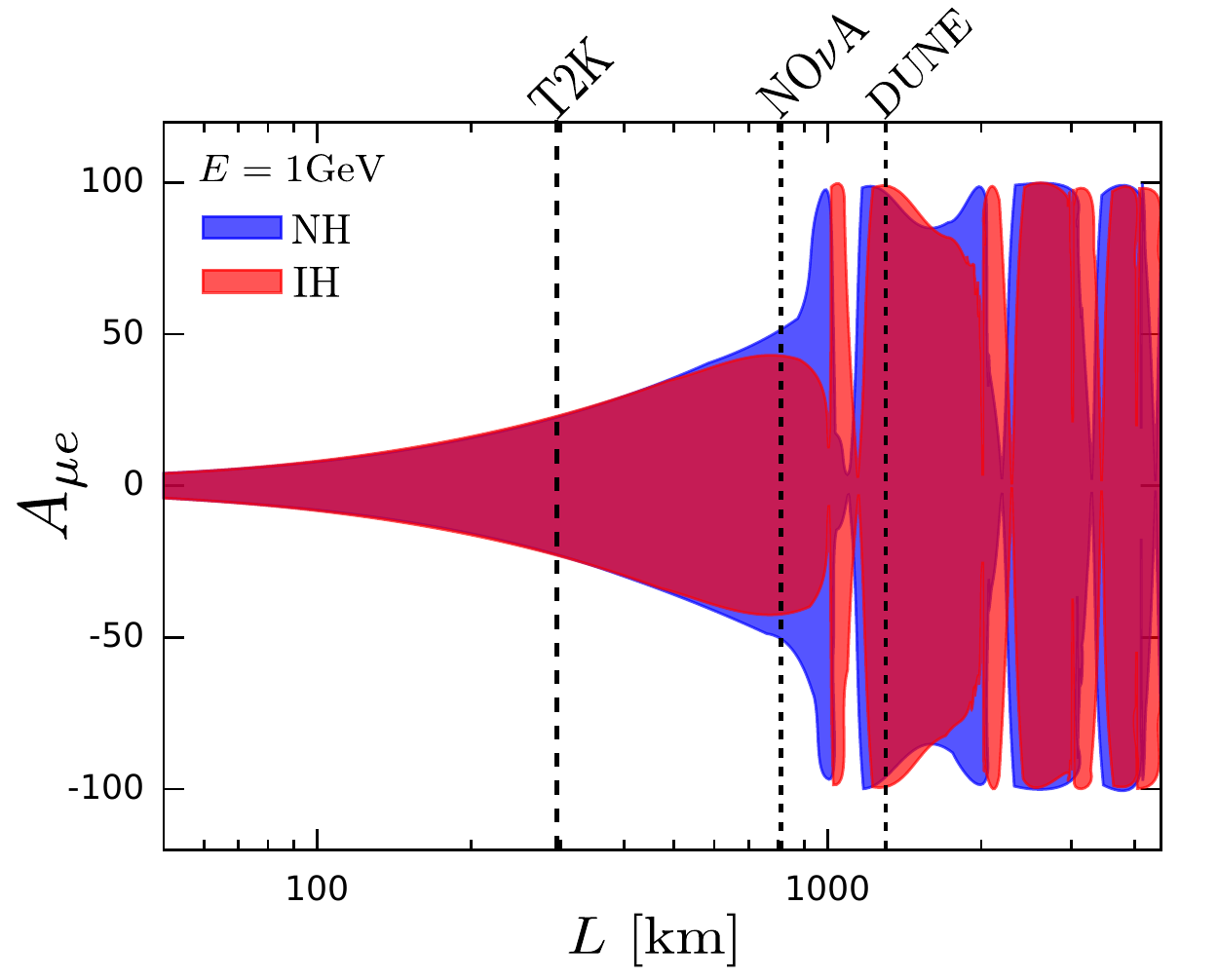}
 \end{tabular}
 \caption{\label{Fig:Delta_mu_e}In the left panel we show the
  $\nu_{\mu} \to \nu_{e}$ transition probability in matter for a
  neutrino energy of $E=1$GeV. The right panel displays the
  neutrino-anti-neutrino asymmetry $\mathcal{A}_{\mu e}$ in
  matter. The oscillation parameters are taken within their currently
  allowed 3$\sigma$ regions~\cite{Forero:2014bxa}. The plot correponds
  to type-VI residual CP symmetry. }
 \end{center}
\end{figure}
\begin{figure}[!hptb]
 \begin{center}
 \begin{tabular}{cc}
  \includegraphics[width=0.44\linewidth]{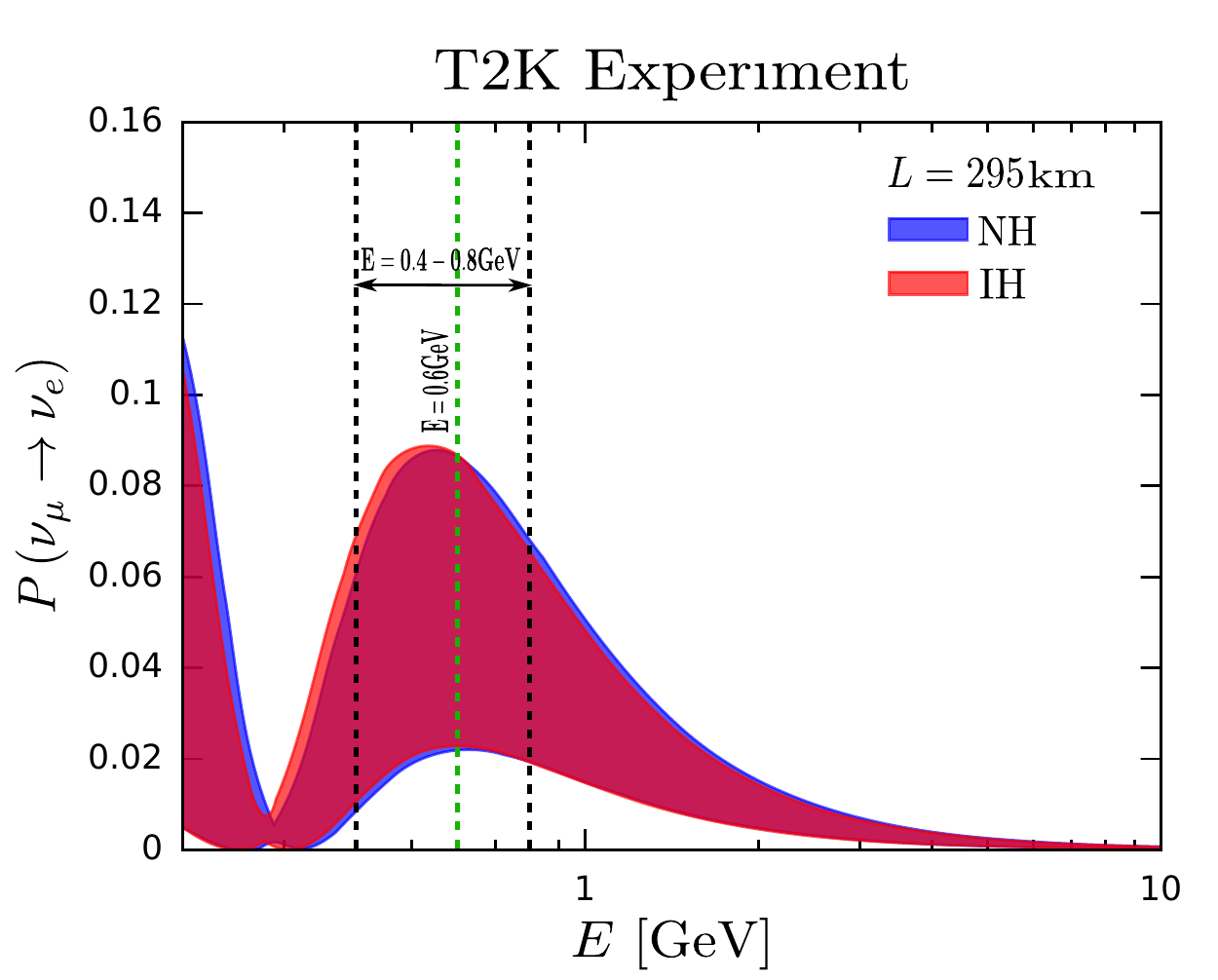} &
  \includegraphics[width=0.44\linewidth]{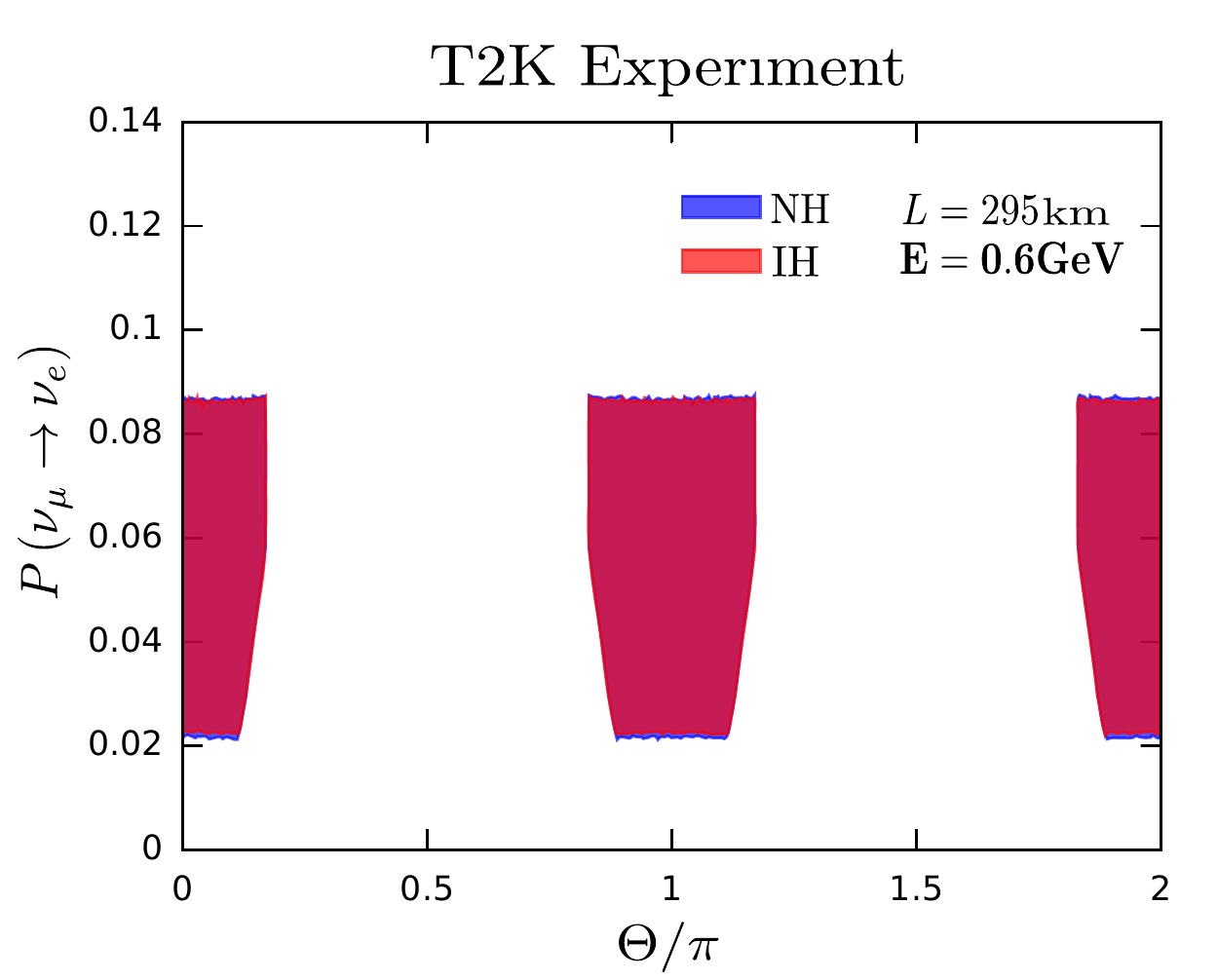}
 \end{tabular}
 \caption{The transition probability
  $P\left({\nu_{\mu}\to\nu_{e}}\right)$ at a baseline of 295km which
  corresponds to the T2K experiment. The neutrino oscillation
  parameters are taken within the currently allowed 3$\sigma$
  regions~\cite{Forero:2014bxa}. The plot correponds to the case of
  type-VI residual CP symmetry.\label{Fig:Exp:T2K}}
 \end{center}
\end{figure}
The existence of leptonic CP violation would manifest itself as the
differences in the oscillation probabilities involving neutrinos and
anti-neutrinos in vacuum~\cite{Nunokawa2008338}:
$$\Delta P_{\alpha \beta} \equiv P \left( \nu_{\alpha} \to \nu_{ \beta }
\right) - P \left( \bar{\nu}_{\alpha} \to \bar{\nu}_{ \beta } \right) =
- 16 \, J_{\alpha \beta} \, \sin \Delta_{ 21 } \sin \Delta_{ 23 }
\sin \Delta_{ 31 },$$
where we have adopted standard definitions
$\Delta_{kj} \equiv \Delta m_{kj}^{2} L /(4E)$ and
$\Delta m_{kj}^{2} = m_{k}^{2} - m_{j}^{2}$, $L$ is the baseline and $E$
stands for the energy of neutrino beam. The Jarlskog invariant is identified
as
\begin{equation}
 J_{\alpha \beta} =
 \Im \left( U_{\alpha 1} U_{\beta 2} U^{\ast}_{\alpha 2}
 U^{\ast}_{\beta 1}\right) = \pm J_{CP}\,,
\end{equation}
where the positive (negative) sign holds for (anti-)cyclic permutation
of the flavour indices $e$, $\mu$ and $\tau$. For the oscillation
between electron and muon neutrinos, the transition probability of
$\nu_{\mu}\to\nu_{e}$ in vacuum is given by~\cite{Nunokawa2008338}
\begin{equation}
 P \left( \nu_{\mu} \to \nu_{e} \right) \simeq
 P_{ \mathrm{atm} } + 2 \sqrt{ P_{ \mathrm{atm} } }
 \sqrt{ P_{ \mathrm{sol} } } \cos \left( \Delta_{32} + \delta_{ \mathrm{CP} }
 \right) + P_{ \mathrm{sol} }\,,
\end{equation}
where
$\sqrt{ P_{ \mathrm{atm} } } = \sin \theta_{23} \sin 2 \theta_{13}
\sin \Delta_{31}$ and $ \sqrt{ P_{ \mathrm{sol} } } = \cos \theta_{23}
\cos \theta_{13} \sin 2\theta_{12} \sin \Delta_{21}$.
As a result, the oscillation probability asymmetry between neutrinos
and anti-neutrinos in vacuum is of the form:
\begin{equation} \label{Eq:Asym:e_mu:1}
 A_{\mu e} =
 \frac{ P \left( \nu_{\mu} \to \nu_{e} \right) -
  P \left( \bar{\nu}_{\mu} \to \bar{\nu}_{e} \right)
 }{
  P \left( \nu_{\mu} \to \nu_{e} \right) +
  P \left( \bar{\nu}_{\mu} \to \bar{\nu}_{e} \right) }
  = \frac{ 2 \sqrt{ P_{ \mathrm{atm} } }
  \sqrt{ P_{ \mathrm{sol} } } \sin \Delta_{32} \sin \delta_{ \mathrm{CP} }
  }{ P_{\mathrm{atm}} + 2 \sqrt{ P_{ \mathrm{atm} } }
  \sqrt{ P_{ \mathrm{sol} } } \cos \Delta_{32} \cos \delta_{ \mathrm{CP} } +
  P_{ \mathrm{sol} } }\,.
\end{equation}
In order to accurately describe realistic long baseline neutrino
oscillation experiments such as T2K, NO$\nu$A or the DUNE proposal, it
is important to include the matter effect associated with neutrino
propagation inside the Earth. Indeed the latter could induce a fake CP
violation effect. In this case the expressions for
$\sqrt{P_{\mathrm{atm}}}$ and $\sqrt{P_{\mathrm{sol}}}$ in matter take
the form:
\begin{equation}
 \sqrt{ P_{ \mathrm{atm} } }
 = \sin \theta_{23} \sin 2\theta_{13}
  \frac{ \sin \left( \Delta_{31}-aL \right) }{ \Delta_{31} - aL } \,
  \Delta_{31}\,,\quad
 \sqrt{ P_{ \mathrm{sol} } } =
  \cos \theta_{23} \sin 2\theta_{12} \frac{ \sin(aL) }{ aL } \,\Delta_{21}\,,
\end{equation}
where $a=G_{F}N_{e}/\sqrt{2}$, $G_{F}$ is the Fermi constant and
$N_{e}$ is the density of electrons. The parameter $a$ is
approximately equal to $(3500\mathrm{km})^{-1}$ for $\rho Y_{e} =
3.0\textrm{g\,cm}^{-3}$, where $Y_{e}$ is the electron
fraction~\cite{Nunokawa2008338}.

In Fig.~\ref{Fig:Delta_mu_e} we show the $\nu_{\mu} \to \nu_{e}$
transition probability as well as the neutrino-anti-neutrino asymmetry
in matter, when the residual CP transformation matrix $\mathbf{X}$ is
assumed to be type-VI. In this figure we require the oscillation
mixing angles lie within their currently allowed 3$\sigma$
region~\cite{Forero:2014bxa}.
\begin{figure}[!htbp]
 \begin{center}
 \begin{tabular}{cc}
  \includegraphics[width=0.44\linewidth]{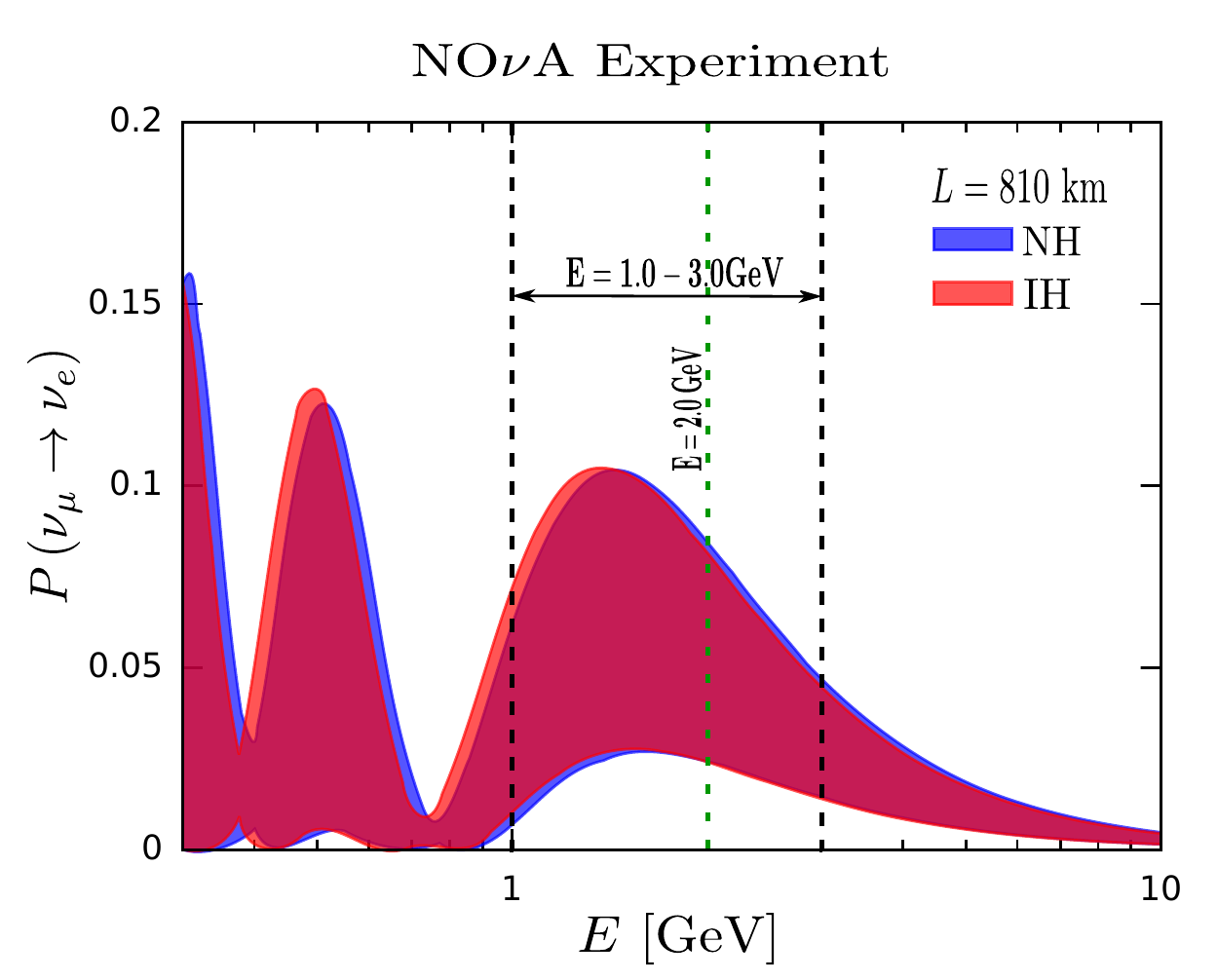} &
  \includegraphics[width=0.44\linewidth]{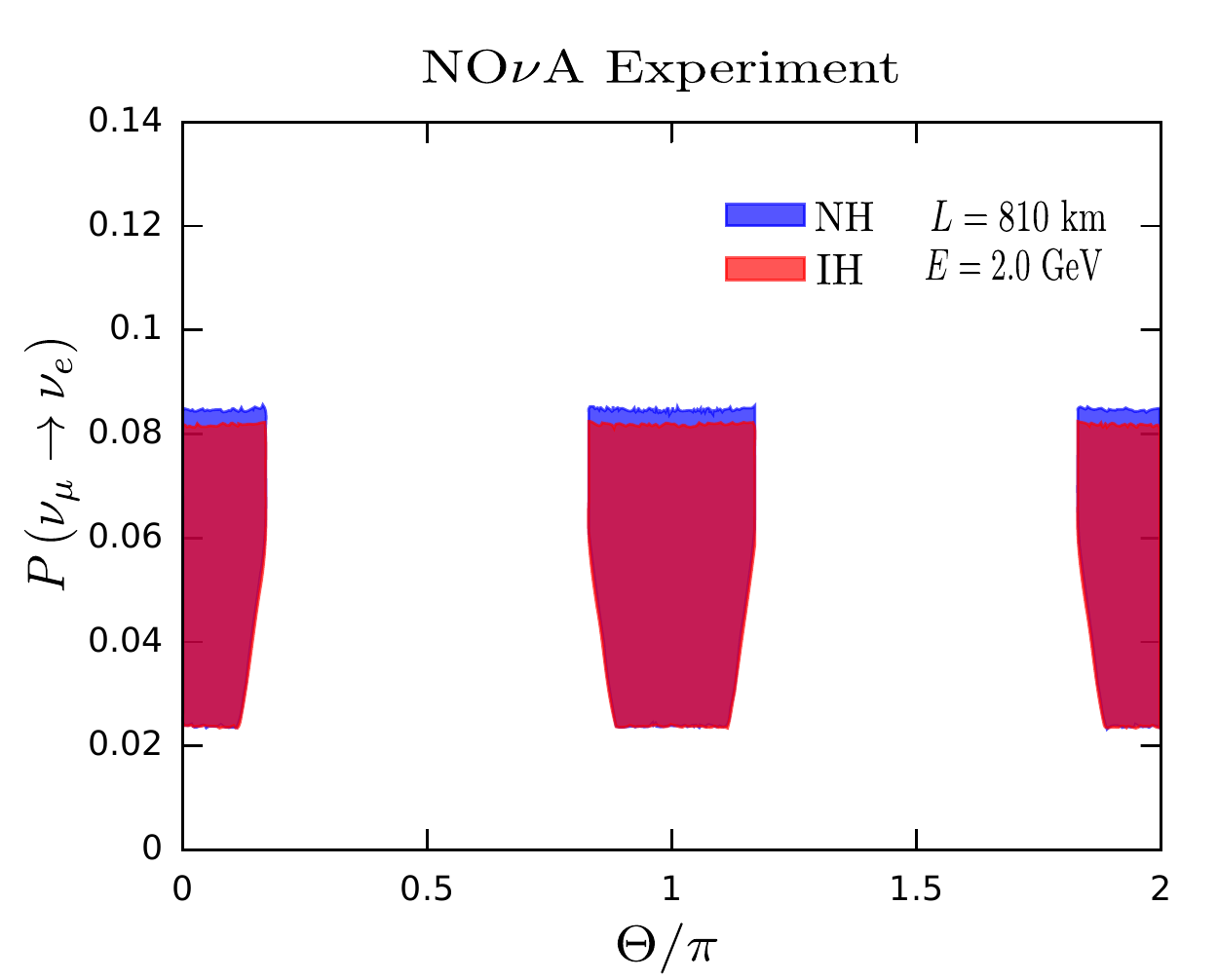}
 \end{tabular}
 \caption{\label{Fig:Exp:NOvA}The transition probability
  $P\left({\nu_{\mu}\to \nu_{e}}\right)$ for type-VI residual CP
  symmetry case at a baseline of 810km which corresponds to the
  NO$\nu$A experiment. The neutrino oscillation parameters are taken
  within their currently allowed 3$\sigma$
  regions~\cite{Forero:2014bxa}. }
 \end{center}
\end{figure}

In Figs.~\ref{Fig:Exp:T2K},~\ref{Fig:Exp:NOvA} we show the
corresponding behavior of the transition probability
$P\left({\nu_{\mu}\to\nu_{e}} \right)$ in terms of neutrino energy
$E$, as well as of the CP parameter $\Theta$ describing our approach,
when the CP symmetry matrix ${\bf X}$ to be type-VI, for baseline
values 295 and 810~km, which correspond to the current T2K and
NO$\nu$A experiments, respectively. One sees that the allowed values
of the CP parameter $\Theta$ describing our approach are quite
restricted.
%
\subsection{Neutrinoless double decay}
%
The rare decay $\left(A, Z\right)\to\left(A, Z+2\right)+e^{-}+e^{-}$
is the lepton number violating process ``par excellence''. Its
observation would establish the Majorana nature of neutrinos
irrespective of their underlying mass generation
mechanism~\cite{Schechter:1981bd,Duerr:2011zd}. Within the simplest
``long-range'' light neutrino exchange mechanism its amplitude is
sensitive to the Majorana phases. As discussed
in~\cite{Rodejohann:2011vc} the most convenient parametrization of the
lepton mixing matrix for the description of neutrinoless double decay
is the fully symmetric one~\cite{Schechter:1980gr}. However, instead
of using the ``symmetrical'' description as in~\cite{Chen:2015siy},
here we stick to the PDG form~\cite{Agashe:2014kda}.

\begin{figure}[!htbp]
 \begin{center}
 \begin{tabular}{cc}\hskip-0.6in
  \includegraphics[width=0.56\linewidth]{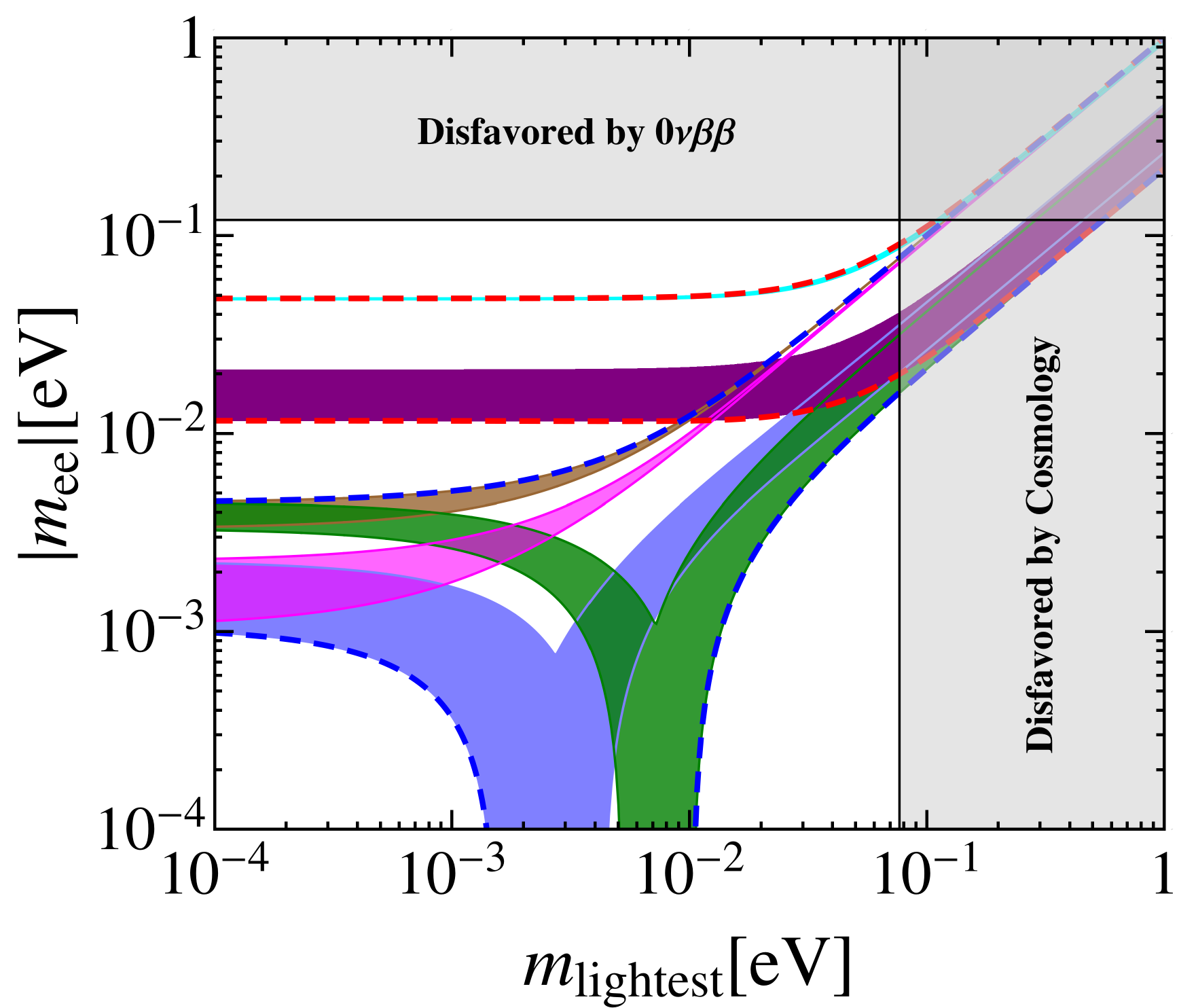}
 \end{tabular}
 \caption{\label{fig:2beta_I}The effective mass $\left| m_{ee}\right|$
  describing neutrinoless double beta decay for type-I CP
  symmetry. Both reactor and solar mixing angles are required to be
  within the experimental $3\sigma$ interval~\cite{Forero:2014bxa},
  while the atmospheric mixing angle is predicted to be maximal with
  $\theta_{23}=\pi/4$. For the inverted neutrino mass ordering, the
  cyan region corresponds to $(k_{1} , k_{2})=(0,0), (0,1)$, and the
  purple area corresponds to $(k_{1}, k_{2})=(1, 0)$, $(1, 1)$. For
  the normal ordering, the brown, magenta, blue and dark green regions
  correspond to $(k_{1}, k_{2})=(0, 0)$, $(0, 1)$, $(1, 0)$ and $(1,
  1)$ respectively. The red and blue dashed lines indicate the
  $3\sigma$ boundaries allowed by current neutrino oscillation
  data~\cite{Forero:2014bxa} for inverted and normal neutrino mass
  ordering, respectively. For comparison we show also the most
  stringent upper bound from \znbb searches, as well as current Planck
  sensitivity.}
 \end{center}
\end{figure}

\begin{figure}[!h]
 \begin{center}
 \begin{tabular}{cc}\hskip-0.6in
  \includegraphics[width=0.8\linewidth]{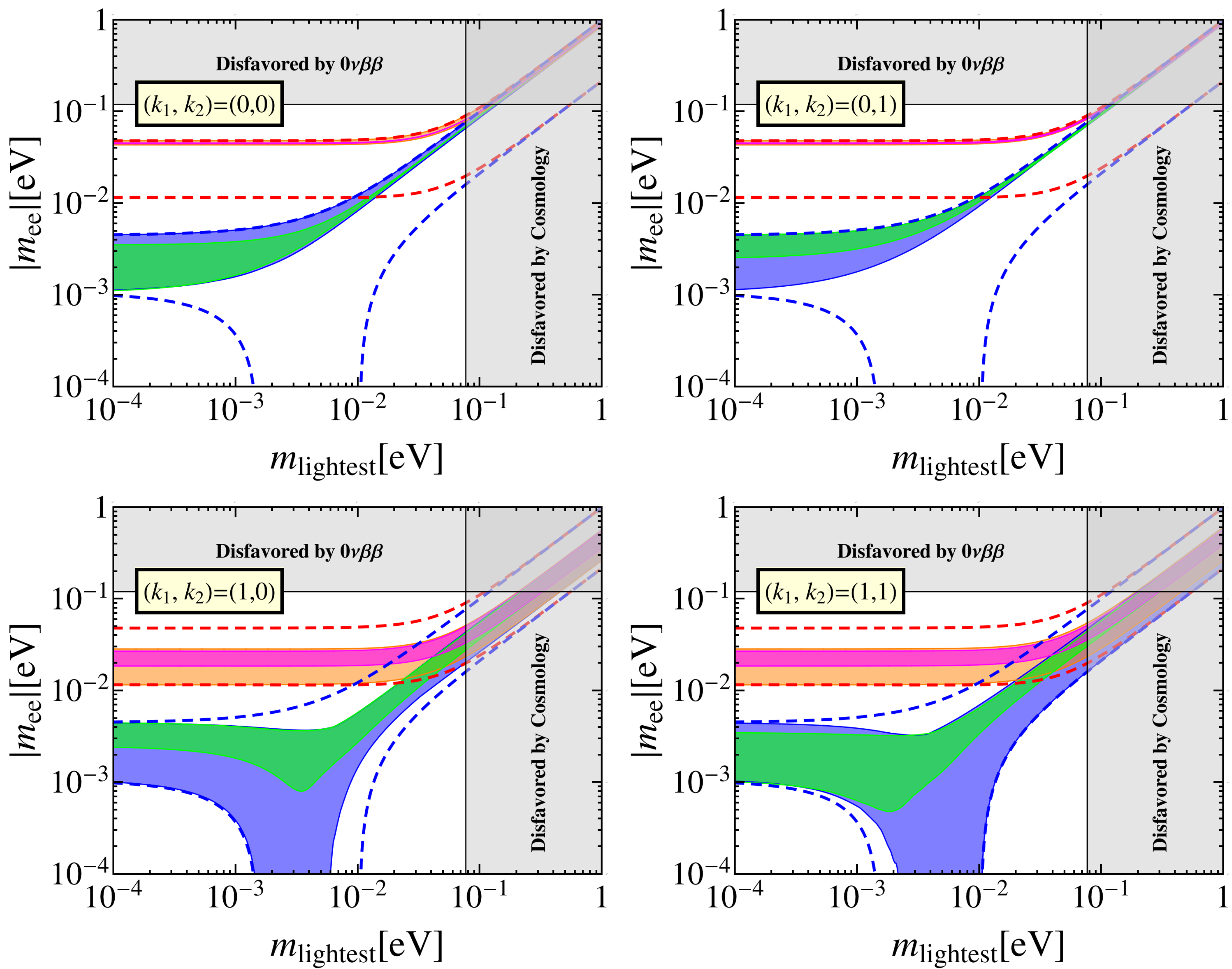}
 \end{tabular}
 \caption{\label{fig:2beta_V}The effective mass $\left| m_{ee}\right|$
  describing the neutrinoless double beta decay amplitude. Using the
  current neutrino oscillation parameters at
  $3\sigma$~\cite{Forero:2014bxa} one obtains the regions delimited by
  the red and blue dashed lines for inverted and normal neutrino mass
  ordering, respectively. In contrast to such generic case, the blue
  and orange regions correspond to letting $\Theta$ and
  $\theta_{1,2,3}$ as free parameters in the type-VI case, while the
  green and magenta regions correspond to $\Theta=\frac{\pi}{7}$ with
  $\theta_{1,2,3}$ free.  }
 \end{center}
\end{figure}

\begin{figure}[!h]
 \begin{center}
 \begin{tabular}{cc}\hskip-0.6in
  \includegraphics[width=0.8\linewidth]{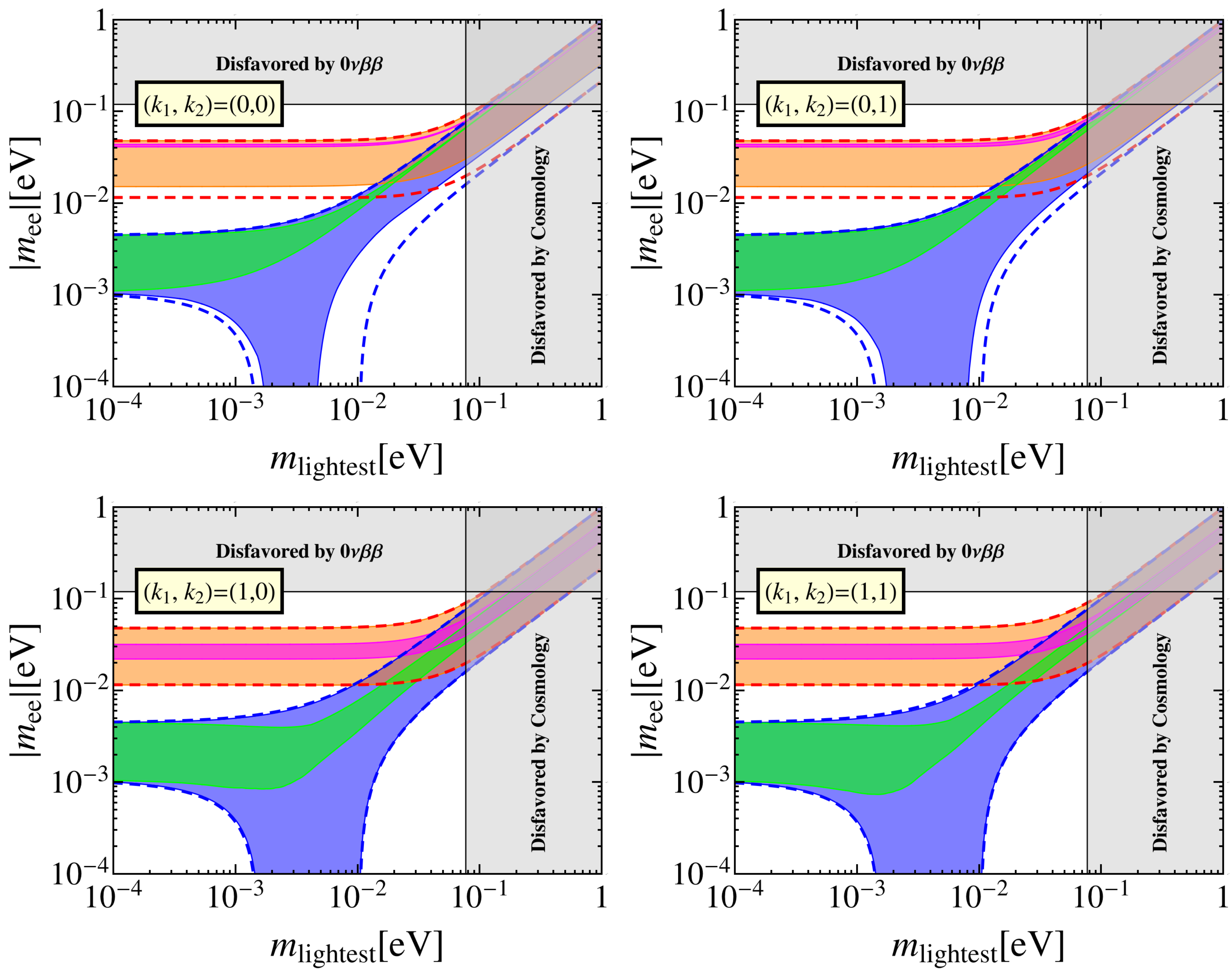}
 \end{tabular}
 \caption{\label{fig:2beta_IX}The effective mass $\left| m_{ee}\right|$
 describing the neutrinoless double beta decay amplitude. Using the current
 neutrino oscillation parameters at $3\sigma$~\cite{Forero:2014bxa} one
 obtains the regions delimited by the red and blue dashed lines for inverted
 and normal neutrino mass ordering, respectively. In contrast to such generic
 case, the blue and orange regions correspond to the type-IX case, with
 $\Theta$ and $\theta_{1,2,3}$ taken as free parameters, while the green and
 magenta regions correspond to $\Theta=\frac{\pi}{7}$ with $\theta_{1,2,3}$
 free.}
 \end{center}
\end{figure}

Up to relatively uncertain nuclear matrix elements~\cite{Simkovic:2012hq} as
well as experimental factors~\cite{Avignone:2007fu,Barabash:2011fn} the decay
amplitude is proportional to the effective mass parameter
\begin{equation}\label{eq:mee}
 \left| m_{ee} \right| =
 \left| m_1 \cos^2 \theta_{12} \cos^2 \theta_{13}
  + m_2 \sin^2 \theta_{12} \cos^2 \theta_{13} e^{i\alpha_{21}}
  + m_3 \sin^2 \theta_{13} e^{i ( \alpha_{31} - 2 \delta_{CP} ) } \right|\,.
\end{equation}
Notice that both Majorana phases $\alpha_{21}$ and $\alpha_{31}$ can
be shifted by $\pi$ by the matrix $\hat{X}^{-1/2}$ in
Eq.~\eqref{eq:X_nu}. Under the transformation $k_1 \to k_1 + 1$
($k_2\to k_2+1$), we have $\alpha_{21} \to \alpha_{21} + \pi$
($\alpha_{31} \to \alpha_{31} + \pi$). Hence without loss of
generality, we can focus on four different cases $(k_1, k_2)=(0, 0)$,
$(0, 1)$, $(1, 0)$ and $(1, 1)$.

We illustrate our results for the effective \znbb mass parameter
$\left| m_{ee}\right|$ by considering the type-I CP symmetric scheme,
given in Fig~\ref{fig:2beta_I}, the type-VI case, given in
Fig.~\ref{fig:2beta_V}, as well as the results for type-IX given in
Fig.~\ref{fig:2beta_IX}. The residual CP transformations of type-V,
type-VII and type-X don't lead to new results for the effective mass
$|m_{ee}|$ since $\theta_{23}$ is not involved in
Eq.~\eqref{eq:mee}. The experimental errors on the mass-squared
splittings are not considered, and the best fit values
from~\cite{Forero:2014bxa} are used with $\Delta
m^2_{21}=7.60\times10^{-5}\,\mathrm{eV}^2$ and $|\Delta
m^2_{31}|=2.48\times10^{-3}\,\mathrm{eV}^2$ for normal ordering and
$|\Delta m^2_{31}|=2.38\times10^{-3}\,\mathrm{eV}^2$ for inverted
ordering. Notice that the red and blue dashed lines (for inverted and
normal neutrino mass ordering, respectively) denote the regions
allowed at $3\sigma$ level by current neutrino oscillation
data~\cite{Forero:2014bxa} for a generic model, without any special
residual CP symmetry. For comparison we display the results for
various CP symmetric cases. We also indicate the disfavored band
associated to the most stringent upper bound
\begin{equation}
 \left| m_{ee} \right| < 0.120~\mathrm {eV}
\end{equation}
which follows from the EXO-200 experiment~\cite{Auger:2012ar,Albert:2014awa}
in combination with results from the first phase of the KamLAND-ZEN
experiment~\cite{Gando:2012zm}. On the other hand the cosmological
upper limit on the mass of the lightest neutrino corresponding to the
latest Planck result is $$\sum_{i} m_{i} < 0.230~\mathrm {eV}$$ at the
$95\%$ confidence level~\cite{Planck:2015xua}.

The results of this section are summarized in
Figs.~\ref{fig:2beta_I},~\ref{fig:2beta_V} and~\ref{fig:2beta_IX}
corresponding to the schemes based on type-I, type-VI and type-IX
remnant CP symmetries respectively. They clearly show that the
attainable values for the effective mass parameter
$\left|m_{ee}\right|$ cover more restrictive ranges than those
expected in generic, non-CP symmetric schemes.

In particular, as illustrated by the green regions in the lower panels
of Figs.~\ref{fig:2beta_V} and~\ref{fig:2beta_IX}, the generalized CP
symmetry assumption may prevent the destructive interference amongst
individual neutrino contributions. This leads to lower bounds for the
\znbb decay rates even for normal hierarchical neutrino mass spectra.
This behavior is a reminiscent of situations already encountered in
the framework of specific flavour symmetry based
models~\cite{Dorame:2011eb,Dorame:2012zv,King:2013hj,Bonilla:2014xla}.

\begin{figure}[!h]
 \begin{center}
 \begin{tabular}{cc}\hskip-0.6in
  \includegraphics[width=0.8\linewidth]{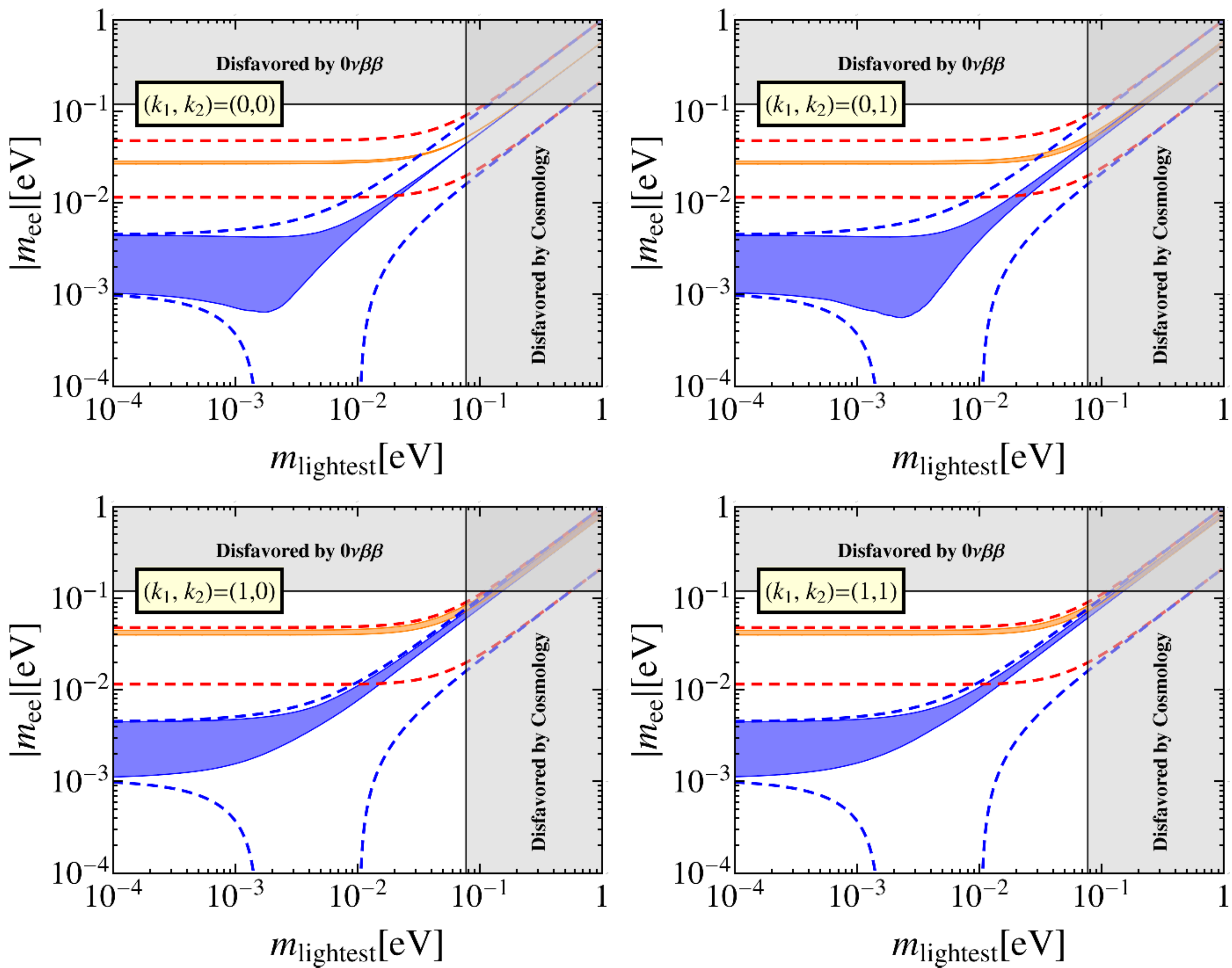}
 \end{tabular}
 \caption{\label{fig:2beta_XI}The effective mass $\left| m_{ee}\right|$
  describing neutrinoless double beta decay for the democratic CP
  symmetry. The red and blue dashed lines indicate the 3$\sigma$
  regions allowed by current neutrino oscillation
  data~\cite{Forero:2014bxa} for inverted and normal neutrino mass
  ordering, respectively. The blue and orange areas denote the
  possible values of $|m_{ee}|$ where $\theta_{1,2,3}$ are treated as
  free parameters and the three mixing angles are in the
  experimentally preferred $3\sigma$ range.}
 \end{center}
\end{figure}

In the case of ${\bf \Sigma}$ matrix type-XI, the predictions for the
effective mass of the neutrinoless double beta decay are shown in
Fig.~\ref{fig:2beta_XI}. As one can read off from this figure, the
effective mass $|m_{ee}|$ is around $0.026$~eV or $0.040$~eV for IH
neutrino mass spectrum, which are within the sensitivity of planned
\znbb decay experiments.  In the case of NH spectrum, the value of
$|m_{ee}|$ is bounded from below: $|m_{ee}|\geq0.00065$ eV for $(k_1,
k_2)=(0, 0)$, $|m_{ee}| \geq 0.00056$~eV for $(k_1, k_2)=(0, 1)$, and
$|m_{ee}|\geq0.0011$ eV for $(k_1, k_2)=(1, 0), (1, 1)$.
%
\subsection{Leptogenesis}
%
The origin of matter-antimatter asymmetry in the Universe is a
puzzling and unexplained phenomenon. Although Sakharov discovered that
CP violation is a necessary condition for explaining the
matter-antimatter asymmetry of the Universe~\cite{Sakharov:1967dj},
the the observed quark CP violation is insufficient for this
purpose~\cite{Kuzmin:1985mm}. The idea that the generation of a
primordial lepton-antilepton asymmetry early in the history of the
Universe induces the observed cosmological baryon asymmetry has been
studied extensively in recent
years~\cite{Fukugita:1986hr,buchmuller:2005eh,Davidson:2008bu}. Although
such ``leptogenesis paradigm'' is closely related with CP violation,
the relation between generalized CP symmetries and leptogenesis
remains, to a large extent, an open research topic. The scenario of
two residual CP transformations preserved by the neutrino mass term
has been analyzed in Ref.~\cite{Chen:2016ptr}, and all the
leptogenesis CP asymmetries are found to only depend on one single
real parameter besides the light neutrino masses and the parameters
characterizing the remnant CP transformations. In this section, we
shall study the phenomenological consequence for leptogenesis if there
is only one residual CP transformation in the neutrino sector. We
shall consider the classical scenario of leptogenesis from the
lightest right-handed (RH) neutrino $N_1$ decay in the type-I seesaw
model. In the RH neutrino and charged lepton mass basis, the type-I
seesaw lagrangian can be written as
\begin{equation}\label{eq:seesaw_lagrangian}
 - \mathcal{L} =
 y_{\alpha} \bar{L}_\alpha H l_{\alpha  R}
 + \bar{N}_{iR} \lambda_{i \alpha} \tilde{H}^{\dagger} L_{\alpha}
 + \frac12 M_i \bar{N}_{iR} N^c_{iR} + h.c.~\,,
\end{equation}
where $L_{\alpha}$ and $l_{\alpha R}$ denote the standard model left-handed
(LH) lepton doublet and RH lepton singlet fields with $\alpha=e,\mu,\tau$ and
$H$ is the Higgs doublet field with the vacuum expectation value
$v = \langle H^{0} \rangle = 174$~GeV. The light neutrino mass matrix is given
by the well-known seesaw formula
\begin{equation}\label{eq:eff_mnu}
 m_{\nu}
 = v^2 \lambda^{T} M^{-1} \lambda
 = \mathbf{U}^{\ast} m \mathbf{U}^{\dag}\,,
\end{equation}
where we denote $M = \mathrm{diag}(M_1, M_2, M_3)$ and
$m = \mathrm{diag}(m_1, m_2, m_3)$, and $m_i$ are the light neutrino mass
eigenvalues.
The most general neutrino Yukawa coupling matrix compatible with the low
energy data is given by~\cite{Casas:2001sr}:
\begin{equation}\label{eq:lambda}
 \lambda = \sqrt{M}\, \mathbf{R} \sqrt{m}\, \mathbf{U}^{\dagger}/v\,,
\end{equation}
where $R$ is generally a complex orthogonal matrix fulfilling
$\mathbf{R} \mathbf{R}^{T} = \mathbf{R}^{T} \mathbf{R} = 1$.

The temperature of the universe at the very early time was extremely high and
the lightest RH neutrino $N_1$ is in thermal equilibrium. As the temperature
drops down to $M_1$, the $N_1$ decay process
$N_1 \to Hl_\alpha( \overline{H} \overline{l_\alpha} )$ and it's inverse
process start to go out of equilibrium and an asymmetry between leptons and
antileptons
is induced accordingly. As the temperature of universe goes down to the
critical temperature of the electroweak phase transition, the sphaleron
interactions convert lepton asymmetry to baryon asymmetry.
One can define the CP asymmetry generated by $N_1$ decays
as~\cite{Covi:1996wh,Endoh:2003mz,Abada:2006ea,Abada:2006fw,Fong:2013wr}
\begin{eqnarray}
 \epsilon_{\alpha} & \equiv &
 \frac{ \Gamma( N_1 \rightarrow H l_{\alpha} )
  - \Gamma( N_1 \rightarrow \overline{H} \overline{l}_{\alpha} )
 }{
  \sum_{\alpha} \Gamma( N_1 \rightarrow Hl_{\alpha} )
  + \Gamma( N_1 \rightarrow \overline{H} \overline{l}_{\alpha} ) }
  \\\label{eq:epsilon_alpha}
 & = & \frac{1}{ 8 \pi( \lambda \lambda^\dag )_{11} }
  \sum_{ j\neq 1 } \bigg\{ \mathrm{Im} \big[ ( \lambda \lambda^\dag )_{1j}
  \lambda_{1 \alpha} \lambda^*_{j\alpha} \big] g(x_j)
  + \mathrm{Im} \big[ ( \lambda \lambda^\dag )_{j1} \lambda_{1 \alpha}
  \lambda^*_{j \alpha} \big] \frac{1}{1 - x_j} \bigg\}\,,
\end{eqnarray}
where $x_j=M^2_j/M_1^2$ and $g(x)$ is the loop function with
\begin{equation}
 g(x) = \sqrt{x} \big[ \frac{1}{ 1 - x } + 1 -( 1 + x )
 \ln \big( \frac{ 1+x }{x} \big) \big]
 \stackrel{x\gg 1}{\longrightarrow}-\frac{3}{2\sqrt{x}}\,.
\end{equation}
As usual, we assume a hierarchical RH neutrinos mass spectrum
$M_1\ll M_2\ll M_3$ which implies $x_j\gg1$.
As a consequence, the flavored CP asymmetries are approximately given
by~\cite{Pascoli:2006ie,Pascoli:2006ci,Branco:2006ce,Davidson:2007va,Davidson:2008bu,Fong:2013wr,Blanchet:2012bk}:
\begin{equation}\label{eq:epsilon_simp}
 \epsilon_{\alpha} =
 -\frac{3M_1}{16\pi v^2}
 \frac{ \Im \left( \sum_{ij} \sqrt{m_i m_j} \,
  m_j \mathbf{R}_{1i} \mathbf{R}_{1j}
  \mathbf{U}^*_{\alpha i} \mathbf{U}_{\alpha j} \right)
 }{
 \sum_j m_j| \mathbf{R}_{1j}|^2 }\,,
\end{equation}
Besides the CP parameter $\epsilon_{\alpha}$, the final baryon asymmetry
depends on the flavour-dependent washout mass parameters,
\begin{equation} \label{eq:mtilde}
 \widetilde{m}_\alpha =
 \frac{ | \lambda_{1 \alpha} |^2 v^2 }{ M_1 }
 = \Big| \sum_jm_j^{1/2} \, \mathbf{R}_{1j}^{}
 \mathbf{U}_{\alpha j}^{\ast}\Big|^2\,.
\end{equation}
At temperatures $T\sim M_1>10^{12}$ GeV where all lepton flavors are out
of equilibrium, the total lepton asymmetry $\epsilon_{1}$ is the sum of the
$\epsilon_{\alpha}$,
\begin{equation} \label{eq:epsilon1}
 \epsilon_1 = \sum_\alpha \epsilon_\alpha
 = -\frac{3M_1}{16\pi v^2} \frac{ \sum_i m_i^2 \Im \left( \mathbf{R}_{1i}^2
  \right) }{ \sum_j m_j| \mathbf{R}_{1j} |^2 }\,,
\end{equation}
which exactly the standard one-flavor
result~\cite{buchmuller:2005eh,Davidson:2008bu}. In the present work we shall
be concerned with temperatures $(10^9\leq T\sim M_1\leq10^{12})$ GeV. In this
mass window only the interactions mediated by the $\tau$ Yukawa
coupling are in equilibrium and the final baryon asymmetry is well
approximated by~\cite{Abada:2006ea,Pascoli:2006ci}
\begin{equation} \label{eq:Yb-1}
 Y_{B} \simeq -\frac{12}{37\,g^*} \left[ \epsilon_{2} \eta \left(
 \frac{417}{589}{\widetilde{m}_2} \right)
 \,+\,
 \epsilon_{\tau} \eta \left( \frac{390}{589}
 { \widetilde{m}_{\tau} } \right) \right]\,,
\end{equation}
where the number of relativistic degrees of freedom $g^{\ast}$ is taken to be
$g^\ast=106.75$ as in the standard model. The combined asymmetry
$\epsilon_2=\epsilon_{e}+\epsilon_{\mu}$ comes from the indistinguishable $e$
and $\mu$ flavored leptons and
$\widetilde{m}_2=\widetilde{m}_{e}+\widetilde{m}_{\mu}$. The efficiency factor
$\eta(\widetilde{m}_{\alpha})$ accounts for the washing out of the total
lepton asymmetry due to inverse decays,
\begin{equation}\label{eq:efficency_factor}
 \eta( \widetilde{m}_{\alpha} ) \simeq \left[ \left(
 \frac{ \widetilde{m}_{\alpha} }{ 8.25\times 10^{-3} \,
 {\rm eV} } \right)^{-1}
 + \left( \frac{0.2\times 10^{-3}\,{\rm eV}}{
 \widetilde{m}_{\alpha}}\right)^{-1.16}\ \right]^{-1}\,.
\end{equation}
For the mass range of $M_1<10^{9}$ GeV, all the three flavours are
distinguishable. and the final value of the baryon asymmetry can be
approximated by~\cite{Abada:2006ea}.
\begin{equation}\label{eq:Yb-2}
 Y_{B} \simeq
 -\frac{12}{37\, g^*}
 \left[ \epsilon_{e} \eta \left( \frac{151}{179}{\widetilde{m}_e} \right)\,
 + \epsilon_{\mu} \eta \left( \frac{344}{537}{\widetilde{m}_{\mu}} \right)\,
 + \,\epsilon_{\tau} \eta \left( \frac{344}{537}
 {\widetilde{m}_{\tau}} \right) \right]\,,
\end{equation}
The baryon asymmetry will be typically too small to account for the observed
value in this case.

In the same fashion as studying lepton flavor mixing in previous sections, we
assume that the seesaw Lagrangian of Eq.~\eqref{eq:seesaw_lagrangian} is
invariant under a CP transformation. We suppose that the lagrangian in
Eq.~\eqref{eq:seesaw_lagrangian} is invariant under the following CP
transformation
\begin{eqnarray}\label{eq:res_CP}
 \text{CP}:~~\nu_{L} \stackrel{}{\longmapsto}
 i \mathbf{X} \gamma_0 \nu^{c}_{L} \,, \quad
 N_{R} \stackrel{}{\longmapsto} i \widehat{X}_{N}\gamma_0N^{c}_{R}\,.
\end{eqnarray}
For the symmetry to hold, the neutrino Yukawa coupling matrix $\lambda$ and
the RH neutrino mass matrix $M$ have to fulfill
\begin{equation}\label{eq:cons_res_CP}
 \widehat{X}^{\dagger}_{N} \lambda \mathbf{X}
 = \lambda^{\ast},\qquad
 \widehat{X}^{\dagger}_{N} M \widehat{X}^{\ast}_{N} = M^{\ast}\,.
\end{equation}
One can immediately see that $\hat{X}_{N}$ should be a diagonal matrix with
elements equal to $\pm1$, i.e.
$\widehat{X}_{N}=\text{diag}(\pm1, \pm1, \pm1)$, where the $\pm$ signs can be
chosen independently. From Eq.~\eqref{eq:cons_res_CP} we can derive that the
postulated residual CP symmetry leads to the following constraints on the
$\mathbf{R}-$matrix and lepton mixing matrix $\mathbf{U}$~\cite{Chen:2016ptr}:
\begin{equation} \label{eq:R_U_cons}
 \mathbf{R}^{\ast} = \hat{X}_{N} \mathbf{R} \hat{X},\qquad
 \mathbf{U}^{\dagger} \mathbf{X} \mathbf{U}^{\ast} = \widehat{X}\,,
\end{equation}
where $\widehat{X}=\text{diag}\left(\pm1, \pm1, \pm1\right)$. Note
that the same constraint on the $\mathbf{R}-$matrix from the CP
invariance was found in Refs.~\cite{Pascoli:2006ci,Mohapatra:2015gwa}
As a consequence, the lepton mixing matrix is determined up to an
orthogonal matrix
$\mathbf{U}=\mathbf{\Sigma}\mathbf{O}_{3\times3}\hat{X}^{-1/2}$ as
shown in Eq.~\eqref{eq:PMNS_master}. On the other hand, depending on
the values of $\hat{X}_{N}$ and $\hat{X}$, each element of the
$\mathbf{R}-$matrix satisfies
$\mathbf{R}_{ij}^\ast=\pm\mathbf{R}_{ij}$ so that $\mathbf{R}_{ij}$ is
either real or pure imaginary while $\mathbf{R}^2_{ij}$ must be
real. Hence the total lepton asymmetry $\epsilon_1$ is always
predicted to be vanishing $\epsilon_1=0$, no matter what form the
residual CP $\mathbf{X}$ takes. Thus, at temperatures where all lepton
flavors are out of equilibrium and the one--flavor approximation is
valid, no baryon asymmetry can be generated in the present
framework. In the rest of the paper, we shall focus on the flavor
dependent leptogenesis, with $M_1$ having a value in the interval of
interest $10^9\;\text{GeV}\leq M_1\leq 10^{12}$ GeV.

Notice that both diagonal matrices $\hat{X}$ and $\hat{X}_{N}$ are not
constrained by the symmetry.
In order to classify different possible cases in a concise and systematical
way, we shall separate out $\hat{X}$ and $\hat{X}_{N}$ explicitly and
introduce the following notations:
\begin{equation}
 \mathbf{U}^{\prime} = \mathbf{U} \hat{X}^{1/2},\qquad
 \mathbf{R}^{\prime} = \hat{X}_{N}^{1/2} \mathbf{R} \hat{X}^{1/2},\qquad
 K_j = ( \hat{X}_{N})_{11} ( \hat{X})_{jj},\quad\text{with}\quad j=1, 2, 3\,.
\end{equation}
Then $\mathbf{U}^{\prime}$ would be a real matrix, and $K_j$ is either $+1$ or
$-1$. Furthermore, the CP asymmetry $\epsilon_\alpha$ and the washout mass
parameter $\widetilde{m}_\alpha$ can be written as
\begin{equation}
 \epsilon_\alpha =
 -\frac{3M_1}{16\pi v^2}
 \frac{ \Im \left( \sum_{ij} \sqrt{m_im_j} \, m_j \mathbf{R}^{\prime}_{1i}
 \mathbf{R}^{\prime}_{1j}
 \mathbf{U}^{\prime \ast}_{\alpha i} \mathbf{U}^{\prime}_{\alpha j}
 K_j\right) }{ \sum_{j} m_j ( \mathbf{R}_{1j}^{\prime})^2 }, \qquad
 \widetilde{m}_{\alpha} =
 \Big| \sum_j m_j^{1/2} \mathbf{R}_{1j}^{\prime}
 \mathbf{U}_{\alpha j}^{\prime\ast}\Big|^2\,.
\end{equation}
Obviously only the elements $\mathbf{R}^{\prime}_{1i}$ of the first row of
$\mathbf{R}^{\prime}$ are relevant to $\epsilon_\alpha$ and
$\widetilde{m}_{\alpha}$. The orthogonal condition
$\mathbf{R}\mathbf{R}^{T}=1$ gives rise to
\begin{equation}\label{eq:R1orth}
 \mathbf{R}^{\prime\,2}_{11}K_1
 + \mathbf{R}^{\prime\,2}_{12} K_2
 + \mathbf{R}^{\prime\,2}_{13} K_3 = 1\,.
\end{equation}
The most general parametrization of the elements $\mathbf{R}^{\prime}_{11}$,
$\mathbf{R}^{\prime}_{12}$ and $\mathbf{R}^{\prime}_{13}$ for different
possible values of $K_1$, $K_2$, $K_3$ are listed in Table~\ref{tab:orth_R}.
Note that the values $(K_1, K_2, K_3)=(-, -, -)$ is not admissible since the
constraint in Eq.~\eqref{eq:R1orth} can not be fulfilled in that case. 

\begin{table}[!hptb] \renewcommand{\arraystretch}{1.5}
 \begin{center}
 {\footnotesize
 \begin{tabular}{|c|c|}\hline\hline
  $(K_1, K_2, K_3)$  &
  $(\mathbf{R}^{\prime}_{11}, \mathbf{R}^{\prime}_{12},
  \mathbf{R}^{\prime}_{13})$  \\\hline
  $(+, +, +)$ & $\left( \cos \rho \cos \varphi, \cos \rho \sin \varphi,
  \sin \rho \right)$ \\\hline
  $(-, +, +)$ & $\left(\sinh\rho, \cosh \rho \cos \varphi,
  \cosh \rho \sin \varphi \right)$ \\\hline
  $(+, -, +)$ &
  $\left(\cosh\rho\sin\varphi, \sinh\rho, \cosh\rho\cos\varphi\right)$
  \\\hline
  $(+, +, -)$ & $\left( \cosh \rho \cos \varphi, \cosh \rho \sin \varphi,
  \sinh \rho \right)$ \\\hline
  $(+, -, -)$ &
  ~$\left( \cosh \rho, \sinh \rho \cos \varphi,
  \sinh \rho \sin \varphi \right)$~ \\\hline
  $(-, +, -)$ &
  ~$\left( \sinh \rho \sin \varphi,
  \cosh \rho, \sinh \rho \cos \varphi \right)$~\\\hline
  $(-, -, +)$ &
  $\left( \sinh \rho \cos \varphi,
  \sinh \rho \sin \varphi, \cosh \rho \right)$ \\\hline\hline
 \end{tabular}}
 \end{center} \renewcommand{\arraystretch}{1.0}
 \caption{\label{tab:orth_R}
 The parametrization of the first row of the $\mathbf{R}^{\prime}$ matrix for
 the possible values of $K_1$, $K_2$ and $K_3$, where both $\varphi$ and
 $\rho$ are real parameters.}
\end{table}

In the present formalism, we show that in general the lepton mixing
angles and CP phases depend on three parameters $\theta_{1,2,3}$, and
two more parameters $\rho$ and $\varphi$ are involved in prediction
for the baryon asymmetry. In what follows, we shall apply the above
general results to the cases of type-V and type-VI residual CP
symmetries with $\Theta=\frac{2\pi}{5}$ and $\frac{\pi}{7}$
respectively. The values of $\theta_{1, 2, 3}$ are determined to
reproduce the best fit values of the three lepton mixing
angles~\cite{Forero:2014bxa}. As a typical example, we choose the RH
neutrino mass $M_1=5\times10^{11}$ GeV, the lightest neutrino mass is
taken to be $m_1 (\text{or}\, m_3)=0.01\,\mathrm{eV}$, and the
mass-squared splittings $\Delta m^2_{21}$ and $|\Delta m^2_{31}|$ are
fixed at their best fit values~\cite{Forero:2014bxa}.

\begin{itemize}[labelindent=-0.7em, leftmargin=1.2em]

\item{Type-V CP symmetry with $\Theta=\frac{2\pi}{5}$}

It is the so-called generalized $\mu-\tau$ reflection
symmetry~\cite{Chen:2015siy}. The explicit form of the $\mathbf{X}$ matrix,
its Takagi factorization matrix $\mathbf{\Sigma}$ and the corresponding
predictions for mixing parameters are collected in Table~\ref{tab:two_zeros}.
From Eq.~\eqref{eq:R_U_cons} we know that the $\mathbf{R}-$matrix and the
mixing matrix $\mathbf{U}$ have the following properties
\begin{equation}
 \mathbf{R}_{1j} = \mathbf{R}^{\ast}_{1j} K_{j},\qquad
 \mathbf{U}_{1j} = e^{i\alpha} \mathbf{U}_{1j}^{\ast} (\hat{X})_{jj}\,.
\end{equation}
It follows that the CP asymmetry $\epsilon_{e}$ is vanishing $\epsilon_{e}=0$
independent of the value of $\Theta$ in this case. The remaining two CP
asymmetries $\epsilon_\mu$ and $\epsilon_\tau$ are related as
$\epsilon_\mu=-\epsilon_\tau$ which is inferred from general prediction
$\epsilon_1=\sum_\alpha\epsilon_\alpha=0$.
In order to accommodate the best fit values of the mixing
angles~\cite{Forero:2014bxa}, we take
$\theta_1=58.026^{\circ}\,[59.106^{\circ}]$,
$\theta_2=8.60^{\circ}\,[8.70^{\circ}]$ and
$\theta_3=145.4^{\circ}\,[145.4^{\circ}]$ for NH and in square brackets for IH
of the neutrino masses, respectively. Then one can predict the CP violating
phases $\delta_{CP}=253.727^{\circ}\,[254.022^{\circ}]$,
$\alpha_{21}~(\text{mod}~\pi)=0^{\circ}\,[0^{\circ}]$ and
$\alpha_{31}~(\text{mod}~\pi)=147.454^{\circ}\,[148.045^{\circ}]$. Note that
the Dirac phase $\delta_{CP}$ is rather close to its present best fit
value~\cite{Forero:2014bxa}, although the statistical significance is quite
low. Since the baryon asymmetry $Y_{B}$ depends on $\rho$ and $\varphi$, we
display the contour regions of $Y_{B}/Y^{obs}_{B}$ in the plane $\varphi$
versus $\rho$ in Fig.~\ref{fig:lep_case_V}.
We see that successful leptogenesis can happen except for NH neutrino mass
spectrum with $(K_1, K_2, K_3)=(-, -, +)$.
\begin{figure}[!hptb]
\begin{center} \hskip-0.23in
\includegraphics[width=0.98\textwidth]{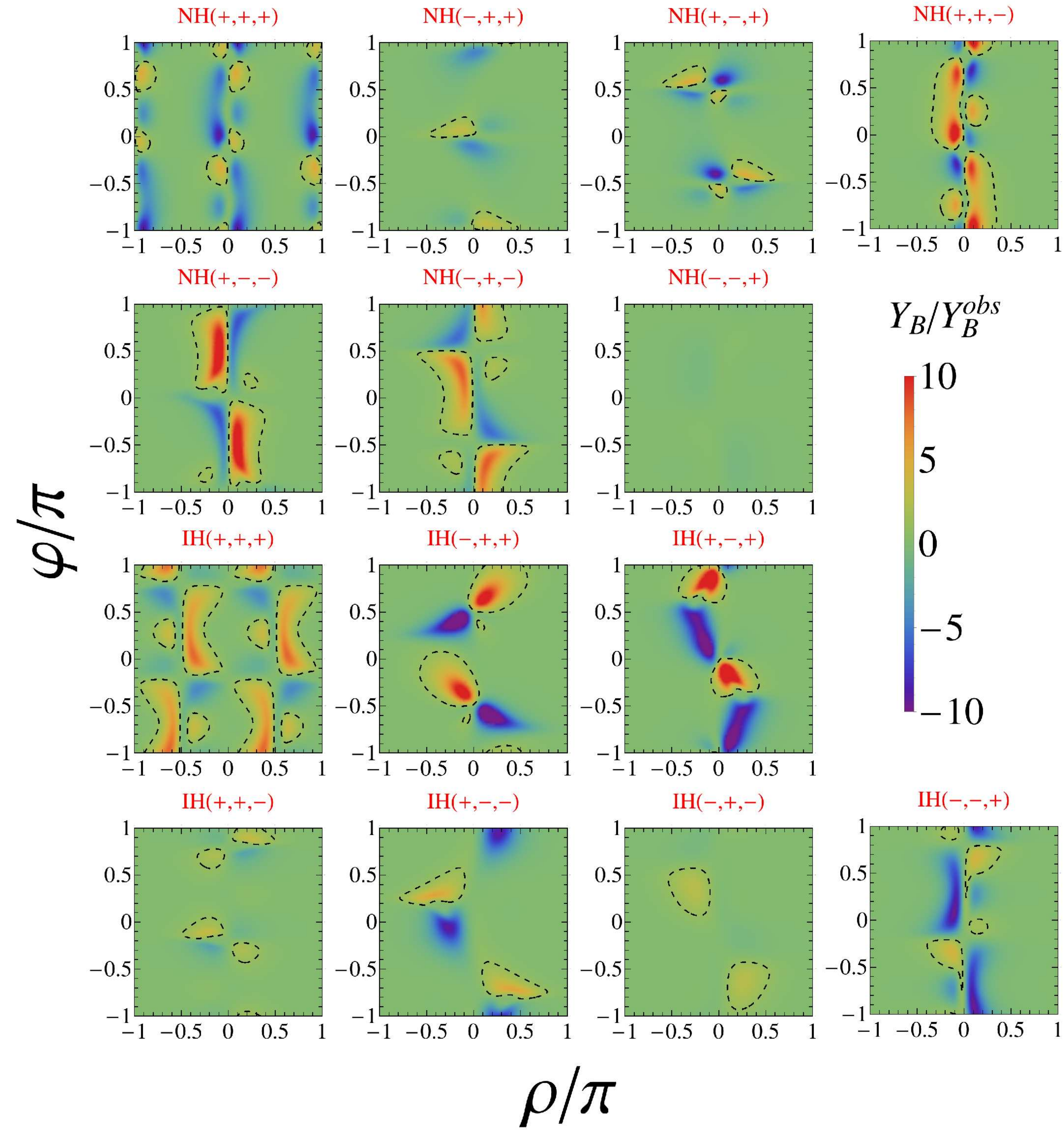}
\caption{\label{fig:lep_case_V} Predictions for $Y_B/Y_B^{obs}$ as a
function of $\rho$ and $\varphi$ in the case of type-V residual CP transformation with $\Theta=2\pi/5$. We have chosen
$M_1=5\times10^{11}$~GeV, $m_1 (\text{or}\, m_3)=0.01\,\mathrm{eV}$. The mass-squared differences $\Delta m^2_{21}$ and $|\Delta m^2_{31}|$ are taken to be the best fit values~\cite{Forero:2014bxa}. We set
   $\theta_1=58.026^{\circ}\,[59.106^{\circ}]$,
   $\theta_2=8.60^{\circ}\,[8.70^{\circ}]$ and
   $\theta_3=145.4^{\circ}\,[145.4^{\circ}]$ to reproduce the best fitting
   values of the mixing angles~\cite{Forero:2014bxa}. The dashed lines denote
   the precisely measured value of the baryon asymmetry
   $Y_B^{obs}=8.66\times10^{-11}$~\cite{Ade:2015xua}.
   Note that successful leptogenesis is not possible for NH neutrino masses
   with $(K_1, K_2, K_3)=(-, -, +)$.}
 \end{center}
\end{figure}

\item{Type-VI CP symmetry with $\Theta=\frac{\pi}{7}$}

In this case, we take $\theta_1=178.345^{\circ}\,[176.995^{\circ}]$,
$\theta_2=48.167^{\circ}\,[48.442^{\circ}]$ and
$\theta_3=57.255^{\circ}\,[58.255^{\circ}]$ so that the best fitting values of
the lepton mixing angles are reproduced exactly. Accordingly, the CP phases
are determined to be $\delta_{CP}=255.105^{\circ}\,[249.292^{\circ}]$,
$\alpha_{21}~(\text{mod}~\pi)=138.569^{\circ}\,[138.360^{\circ}]$ and
$\alpha_{31}~(\text{mod}~\pi)=151.226^{\circ}\,[150.738^{\circ}]$. We plot the
contour regions for $Y_{B}/Y^{obs}_{B}$ in the $\rho-\varphi$ plane in
Fig.~\ref{fig:lep_case_VI}. As can be seen, we can have successful
leptogenesis except for the cases of NH neutrino masses with
$(K_1, K_2, K_3)=(-, -, +)$ and IH with $(K_1, K_2, K_3)=(+, -, -)$.

\begin{figure}[!hptb]
 \begin{center} \hskip-0.23in
  \includegraphics[width=0.98\textwidth]{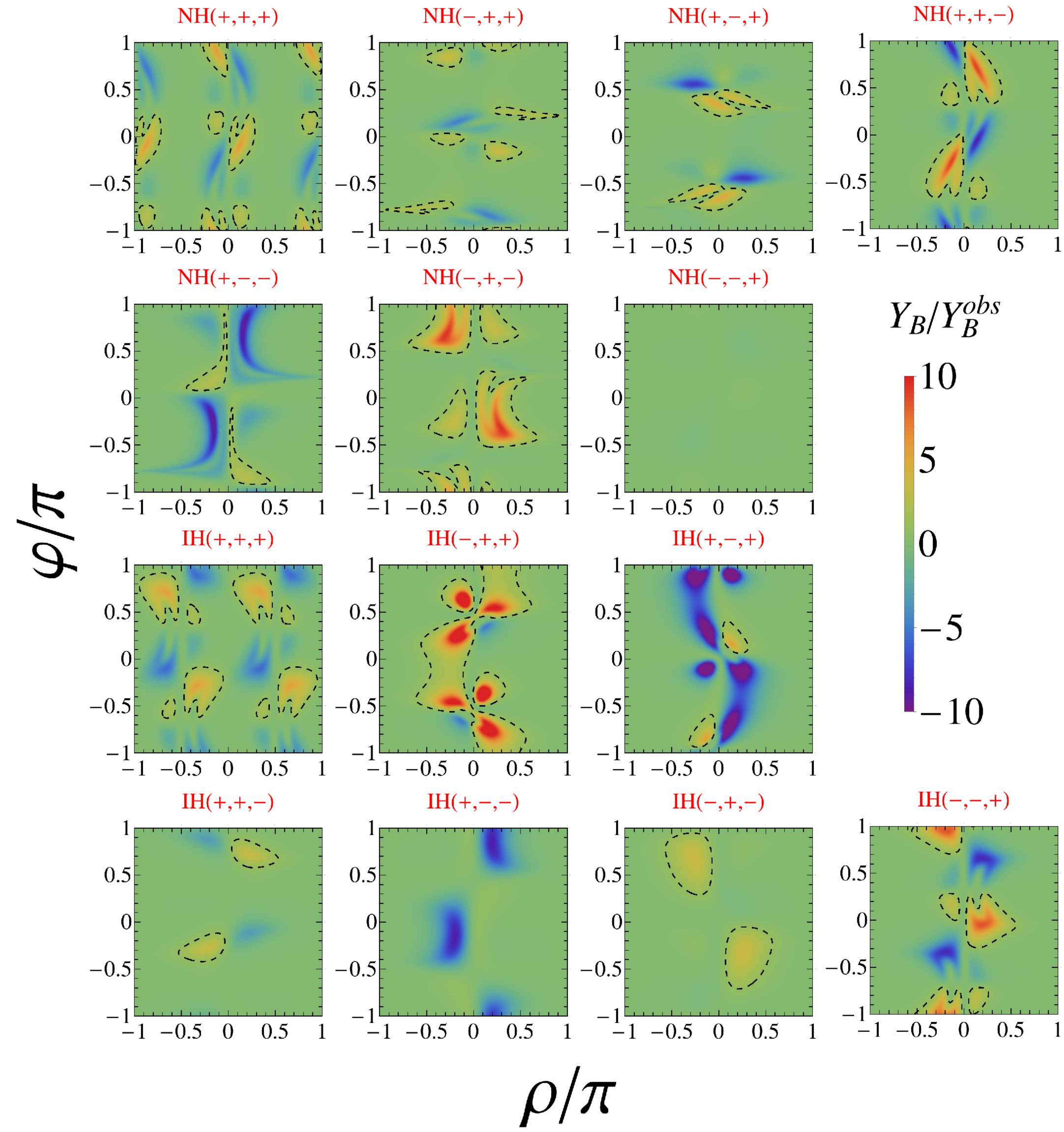}
  \caption{\label{fig:lep_case_VI} Predictions for $Y_B/Y_B^{obs}$ as a
   function of $\rho$ and $\varphi$ in the case of type-VI residual CP
   transformation with $\Theta=\pi/7$. We have chosen
   $M_1=5\times10^{11}$~GeV, $m_1 (\text{or}\, m_3)=0.01\,\mathrm{eV}$. The
   mass-squared differences $\Delta m^2_{21}$ and $|\Delta m^2_{31}|$ are
   taken to be the best fit values~\cite{Forero:2014bxa}. We set
   $\theta_1=178.345^{\circ}\,[176.995^{\circ}]$, $\theta_2=48.167^{\circ}\,
   [48.442^{\circ}]$ and $\theta_3=57.255^{\circ}\,[58.255^{\circ}]$ to
   reproduce the best fitting values of the mixing
   angles~\cite{Forero:2014bxa}. The dashed lines denote the precisely
   measured value of the baryon asymmetry
   $Y_B^{obs}=8.66\times10^{-11}$~\cite{Ade:2015xua}.
   Note that successful leptogenesis is not possible for NH neutrino masses
   with $(K_1, K_2, K_3)=(-, -, +)$ and IH case with
   $(K_1, K_2, K_3)=(+, -, -)$.}
 \end{center}
\end{figure}

\end{itemize}

\begin{figure}[!hptb]
 \begin{center} \hskip-0.23in
  \includegraphics[width=0.98\textwidth]{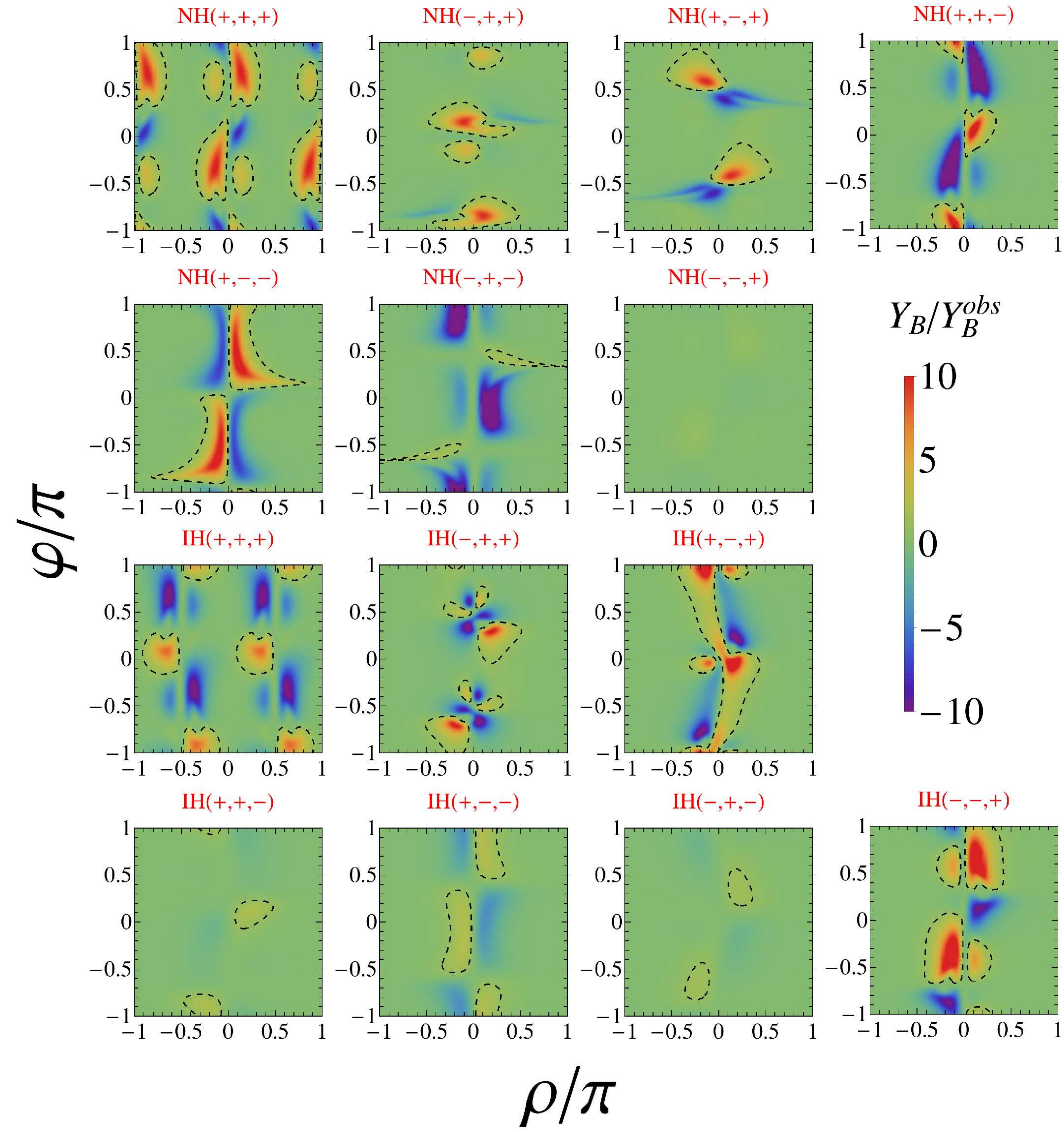}
  \caption{\label{fig:lep_case_XI}Predictions for $Y_B/Y_B^{obs}$ as a
   function of $\rho$ and $\varphi$ for the case of democratic CP
   transformation. We have chosen $M_1=5\times10^{11}$ GeV,
   $m_1( \text{or}\, m_3) = 0.01 \,\mathrm{eV}$. The mass-squared differences
   $\Delta m^2_{21}$ and $|\Delta m^2_{31}|$ are taken to be the best
   fit values~\cite{Forero:2014bxa}. We set
   $\theta_1=176.556^{\circ}\,[176.886^{\circ}]$,
   $\theta_2=9.122^{\circ}\,[9.403^{\circ}]$ and
   $\theta_3=53.464^{\circ}\,[53.508^{\circ}]$ so as to reproduce the
   best fit values of the neutrinos mixing
   angles~\cite{Forero:2014bxa}. The dashed lines denote the precisely
   measured value of the baryon asymmetry
   $Y_B^{obs}=8.66\times10^{-11}$~\cite{Ade:2015xua}.  Note that
   successful leptogenesis is not possible for NH neutrino masses with
   $(K_1, K_2, K_3)=(-, -, +)$.}
 \end{center}
\end{figure}

We also study the predictions for leptogenesis in the type-XI case. As
an example, we choose $\theta_1=176.556^{\circ}\,[176.886^{\circ}]$,
$\theta_2=9.122^{\circ}\,[9.403^{\circ}]$ and
$\theta_3=53.464^{\circ}\,[53.508^{\circ}]$ to reproduce the best fit
values of the neutrino mixing angles. Then the CP violating phases can
be predicted as $\delta_{CP}=285.771^{\circ}\,[287.100^{\circ}]$,
$\alpha_{21}~(\text{mod}~\pi)=56.593^{\circ}\,[56.580^{\circ}]$ and
$\alpha_{31}~(\text{mod}~\pi)=175.842^{\circ}\,[175.919^{\circ}]$. The
contour region for $Y_{B}/Y^{obs}_{B}$ in the plane $\varphi$ versus
$\rho$ is shown in Fig.~\ref{fig:lep_case_XI}. The existing
matter--antimatter asymmetry can be reproduced for appropriate values
of $\rho$ and $\varphi$ except the case of NH with $(K_1, K_2,
K_3)=(-, -, +)$.
%

\section{Conclusions}

In this paper we have given a full classification of generalized CP
symmetries preserved by the neutrino mass matrix, taking as basis the
number of zero entries in the transformation matrix. We have
determined the corresponding constrained form of the lepton mixing
matrix. We have shown how this results in correlations between the
lepton mixing angles and the Majorana and Dirac CP violating
phases. We have also mapped out the corresponding restrictions that
follow from current neutrino oscillation global fits and found that,
in some cases, the Dirac CP violating phase characterizing neutrino
oscillations is highly constrained. Focussing on the expected CP
asymmetries for the ``golden'' oscillation channel we have derived
implications for current long baseline neutrino oscillation
experiments T2K, NO$\nu$A, forecasting also the corresponding results
for the upcoming long baseline DUNE experiment. We have also discussed
the predicted ranges for the effective neutrino mass parameter
characterizing the neutrinoless double beta decay rates. Finally we
have also studied the cosmological implications of such schemes for
leptogenesis.

The results of this paper are quite general in the sense that they are
independent of how the assumed residual CP symmetry is dynamically
achieved. If the residual CP symmetry $\mathbf{X}$ originates from the
breaking of the generalized CP symmetry compatible with a finite
flavor symmetry group $G_{f}$, the admissible form of the residual CP
transformation would be strongly constrained to satisfy the
consistency condition. If $\mathbf{X}$ has at least one zero entry, it
would belong to one of the cases studied in the present work. The
corresponding prediction for lepton mixing matrix could be
straightforwardly obtained by exploiting the master formula
Eq.~\eqref{eq:PMNS_master}. In this paper we have discussed the
possible mixing patterns which can be achieved in this method, and the
resulting phenomenological predictions for neutrino oscillation,
neutrinoless double beta decay and leptogenesis.
By comparing with the extensively studied scenarios with two residual
CP transformations preserved in the neutrino
sector~\cite{Feruglio:2012cw,Ding:2013hpa,Feruglio:2013hia,Ding:2013bpa,Li:2013jya,Ding:2013nsa,Ding:2014hva,Ding:2014ssa,Li:2014eia,Li:2015jxa,DiIura:2015kfa,Ballett:2015wia,Branco:2015gna,Turner:2015uta,
  Hagedorn:2014wha,Ding:2014ora,Ding:2015rwa,Li:2016ppt}, one expects
to obtain new phenomenologically viable mixing patterns and new
predictions for the CP violation phases.

\section*{Acknowledgements}

This work was supported by the National Natural Science Foundation of
China under Grant Nos. 11275188, 11179007 and 11522546 (P.C. and
G.J.D.); by the Spanish grants FPA2014-58183-P, Multidark
CSD2009-00064 and SEV-2014-0398 (MINECO), and PROMETEOII/2014/084
(Generalitat Valenciana) (J.W.F.V. and F.G.C.), and the Mexican grants
CONACYT 236394 (F.G.C.).

\begin{appendix}
\section{Definition domain of $\mathbf{O}_{3\times3}$}

In this appendix, we would like to discuss the domain of the parameters
$\theta_1$, $\theta_2$ and $\theta_3$ in the $\mathbf{O}_{3 \times 3}$ matrix.
In Eq.~\eqref{eq:Orthogonal_matrix}, ${\bf O}_{3\times3}$ is
parameterized as
\begin{equation}
 \mathbf{O}_{3\times3} (\theta_1, \theta_2, \theta_3)\equiv
 \left(\begin{array}{ccc}
  1 & 0 & 0 \\
  0 &  \cos \theta_1 & \sin \theta_1 \\
  0 & -\sin \theta_1 & \cos \theta_1
 \end{array}\right)
 \left(\begin{array}{ccc}
  \cos \theta_2 & 0 & \sin\theta_2 \\
  0 & 1 & 0 \\
  -\sin\theta_2 & 0 & \cos\theta_2
 \end{array}\right)
 \left(\begin{array}{ccc}
 \cos\theta_3 & \sin\theta_3 & 0 \\
 -\sin\theta_3 & \cos\theta_3 & 0 \\
 0 & 0  & 1
 \end{array} \right)\,,
\end{equation}
where $\theta_1$, $\theta_2$ and $\theta_3$ can freely vary in the
range of $[0,2\pi)$. Notice that $\mathbf{O}_{3\times3}$ has the
following properties
\begin{eqnarray}\nonumber
 & & \mathbf{O}_{3\times3} (\theta_1, \theta_2, \theta_3+\pi)
 = \mathbf{O}_{3\times3} ( \theta_1, \theta_2, \theta_3)\,
 \text{diag} (-1, -1, 1) \,,\\ \nonumber
 & & \mathbf{O}_{3\times3}(\theta_1, \theta_2+\pi, \theta_3) =
 \mathbf{O}_{3\times3}( \theta_1, \theta_2, \pi-\theta_3)\,
 \text{diag}(1, -1, -1)\,,\\
 & & \label{eq:O33_char}
 \mathbf{O}_{3\times3}(\theta_1+\pi, \theta_2, \theta_3) =
 \mathbf{O}_{3\times3}(\theta_1, \pi-\theta_2, \theta_3)\,
 \text{diag}(-1, -1, 1)\,,
\end{eqnarray}
where the diagonal matrices can be absorbed into the matrix
$\hat{X}^{-1/2}_{\nu}$. As a result, the fundamental interval of the
parameters $\theta_{1, 2, 3}$ can be taken to be $[0,\pi)$.
\end{appendix}

\bibliographystyle{bib_style_T1}

\begin{thebibliography}{10}
\providecommand{\url}[1]{\texttt{#1}}
\providecommand{\urlprefix}{URL }
\providecommand{\eprint}[2][]{\url{#2}}

\bibitem{Altarelli:2010gt}
G.~Altarelli and F.~Feruglio, \emph{{Discrete Flavor Symmetries and Models of
  Neutrino Mixing}}, Rev.Mod.Phys. \textbf{82} (2010) 2701--2729,
  \MYhref[eprintLinks]{http://arxiv.org/abs/1002.0211}{{\ttfamily
  arXiv:1002.0211 [hep-ph]}}.

\bibitem{ishimori2012introduction}
H.~Ishimori et~al., \emph{An Introduction to Non-Abelian Discrete Symmetries
  for Particle Physicists}, Lecture Notes in Physics, Springer (2012), ISBN
  9783642308048.

\bibitem{King:2013eh}
S.~F. King and C.~Luhn, \emph{{Neutrino Mass and Mixing with Discrete
  Symmetry}}, Rept.Prog.Phys. \textbf{76} (2013) 056201,
  \MYhref[eprintLinks]{http://arxiv.org/abs/1301.1340}{{\ttfamily
  arXiv:1301.1340 [hep-ph]}}.

\bibitem{King:2014nza}
S.~F. King et~al., \emph{{Neutrino Mass and Mixing: from Theory to
  Experiment}},
  \MYhref[journalLinks]{http://dx.doi.org/10.1088/1367-2630/16/4/045018}{New
  J.Phys.
  }\MYhref[journalLinks]{http://dx.doi.org/10.1088/1367-2630/16/4/045018}{\textbf{16}
  (2014) 045018},
  \MYhref[eprintLinks]{http://arxiv.org/abs/1402.4271}{{\ttfamily
  arXiv:1402.4271 [hep-ph]}}.

\bibitem{King:2015aea}
S.~F. King, \emph{Models of neutrino mass, mixing and cp violation}, J. Phys.
  \textbf{G42} (2015) 123001,
  \MYhref[eprintLinks]{http://arxiv.org/abs/1510.02091}{{\ttfamily
  arXiv:1510.02091 [hep-ph]}}.

\bibitem{Schechter:1980gr}
J.~Schechter and J.~Valle, \emph{{Neutrino Masses in $SU(2) \times U(1)$
  Theories}},
  \MYhref[journalLinks]{http://dx.doi.org/10.1103/PhysRevD.22.2227}{Phys.Rev.
  }\MYhref[journalLinks]{http://dx.doi.org/10.1103/PhysRevD.22.2227}{\textbf{D22}
  (1980) 2227}.

\bibitem{Chen:2015siy}
P.~Chen, G.-J. Ding, F.~Gonzalez-Canales and J.~W.~F. Valle, \emph{{Generalized
  $\mu-\tau$ reflection symmetry and leptonic CP violation}},
  \MYhref[journalLinks]{http://dx.doi.org/10.1016/j.physletb.2015.12.069}{Phys.
  Lett.
  }\MYhref[journalLinks]{http://dx.doi.org/10.1016/j.physletb.2015.12.069}{\textbf{B753}
  (2016) 644--652},
  \MYhref[eprintLinks]{http://arxiv.org/abs/1512.01551}{{\ttfamily
  arXiv:1512.01551 [hep-ph]}}.

\bibitem{Xing:2015fdg}
Z.-z. Xing and Z.-h. Zhao, \emph{{A review of mu-tau flavor symmetry in
  neutrino physics}}, --  (2015),
  \MYhref[eprintLinks]{http://arxiv.org/abs/1512.04207}{{\ttfamily
  arXiv:1512.04207 [hep-ph]}}.

\bibitem{Chen:2014wxa}
P.~Chen, C.-C. Li and G.-J. Ding, \emph{{Lepton Flavor Mixing and CP
  Symmetry}},
  \MYhref[journalLinks]{http://dx.doi.org/10.1103/PhysRevD.91.033003}{Phys.
  Rev.
  }\MYhref[journalLinks]{http://dx.doi.org/10.1103/PhysRevD.91.033003}{\textbf{D91}
  (2015) 3 033003},
  \MYhref[eprintLinks]{http://arxiv.org/abs/1412.8352}{{\ttfamily
  arXiv:1412.8352 [hep-ph]}}.

\bibitem{Chen:2015nha}
P.~Chen, C.-Y. Yao and G.-J. Ding, \emph{{Neutrino Mixing from CP Symmetry}},
  \MYhref[journalLinks]{http://dx.doi.org/10.1103/PhysRevD.92.073002}{Phys.
  Rev.
  }\MYhref[journalLinks]{http://dx.doi.org/10.1103/PhysRevD.92.073002}{\textbf{D92}
  (2015) 7 073002},
  \MYhref[eprintLinks]{http://arxiv.org/abs/1507.03419}{{\ttfamily
  arXiv:1507.03419 [hep-ph]}}.

\bibitem{Agashe:2014kda}
K.~Olive et~al. (Particle Data Group), \emph{{Review of Particle Physics}},
  \MYhref[journalLinks]{http://dx.doi.org/10.1088/1674-1137/38/9/090001}{Chin.Phys.
  }\MYhref[journalLinks]{http://dx.doi.org/10.1088/1674-1137/38/9/090001}{\textbf{C38}
  (2014) 090001}.

\bibitem{Rodejohann:2011vc}
W.~Rodejohann and J.~W.~F. Valle, \emph{{Symmetrical Parametrizations of the
  Lepton Mixing Matrix}}, Phys.Rev. \textbf{D84} (2011) 073011,
  \MYhref[eprintLinks]{http://arxiv.org/abs/1108.3484}{{\ttfamily
  arXiv:1108.3484 [hep-ph]}}.

\bibitem{Branco:1986gr}
G.~C. Branco, L.~Lavoura and M.~N. Rebelo, \emph{{Majorana Neutrinos and {CP}
  Violation in the Leptonic Sector}},
  \MYhref[journalLinks]{http://dx.doi.org/10.1016/0370-2693(86)90307-2}{Phys.
  Lett.
  }\MYhref[journalLinks]{http://dx.doi.org/10.1016/0370-2693(86)90307-2}{\textbf{B180}
  (1986) 264}.

\bibitem{Jenkins:2007ip}
E.~E. Jenkins and A.~V. Manohar, \emph{{Rephasing Invariants of Quark and
  Lepton Mixing Matrices}},
  \MYhref[journalLinks]{http://dx.doi.org/10.1016/j.nuclphysb.2007.09.031}{Nucl.
  Phys.
  }\MYhref[journalLinks]{http://dx.doi.org/10.1016/j.nuclphysb.2007.09.031}{\textbf{B792}
  (2008) 187--205},
  \MYhref[eprintLinks]{http://arxiv.org/abs/0706.4313}{{\ttfamily
  arXiv:0706.4313 [hep-ph]}}.

\bibitem{Branco:2011zb}
G.~C. Branco, R.~G. Felipe and F.~R. Joaquim, \emph{Leptonic cp violation},
  \MYhref[journalLinks]{http://dx.doi.org/10.1103/RevModPhys.84.515}{Rev. Mod.
  Phys.
  }\MYhref[journalLinks]{http://dx.doi.org/10.1103/RevModPhys.84.515}{\textbf{84}
  (2012) 515--565},
  \MYhref[eprintLinks]{http://arxiv.org/abs/1111.5332}{{\ttfamily
  arXiv:1111.5332 [hep-ph]}}.

\bibitem{Fritzsch:1999ee}
H.~Fritzsch and Z.-z. Xing, \emph{{Mass and flavor mixing schemes of quarks and
  leptons}},
  \MYhref[journalLinks]{http://dx.doi.org/10.1016/S0146-6410(00)00102-2}{Prog.
  Part. Nucl. Phys.
  }\MYhref[journalLinks]{http://dx.doi.org/10.1016/S0146-6410(00)00102-2}{\textbf{45}
  (2000) 1--81},
  \MYhref[eprintLinks]{http://arxiv.org/abs/hep-ph/9912358}{{\ttfamily
  arXiv:hep-ph/9912358 [hep-ph]}}.

\bibitem{Harrison:2002kp}
P.~F. Harrison and W.~G. Scott, \emph{Symmetries and generalisations of
  tri-bimaximal neutrino mixing}, Phys. Lett. \textbf{B535} (2002) 163--169,
  \MYhref[eprintLinks]{http://arxiv.org/abs/hep-ph/0203209}{{\ttfamily
  hep-ph/0203209}}.

\bibitem{Grimus:2003yn}
W.~Grimus and L.~Lavoura, \emph{{A non-standard CP transformation leading to
  maximal atmospheric neutrino mixing}}, Phys. Lett. \textbf{B579} (2004)
  113--122,
  \MYhref[eprintLinks]{http://arxiv.org/abs/hep-ph/0305309}{{\ttfamily
  hep-ph/0305309}}.

\bibitem{Farzan:2006vj}
Y.~Farzan and A.~{\relax Yu}. Smirnov, \emph{{Leptonic CP violation: Zero,
  maximal or between the two extremes}},
  \MYhref[journalLinks]{http://dx.doi.org/10.1088/1126-6708/2007/01/059}{JHEP
  }\MYhref[journalLinks]{http://dx.doi.org/10.1088/1126-6708/2007/01/059}{\textbf{01}
  (2007) 059},
  \MYhref[eprintLinks]{http://arxiv.org/abs/hep-ph/0610337}{{\ttfamily
  arXiv:hep-ph/0610337 [hep-ph]}}.

\bibitem{Forero:2014bxa}
D.~Forero, M.~Tortola and J.~Valle, \emph{{Neutrino oscillations refitted}},
  \MYhref[journalLinks]{http://dx.doi.org/10.1103/PhysRevD.90.093006}{Phys.Rev.
  }\MYhref[journalLinks]{http://dx.doi.org/10.1103/PhysRevD.90.093006}{\textbf{D90}
  (2014) 9 093006},
  \MYhref[eprintLinks]{http://arxiv.org/abs/1405.7540}{{\ttfamily
  arXiv:1405.7540 [hep-ph]}}.

\bibitem{Abe:2015awa}
K.~Abe et~al. (T2K), \emph{{Measurements of neutrino oscillation in appearance
  and disappearance channels by the T2K experiment with 6.6$\times$10$^{20}$ protons
  on target}},
  \MYhref[journalLinks]{http://dx.doi.org/10.1103/PhysRevD.91.072010}{Phys.
  Rev.
  }\MYhref[journalLinks]{http://dx.doi.org/10.1103/PhysRevD.91.072010}{\textbf{D91}
  (2015) 7 072010},
  \MYhref[eprintLinks]{http://arxiv.org/abs/1502.01550}{{\ttfamily
  arXiv:1502.01550 [hep-ex]}}.

\bibitem{Adams:2013qkq}
C.~Adams et~al. (LBNE), \emph{{The Long-Baseline Neutrino Experiment: Exploring
  Fundamental Symmetries of the Universe}}, --  (2013),
  \MYhref[eprintLinks]{http://arxiv.org/abs/1307.7335}{{\ttfamily
  arXiv:1307.7335 [hep-ex]}}.

\bibitem{::2013kaa}
S.~K. Agarwalla et~al. (LAGUNA-LBNO), \emph{{The mass-hierarchy and
  CP-violation discovery reach of the LBNO long-baseline neutrino experiment}},
  \MYhref[journalLinks]{http://dx.doi.org/10.1007/JHEP05(2014)094}{JHEP
  }\MYhref[journalLinks]{http://dx.doi.org/10.1007/JHEP05(2014)094}{\textbf{05}
  (2014) 094}, \MYhref[eprintLinks]{http://arxiv.org/abs/1312.6520}{{\ttfamily
  arXiv:1312.6520 [hep-ph]}}.

\bibitem{Kearns:2013lea}
H.-K.~W. Group (Hyper-Kamiokande Working Group), \emph{Hyper-kamiokande physics
  opportunities}, --  (2013),
  \MYhref[eprintLinks]{http://arxiv.org/abs/1309.0184}{{\ttfamily
  arXiv:1309.0184 [hep-ex]}}.

\bibitem{Nunokawa2008338}
H.~Nunokawa, S.~Parke and J.~W. Valle, \emph{CP violation and neutrino
  oscillations},
  \MYhref[journalLinks]{http://dx.doi.org/http://dx.doi.org/10.1016/j.ppnp.2007.10.001}{Progress
  in Particle and Nuclear Physics
  }\MYhref[journalLinks]{http://dx.doi.org/http://dx.doi.org/10.1016/j.ppnp.2007.10.001}{\textbf{60}
  (2008) 2 338 -- 402}, ISSN 0146-6410,
  \urlprefix\url{http://www.sciencedirect.com/science/article/pii/S014664100700083X}.

\bibitem{Schechter:1981bd}
J.~Schechter and J.~Valle, \emph{{Neutrinoless Double beta Decay in SU(2) x
  U(1) Theories}},
  \MYhref[journalLinks]{http://dx.doi.org/10.1103/PhysRevD.25.2951}{Phys.Rev.
  }\MYhref[journalLinks]{http://dx.doi.org/10.1103/PhysRevD.25.2951}{\textbf{D25}
  (1982) 2951}.

\bibitem{Duerr:2011zd}
M.~Duerr, M.~Lindner and A.~Merle, \emph{{On the Quantitative Impact of the
  Schechter-Valle Theorem}}, JHEP \textbf{1106} (2011) 091,
  \MYhref[eprintLinks]{http://arxiv.org/abs/1105.0901}{{\ttfamily
  arXiv:1105.0901 [hep-ph]}}.

\bibitem{Simkovic:2012hq}
F.~Simkovic, S.~M. Bilenky, A.~Faessler and T.~Gutsche, \emph{{Possibility of
  measuring the CP Majorana phases in \onbb decay}},
  \MYhref[journalLinks]{http://dx.doi.org/10.1103/PhysRevD.87.073002}{Phys.
  Rev.
  }\MYhref[journalLinks]{http://dx.doi.org/10.1103/PhysRevD.87.073002}{\textbf{D87}
  (2013) 7 073002},
  \MYhref[eprintLinks]{http://arxiv.org/abs/1210.1306}{{\ttfamily
  arXiv:1210.1306 [hep-ph]}}.

\bibitem{Avignone:2007fu}
F.~T. III~Avignone, S.~R. Elliott and J.~Engel, \emph{Double beta decay,
  majorana neutrinos, and neutrino mass},
  \MYhref[journalLinks]{http://dx.doi.org/10.1103/RevModPhys.80.481}{Rev. Mod.
  Phys.
  }\MYhref[journalLinks]{http://dx.doi.org/10.1103/RevModPhys.80.481}{\textbf{80}
  (2008) 481--516},
  \MYhref[eprintLinks]{http://arxiv.org/abs/0708.1033}{{\ttfamily
  arXiv:0708.1033 [nucl-ex]}}.

\bibitem{Barabash:2011fn}
A.~Barabash, \emph{{75 years of double beta decay: yesterday, today and
  tomorrow}}, --  (2011),
  \MYhref[eprintLinks]{http://arxiv.org/abs/1101.4502}{{\ttfamily
  arXiv:1101.4502 [nucl-ex]}}.

\bibitem{Auger:2012ar}
M.~Auger et~al. (EXO Collaboration), \emph{{Search for Neutrinoless Double-Beta
  Decay in $^{136}$Xe with EXO-200}}, Phys.Rev.Lett. \textbf{109} (2012)
  032505, \MYhref[eprintLinks]{http://arxiv.org/abs/1205.5608}{{\ttfamily
  arXiv:1205.5608 [hep-ex]}}.

\bibitem{Albert:2014awa}
J.~B. Albert et~al. (EXO-200), \emph{{Search for Majorana neutrinos with the
  first two years of EXO-200 data}},
  \MYhref[journalLinks]{http://dx.doi.org/10.1038/nature13432}{Nature
  }\MYhref[journalLinks]{http://dx.doi.org/10.1038/nature13432}{\textbf{510}
  (2014) 229--234},
  \MYhref[eprintLinks]{http://arxiv.org/abs/1402.6956}{{\ttfamily
  arXiv:1402.6956 [nucl-ex]}}.

\bibitem{Gando:2012zm}
A.~Gando et~al. (KamLAND-Zen Collaboration), \emph{Limit on Neutrinoless $\beta\beta$ Decay of $^{136}$Xe from the First Phase of KamLAND-Zen and Comparison with the Positive Claim in $^{76}$Ge}, (2012), \MYhref[eprintLinks]{http://arxiv.org/abs/1211.3863}{{\ttfamily
  arXiv:1211.3863 [hep-ex]}}.

\bibitem{Planck:2015xua}
P.~Ade et~al. (Planck), \emph{{Planck 2015 results. XIII. Cosmological
  parameters}}, arXiv:1502.01589  (2015),
  \MYhref[eprintLinks]{http://arxiv.org/abs/1502.01589}{{\ttfamily
  arXiv:1502.01589 [astro-ph.CO]}}.

\bibitem{Dorame:2011eb}
L.~Dorame et~al., \emph{{Constraining Neutrinoless Double Beta Decay}},
  Nucl.Phys. \textbf{B861} (2012) 259--270,
  \MYhref[eprintLinks]{http://arxiv.org/abs/1111.5614}{{\ttfamily
  arXiv:1111.5614 [hep-ph]}}.

\bibitem{Dorame:2012zv}
L.~Dorame et~al., \emph{{A new neutrino mass sum rule from inverse seesaw}},  Phys.Rev. \textbf{D86} (2012) 056001,
  \MYhref[eprintLinks]{http://arxiv.org/abs/1203.0155}{{\ttfamily
  arXiv:1203.0155 [hep-ph]}}.

\bibitem{King:2013hj}
S.~King, S.~Morisi, E.~Peinado and J.~W.~F. Valle, \emph{{Quark-Lepton Mass
  Relation in a Realistic A4 Extension of the Standard Model}}, Phys. Lett. B
  \textbf{724} (2013) 68--72,
  \MYhref[eprintLinks]{http://arxiv.org/abs/1301.7065}{{\ttfamily
  arXiv:1301.7065 [hep-ph]}}.

\bibitem{Bonilla:2014xla}
C.~Bonilla, S.~Morisi, E.~Peinado and J.~W.~F. Valle, \emph{{Relating quarks
  and leptons with the $T_7$ flavour group}},
  \MYhref[journalLinks]{http://dx.doi.org/10.1016/j.physletb.2015.01.017}{Phys.
  Lett.
  }\MYhref[journalLinks]{http://dx.doi.org/10.1016/j.physletb.2015.01.017}{\textbf{B742}
  (2015) 99--106},
  \MYhref[eprintLinks]{http://arxiv.org/abs/1411.4883}{{\ttfamily
  arXiv:1411.4883 [hep-ph]}}.

\bibitem{Sakharov:1967dj}
A.~Sakharov, \emph{{Violation of CP Invariance, C Asymmetry, and Baryon Asymmetry of the Universe}}, \MYhref[journalLinks]{http://dx.doi.org/10.1070/PU1991v034n05ABEH002497}{Pisma
  Zh.Eksp.Teor.Fiz.
  }\MYhref[journalLinks]{http://dx.doi.org/10.1070/PU1991v034n05ABEH002497}{\textbf{5}
  (1967) 32--35}.

\bibitem{Kuzmin:1985mm}
V.~A. Kuzmin, V.~A. Rubakov and M.~E. Shaposhnikov, \emph{On the Anomalous Electroweak Baryon Number Nonconservation in the Early Universe}, Phys. Lett.
  \textbf{B155} (1985) 36.

\bibitem{Fukugita:1986hr}
M.~Fukugita and T.~Yanagida, \emph{{Baryogenesis Without Grand Unification}},
  \MYhref[journalLinks]{http://dx.doi.org/10.1016/0370-2693(86)91126-3}{Phys.Lett.
  }\MYhref[journalLinks]{http://dx.doi.org/10.1016/0370-2693(86)91126-3}{\textbf{B174}
  (1986) 45}.

\bibitem{buchmuller:2005eh}
W.~Buchmuller, R.~D. Peccei and T.~Yanagida, \emph{Leptogenesis as the origin of matter}, Ann. Rev. Nucl. Part. Sci. \textbf{55} (2005) 311--355,
  \MYhref[eprintLinks]{http://arxiv.org/abs/hep-ph/0502169}{{\ttfamily
  hep-ph/0502169}}.

\bibitem{Davidson:2008bu}
S.~Davidson, E.~Nardi and Y.~Nir, \emph{{Leptogenesis}},
  \MYhref[journalLinks]{http://dx.doi.org/10.1016/j.physrep.2008.06.002}{Phys.Rept.
  }\MYhref[journalLinks]{http://dx.doi.org/10.1016/j.physrep.2008.06.002}{\textbf{466}
  (2008) 105--177},
  \MYhref[eprintLinks]{http://arxiv.org/abs/0802.2962}{{\ttfamily
  arXiv:0802.2962 [hep-ph]}}.

\bibitem{Chen:2016ptr}
P.~Chen, G.-J. Ding and S.~F. King, \emph{Leptogenesis and residual CP symmetry},
\MYhref[journalLinks]{http://dx.doi.org/10.1007/JHEP03(2016)206}{JHEP}\MYhref[journalLinks]{http://dx.doi.org/10.1007/JHEP03(2016)206}{\textbf{03}
  (2016) 206}, \MYhref[eprintLinks]{http://arxiv.org/abs/1602.03873}{{\ttfamily
  arXiv:1602.03873 [hep-ph]}}.

\bibitem{Casas:2001sr}
J.~A. Casas and A.~Ibarra, \emph{Oscillating neutrinos and $\mu\rightarrow e,\gamma$},
  Nucl. Phys. \textbf{B618} (2001) 171--204,
  \MYhref[eprintLinks]{http://arxiv.org/abs/hep-ph/0103065}{{\ttfamily
  hep-ph/0103065}}.

\bibitem{Covi:1996wh}
L.~Covi, E.~Roulet and F.~Vissani, \emph{{CP violating decays in leptogenesis scenarios}}, Phys. Lett. \textbf{B384} (1996) 169--174,
  \MYhref[eprintLinks]{http://arxiv.org/abs/hep-ph/9605319}{{\ttfamily
  hep-ph/9605319}}.

\bibitem{Endoh:2003mz}
T.~Endoh, T.~Morozumi and Z.~Xiong, \emph{Primordial lepton family asymmtries in seesaw model}, Prog. Theor. Phys. \textbf{111} (2004) 123, \MYhref[eprintLinks]{http://arxiv.org/abs/hep-ph/0308276}{{\ttfamily
  arXiv:hep-ph/0308276}}.

\bibitem{Abada:2006ea}
A.~Abada et~al., \emph{Flavour Matters in Leptogenesis}, JHEP \textbf{09} (2006) 010, \MYhref[eprintLinks]{http://arxiv.org/abs/hep-ph/0605281}{{\ttfamily
  arXiv:hep-ph/0605281}}.

\bibitem{Abada:2006fw}
A.~Abada et~al., \emph{Flavor issues in leptogenesis}, JCAP \textbf{0604} (2006) 004,  \MYhref[eprintLinks]{http://arxiv.org/abs/hep-ph/0601083}{{\ttfamily
  arXiv:hep-ph/0601083}}.

\bibitem{Fong:2013wr}
C.~S. Fong, E.~Nardi and A.~Riotto, \emph{Leptogenesis in the Universe}, Adv.\ High Energy Phys. \textbf{2012} (2012) 158303,
  \MYhref[eprintLinks]{http://arxiv.org/abs/1301.3062}{{\ttfamily
  arXiv:1301.3062 [hep-ph]}}.

\bibitem{Pascoli:2006ie}
S.~Pascoli, S.~T. Petcov and A.~Riotto, \emph{Connecting low energy leptonic CP-violation to leptogenesis}, Phys. Rev. \textbf{D75} (2007) 083511, \MYhref[eprintLinks]{http://arxiv.org/abs/hep-ph/0609125}{{\ttfamily
  arXiv:hep-ph/0609125}}.

\bibitem{Pascoli:2006ci}
S.~Pascoli, S.~Petcov and A.~Riotto, \emph{Leptogenesis and Low Energy CP Violation in Neutrino Physics}, Nucl. Phys. \textbf{B774} (2007) 1--52, \MYhref[eprintLinks]{http://arxiv.org/abs/hep-ph/0611338}{{\ttfamily
  arXiv:hep-ph/0611338}}.

\bibitem{Branco:2006ce}
G.~C. Branco, R.~Gonzalez~Felipe and F.~R. Joaquim, \emph{{A New bridge between leptonic CP violation and leptogenesis}}, Phys. Lett. \textbf{B645} (2007)
  432--436,
  \MYhref[eprintLinks]{http://arxiv.org/abs/hep-ph/0609297}{{\ttfamily
  arXiv:hep-ph/0609297}}.

\bibitem{Davidson:2007va}
S.~Davidson, J.~Garayoa, F.~Palorini and N.~Rius, \emph{Insensitivity of flavoured leptogenesis to low energy CP violation}, Phys.
  Rev. Lett. \textbf{99} (2007) 161801,
  \MYhref[eprintLinks]{http://arxiv.org/abs/0705.1503}{{\ttfamily
  arXiv:0705.1503 [hep-ph]}}.

\bibitem{Blanchet:2012bk}
S.~Blanchet and P.~D. Bari, \emph{The minimal scenario of leptogenesis}, New J.\ Phys.\ \textbf{14} (2012) 125012,
  \MYhref[eprintLinks]{http://arxiv.org/abs/1211.0512}{{\ttfamily
  arXiv:1211.0512 [hep-ph]}}.

\bibitem{Mohapatra:2015gwa}
R.~N. Mohapatra and C.~C. Nishi, \emph{{Implications of $\mu-\tau$ flavored CP symmetry of leptons}}, \MYhref[journalLinks]{http://dx.doi.org/10.1007/JHEP08(2015)092}{JHEP
  }\MYhref[journalLinks]{http://dx.doi.org/10.1007/JHEP08(2015)092}{\textbf{08}
  (2015) 092}, \MYhref[eprintLinks]{http://arxiv.org/abs/1506.06788}{{\ttfamily
  arXiv:1506.06788 [hep-ph]}}.

\bibitem{Ade:2015xua}
P.~A.~R. Ade et~al. (Planck), \emph{Planck 2015 results. XIII. Cosmological parameters}, \MYhref[eprintLinks]{http://arxiv.org/abs/1502.01589}{{\ttfamily
  arXiv:1502.01589 [astro-ph]}}.

\bibitem{Feruglio:2012cw}
F.~Feruglio, C.~Hagedorn and R.~Ziegler, \emph{Lepton Mixing Parameters from Discrete and CP Symmetries}, JHEP \textbf{1307} (2013) 027, \MYhref[eprintLinks]{http://arxiv.org/abs/1211.5560}{{\ttfamily
  arXiv:1211.5560 [hep-ph]}}.

\bibitem{Ding:2013hpa}
G.-J. Ding, S.~F. King, C.~Luhn and A.~J. Stuart, \emph{Spontaneous CP violation from vacuum alignment in $S_4$ models of leptons}, JHEP
  \textbf{1305} (2013) 084,
  \MYhref[eprintLinks]{http://arxiv.org/abs/1303.6180}{{\ttfamily
  arXiv:1303.6180 [hep-ph]}}.

\bibitem{Feruglio:2013hia}
F.~Feruglio, C.~Hagedorn and R.~Ziegler, \emph{{A realistic pattern of lepton mixing and masses from $S_4$ and CP}}, \MYhref[journalLinks]{http://dx.doi.org/10.1140/epjc/s10052-014-2753-2}{Eur.
  Phys. J.
  }\MYhref[journalLinks]{http://dx.doi.org/10.1140/epjc/s10052-014-2753-2}{\textbf{C74}
  (2014) 2753}, \MYhref[eprintLinks]{http://arxiv.org/abs/1303.7178}{{\ttfamily
  arXiv:1303.7178 [hep-ph]}}.

\bibitem{Ding:2013bpa}
G.-J. Ding, S.~F. King and A.~J. Stuart, \emph{Generalised CP and $A_4$ Family Symmetry}, JHEP \textbf{1312} (2013) 006,
  \MYhref[eprintLinks]{http://arxiv.org/abs/1307.4212}{{\ttfamily
  arXiv:1307.4212 [hep-ph]}}.

\bibitem{Li:2013jya}
C.-C. Li and G.-J. Ding, \emph{Generalised CP and trimaximal $TM_1$ lepton mixing in $S_4$ family symmetry}, Nucl.Phys. \textbf{B881} (2014) 206--232,
  \MYhref[eprintLinks]{http://arxiv.org/abs/1312.4401}{{\ttfamily
  arXiv:1312.4401 [hep-ph]}}.

\bibitem{Ding:2013nsa}
G.-J. Ding and Y.-L. Zhou, \emph{Predicting lepton flavor mixing from $\Delta$(48) and generalized $CP$ symmetries}, Chin.Phys. \textbf{C39} (2015) 021001, \MYhref[eprintLinks]{http://arxiv.org/abs/1312.5222}{{\ttfamily
  arXiv:1312.5222 [hep-ph]}}.

\bibitem{Ding:2014hva}
G.-J. Ding and Y.-L. Zhou, \emph{Lepton mixing parameters from $\Delta(48)$ family symmetry and generalised CP}, JHEP \textbf{1406} (2014) 023, \MYhref[eprintLinks]{http://arxiv.org/abs/1404.0592}{{\ttfamily
  arXiv:1404.0592 [hep-ph]}}.

\bibitem{Ding:2014ssa}
G.-J. Ding and S.~F. King, \emph{{Generalized $CP$ and $\Delta(96)$ family symmetry}}, \MYhref[journalLinks]{http://dx.doi.org/10.1103/PhysRevD.89.093020}{Phys.
  Rev.}\MYhref[journalLinks]{http://dx.doi.org/10.1103/PhysRevD.89.093020}{\textbf{D89}
  (2014) 9 093020},
  \MYhref[eprintLinks]{http://arxiv.org/abs/1403.5846}{{\ttfamily
  arXiv:1403.5846 [hep-ph]}}.

\bibitem{Li:2014eia}
C.-C. Li and G.-J. Ding, \emph{{Deviation from bimaximal mixing and leptonic CP phases in S$_{4}$ family symmetry and generalized CP}},
  \MYhref[journalLinks]{http://dx.doi.org/10.1007/JHEP08(2015)017}{JHEP
  }\MYhref[journalLinks]{http://dx.doi.org/10.1007/JHEP08(2015)017}{\textbf{08}
  (2015) 017}, \MYhref[eprintLinks]{http://arxiv.org/abs/1408.0785}{{\ttfamily
  arXiv:1408.0785 [hep-ph]}}.

\bibitem{Li:2015jxa}
C.-C. Li and G.-J. Ding, \emph{{Lepton Mixing in $A_5$ Family Symmetry and Generalized CP}}, \MYhref[journalLinks]{http://dx.doi.org/10.1007/JHEP05(2015)100}{JHEP}\MYhref[journalLinks]{http://dx.doi.org/10.1007/JHEP05(2015)100}{\textbf{05}
  (2015) 100}, \MYhref[eprintLinks]{http://arxiv.org/abs/1503.03711}{{\ttfamily
  arXiv:1503.03711 [hep-ph]}}.

\bibitem{DiIura:2015kfa}
A.~D. Iura, C.~Hagedorn and D.~Meloni, \emph{Lepton mixing from the interplay of the alternating group A$_{5}$ and CP}, JHEP \textbf{08} (2015) 037,
  \MYhref[eprintLinks]{http://arxiv.org/abs/1503.04140}{{\ttfamily
  arXiv:1503.04140 [hep-ph]}}.

\bibitem{Ballett:2015wia}
P.~Ballett, S.~Pascoli and J.~Turner, \emph{Mixing angle and phase correlations from A5 with generalized CP and their prospects for discovery}, Phys. Rev. \textbf{D92} (2015) 093008,
  \MYhref[eprintLinks]{http://arxiv.org/abs/1503.07543}{{\ttfamily
  arXiv:1503.07543 [hep-ph]}}.

\bibitem{Branco:2015gna}
G.~C. Branco, I.~de~Medeiros~Varzielas and S.~F. King, \emph{{Invariant
  approach to $\mathcal {CP}$ in unbroken $\Delta(27)$}},
  \MYhref[journalLinks]{http://dx.doi.org/10.1016/j.nuclphysb.2015.07.024}{Nucl.
  Phys.
  }\MYhref[journalLinks]{http://dx.doi.org/10.1016/j.nuclphysb.2015.07.024}{\textbf{B899}
  (2015) 14--36},
  \MYhref[eprintLinks]{http://arxiv.org/abs/1505.06165}{{\ttfamily
  arXiv:1505.06165 [hep-ph]}}; G.~C.~Branco, I.~de Medeiros Varzielas and S.~F.~King,
\emph{Invariant approach to CP in family symmetry models},
 \MYhref[journalLinks]{http://dx.doi.org/10.1103/PhysRevD.92.036007}{Phys.\ Rev.\ }\MYhref[journalLinks]{http://dx.doi.org/10.1103/PhysRevD.92.036007}{\textbf{D92}
  (2015)036007},
  \MYhref[eprintLinks]{http://arxiv.org/abs/1502.03105}{{\ttfamily
  arXiv:1502.03105 [hep-ph]}}.


\bibitem{Turner:2015uta}
J.~Turner, \emph{Predictions for leptonic mixing angle correlations and nontrivial Dirac CP violation from A$_5$ with generalized CP symmetry}, Phys. Rev. \textbf{D92} (2015) 116007,
  \MYhref[eprintLinks]{http://arxiv.org/abs/1507.06224}{{\ttfamily
  arXiv:1507.06224 [hep-ph]}}.

\bibitem{Hagedorn:2014wha}
C.~Hagedorn, A.~Meroni and E.~Molinaro, \emph{Lepton mixing from $\Delta$(3$n^2$) and $\Delta$(6$n^2$) and CP}, Nucl.Phys. \textbf{B891} (2015) 499--557,
  \MYhref[eprintLinks]{http://arxiv.org/abs/1408.7118}{{\ttfamily
  arXiv:1408.7118 [hep-ph]}}.

\bibitem{Ding:2014ora}
G.-J. Ding, S.~F. King and T.~Neder, \emph{{Generalised CP and $\Delta(6n^2)$ family symmetry in semi-direct models of leptons}},
  \MYhref[journalLinks]{http://dx.doi.org/10.1007/JHEP12(2014)007}{JHEP
  }\MYhref[journalLinks]{http://dx.doi.org/10.1007/JHEP12(2014)007}{\textbf{12}
  (2014) 007}, \MYhref[eprintLinks]{http://arxiv.org/abs/1409.8005}{{\ttfamily
  arXiv:1409.8005 [hep-ph]}}.

\bibitem{Ding:2015rwa}
G.-J. Ding and S.~F. King, \emph{Generalized CP and $\Delta (3n^2)$ Family Symmetry for Semi-Direct Predictions of the PMNS Matrix}, Phys. Rev. \textbf{D93} (2016) 025013, \MYhref[eprintLinks]{http://arxiv.org/abs/1510.03188}{{\ttfamily
  arXiv:1510.03188 [hep-ph]}}.

\bibitem{Li:2016ppt}
C.-C. Li, C.-Y. Yao and G.-J. Ding, \emph{Lepton Mixing Predictions from Infinite Group Series $D^{(1)}_{9n, 3n}$ with Generalized CP}, \MYhref[eprintLinks]{http://arxiv.org/abs/1601.06393}{{\ttfamily
  arXiv:1601.06393 [hep-ph]}}.

\end{thebibliography}

\end{document}